\pdfoutput=1
\documentclass[12pt,a4paper]{article}
\usepackage{ifthen} 
\newboolean{pdflatex}
\setboolean{pdflatex}{true} 
\newboolean{articletitles}
\setboolean{articletitles}{true} 
\newboolean{uprightparticles}
\setboolean{uprightparticles}{false} 
\newboolean{inbibliography}
\setboolean{inbibliography}{false} 
\usepackage{booktabs}
\usepackage{multirow}
\usepackage{diagbox}
\usepackage[section]{placeins}

\usepackage{microtype}
\usepackage{lineno}  
\usepackage{xspace} 
\usepackage{caption} 

\usepackage{graphicx}  
\usepackage{color}
\usepackage{colortbl}
\graphicspath{{./figs/}} 

\usepackage{amsmath} 
\usepackage{amssymb}
\usepackage{amsfonts}
\usepackage{upgreek} 

\newcommand*\patchAmsMathEnvironmentForLineno[1]{%
\expandafter\let\csname old#1\expandafter\endcsname\csname #1\endcsname
\expandafter\let\csname oldend#1\expandafter\endcsname\csname
end#1\endcsname
 \renewenvironment{#1}%
   {\linenomath\csname old#1\endcsname}%
   {\csname oldend#1\endcsname\endlinenomath}%
}
\newcommand*\patchBothAmsMathEnvironmentsForLineno[1]{%
  \patchAmsMathEnvironmentForLineno{#1}%
  \patchAmsMathEnvironmentForLineno{#1*}%
}
\AtBeginDocument{%
\patchBothAmsMathEnvironmentsForLineno{equation}%
\patchBothAmsMathEnvironmentsForLineno{align}%
\patchBothAmsMathEnvironmentsForLineno{flalign}%
\patchBothAmsMathEnvironmentsForLineno{alignat}%
\patchBothAmsMathEnvironmentsForLineno{gather}%
\patchBothAmsMathEnvironmentsForLineno{multline}%
\patchBothAmsMathEnvironmentsForLineno{eqnarray}%
}

\usepackage{hyperref}    
\usepackage[all]{hypcap} 

\def\lhcb {\mbox{LHCb}\xspace}
\def\atlas  {\mbox{ATLAS}\xspace}
\def\cms    {\mbox{CMS}\xspace}
\def\alice  {\mbox{ALICE}\xspace}

\def\cdf    {\mbox{CDF}\xspace}

\def\lhc    {\mbox{LHC}\xspace}

\def\MagUp {\mbox{\em Mag\kern -0.05em Up}\xspace}

\ifthenelse{\boolean{uprightparticles}}%
{

 \def\Pmu         {\ensuremath{\upmu}\xspace}

 \def\Ppsi        {\ensuremath{\uppsi}\xspace}

 \def\PDelta      {\ensuremath{\Delta}\xspace}                 
 \def\PXi      {\ensuremath{\Xi}\xspace}                 
 \def\PLambda      {\ensuremath{\Lambda}\xspace}                 
 \def\PSigma      {\ensuremath{\Sigma}\xspace}                 
 \def\POmega      {\ensuremath{\Omega}\xspace}                 
 \def\PUpsilon      {\ensuremath{\Upsilon}\xspace}                 
 
 \def\PB      {\ensuremath{\mathrm{B}}\xspace}                 
                  
 \def\PD      {\ensuremath{\mathrm{D}}\xspace}

 \def\PJ      {\ensuremath{\mathrm{J}}\xspace}                 
 \def\PK      {\ensuremath{\mathrm{K}}\xspace}

 \def\Pb      {\ensuremath{\mathrm{b}}\xspace}                 
 \def\Pc      {\ensuremath{\mathrm{c}}\xspace}

 \def\Pi      {\ensuremath{\mathrm{i}}\xspace}

}
{

 \def\Pmu         {\ensuremath{\mu}\xspace}

 \def\Ppsi        {\ensuremath{\psi}\xspace}                 
                  
 \mathchardef\PDelta="7101
 \mathchardef\PXi="7104
 \mathchardef\PLambda="7103
 \mathchardef\PSigma="7106
 \mathchardef\POmega="710A
 \mathchardef\PUpsilon="7107
                  
 \def\PB      {\ensuremath{B}\xspace}                 
                  
 \def\PD      {\ensuremath{D}\xspace}

 \def\PJ      {\ensuremath{J}\xspace}                 
 \def\PK      {\ensuremath{K}\xspace}

 \def\Pb      {\ensuremath{b}\xspace}                 
 \def\Pc      {\ensuremath{c}\xspace}

 \def\Pi      {\ensuremath{i}\xspace}

}

\makeatletter
\ifcase \@ptsize \relax
  \newcommand{\miniscule}{\@setfontsize\miniscule{4}{5}}
\or
  \newcommand{\miniscule}{\@setfontsize\miniscule{5}{6}}
\or
  \newcommand{\miniscule}{\@setfontsize\miniscule{5}{6}}
\fi
\makeatother

\DeclareRobustCommand{\optbar}[1]{\shortstack{{\miniscule (\rule[.5ex]{1.25em}{.18mm})}
  \\ [-.7ex] $#1$}}

\def\mup        {{\ensuremath{\Pmu^+}}\xspace}
\def\mun        {{\ensuremath{\Pmu^-}}\xspace} 
\def\mumu       {{\ensuremath{\Pmu^+\Pmu^-}}\xspace}

\def\cquark    {{\ensuremath{\Pc}}\xspace}

\def\bquark    {{\ensuremath{\Pb}}\xspace}
\def\bquarkbar {{\ensuremath{\overline \bquark}}\xspace}
\def\bbbar     {{\ensuremath{\bquark\bquarkbar}}\xspace}

  \def\Kbar    {{\kern 0.2em\overline{\kern -0.2em \PK}{}}\xspace}

\def\KorKbar    {\kern 0.18em\optbar{\kern -0.18em K}{}\xspace}

  \def\Dbar    {{\kern 0.2em\overline{\kern -0.2em \PD}{}}\xspace}

\def\DorDbar    {\kern 0.18em\optbar{\kern -0.18em D}{}\xspace}

\def\Bbar    {{\ensuremath{\kern 0.18em\overline{\kern -0.18em \PB}{}}}\xspace}

\def\BorBbar    {\kern 0.18em\optbar{\kern -0.18em B}{}\xspace}

\def\jpsi     {{\ensuremath{{\PJ\mskip -3mu/\mskip -2mu\Ppsi\mskip 2mu}}}\xspace}

  \def\Y#1S{\ensuremath{\PUpsilon{(#1S)}}\xspace}

\def\Lbar        {{\ensuremath{\kern 0.1em\overline{\kern -0.1em\PLambda}}}\xspace}
\def\LorLbar    {\kern 0.18em\optbar{\kern -0.18em \PLambda}{}\xspace}

\def\BF         {{\ensuremath{\cal B}}\xspace}

\def\BR         {\BF}
\newcommand{\decay}[2]{\ensuremath{#1\!\to #2}\xspace}         

\def\to                 {\ensuremath{\rightarrow}\xspace}

\newcommand{\tpm}{$ & $\,\pm\,$ & $}

\newcommand{\etot}{{\ensuremath{\varepsilon_{\rm tot}}}\xspace}

\def\AT#1     {\ensuremath{A_{\mathrm{T}}^{#1}}\xspace}           

\def\C#1      {\ensuremath{\mathcal{C}_{#1}}\xspace}                       
\def\Cp#1     {\ensuremath{\mathcal{C}_{#1}^{'}}\xspace}                    
\def\Ceff#1   {\ensuremath{\mathcal{C}_{#1}^{\mathrm{(eff)}}}\xspace}        
\def\Cpeff#1  {\ensuremath{\mathcal{C}_{#1}^{'\mathrm{(eff)}}}\xspace}       
\def\Ope#1    {\ensuremath{\mathcal{O}_{#1}}\xspace}                       
\def\Opep#1   {\ensuremath{\mathcal{O}_{#1}^{'}}\xspace}                    

\newcommand{\tev}{\ifthenelse{\boolean{inbibliography}}{\ensuremath{~T\kern -0.05em eV}\xspace}{\ensuremath{\mathrm{\,Te\kern -0.1em V}}}\xspace}
\newcommand{\gev}{\ensuremath{\mathrm{\,Ge\kern -0.1em V}}\xspace}
\newcommand{\mev}{\ensuremath{\mathrm{\,Me\kern -0.1em V}}\xspace}
\newcommand{\kev}{\ensuremath{\mathrm{\,ke\kern -0.1em V}}\xspace}
\newcommand{\ev}{\ensuremath{\mathrm{\,e\kern -0.1em V}}\xspace}
\newcommand{\gevc}{\ensuremath{{\mathrm{\,Ge\kern -0.1em V\!/}c}}\xspace}
\newcommand{\mevc}{\ensuremath{{\mathrm{\,Me\kern -0.1em V\!/}c}}\xspace}
\newcommand{\gevcc}{\ensuremath{{\mathrm{\,Ge\kern -0.1em V\!/}c^2}}\xspace}
\newcommand{\gevgevcccc}{\ensuremath{{\mathrm{\,Ge\kern -0.1em V^2\!/}c^4}}\xspace}
\newcommand{\mevcc}{\ensuremath{{\mathrm{\,Me\kern -0.1em V\!/}c^2}}\xspace}

\def\mub{\ensuremath{{\rm \,\upmu b}}\xspace}

\def\invpb {\ensuremath{\mbox{\,pb}^{-1}}\xspace}

\def\ps   {\ensuremath{{\rm \,ps}}\xspace}

\def\deriv {\ensuremath{\mathrm{d}}}

\def\gsim{{~\raise.15em\hbox{$>$}\kern-.85em
          \lower.35em\hbox{$\sim$}~}\xspace}
\def\lsim{{~\raise.15em\hbox{$<$}\kern-.85em
          \lower.35em\hbox{$\sim$}~}\xspace}

\def\sPlot{\mbox{\em sPlot}\xspace}

\def\ptot       {\mbox{$p$}\xspace}
\def\pt         {\mbox{$p_{\rm T}$}\xspace}

\def\evtgen     {\mbox{\textsc{EvtGen}}\xspace}

\def\geant      {\mbox{\textsc{Geant4}}\xspace}

\def\photos     {\mbox{\textsc{Photos}}\xspace}

\def\pythia     {\mbox{\textsc{Pythia}}\xspace}

\def\tell1  {TELL1\xspace}
\def\ukl1   {UKL1\xspace}

\usepackage{cite} 
\usepackage{mciteplus}

\usepackage{longtable} 
\usepackage{enumitem}

\def\jpsifromb       {{\ensuremath{\jpsi\text{-from-}b}}\xspace}

\def\jpsimumu  {\decay{\jpsi}{\mumu}}
\newcommand{\TEV}{{\ensuremath{\mathrm{\,Te\kern -0.1em V}}}\xspace}
\newcommand{\GEV}{{\ensuremath{\mathrm{\,Ge\kern -0.1em V}}}\xspace}

\begin{document}
\renewcommand{\thefootnote}{\fnsymbol{footnote}}
\setcounter{footnote}{1}
\begin{titlepage}
\pagenumbering{roman}

\vspace*{-1.5cm}
\centerline{\large EUROPEAN ORGANIZATION FOR NUCLEAR RESEARCH (CERN)}
\vspace*{1.5cm}
\noindent
\begin{tabular*}{\linewidth}{lc@{\extracolsep{\fill}}r@{\extracolsep{0pt}}}
\ifthenelse{\boolean{pdflatex}}
{\vspace*{-2.7cm}\mbox{\!\!\!\includegraphics[width=.14\textwidth]{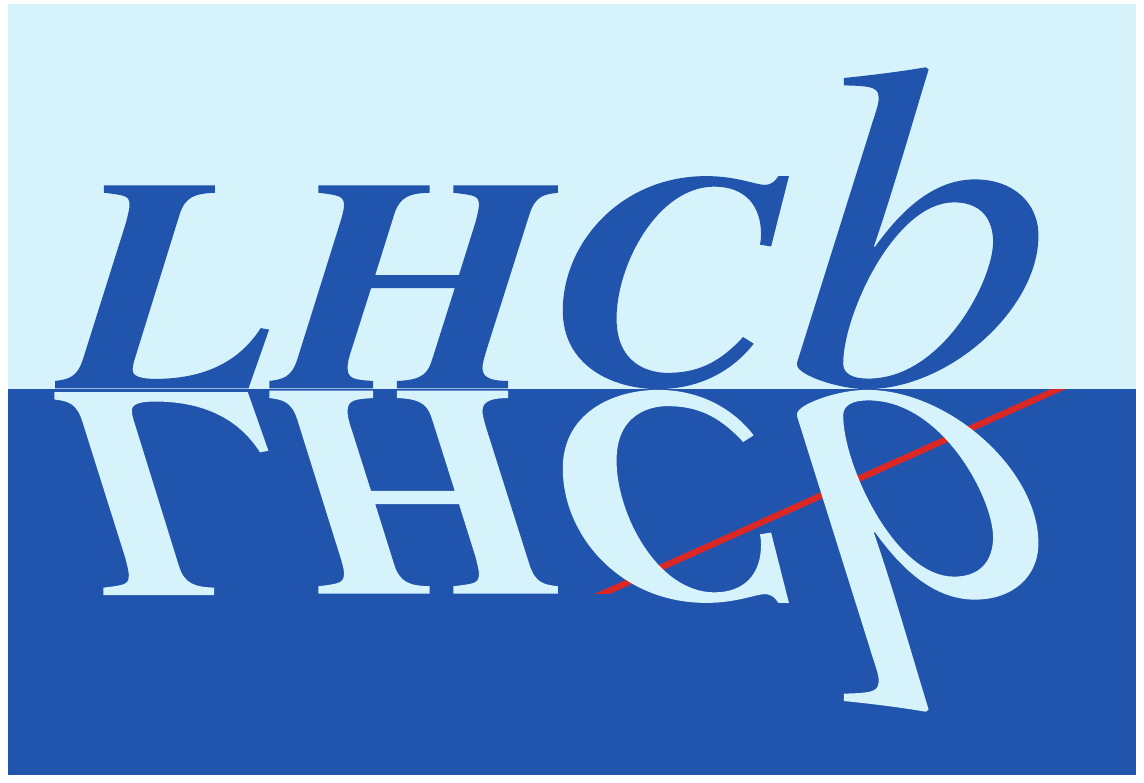}} & &}%
{\vspace*{-1.2cm}\mbox{\!\!\!\includegraphics[width=.12\textwidth]{lhcb-logo.eps}} & &}%
\\
 & & CERN-PH-EP-2015-222\\  
 & & LHCb-PAPER-2015-037\\  
 & & \\
\end{tabular*}
\vspace*{2.1cm}

{\bf\boldmath\huge
\begin{center}
Measurement of forward \jpsi production cross-sections in $pp$ collisions at $\sqrt{s}=13\tev$
\end{center}
}

\vspace*{1.0cm}

\begin{center}
The LHCb collaboration\footnote{Authors are listed at the end of this paper.}
\end{center}

\vspace{\fill}

\begin{abstract}
\noindent
The production of \jpsi mesons in proton-proton collisions at a centre-of-mass energy of $\sqrt{s}=13\tev$
is studied with the \lhcb detector. 
Cross-section measurements are performed as a function of the transverse momentum \pt and the rapidity $y$ of the
\jpsi meson in the region $\pt<14\gevc$ and $2.0<y<4.5$, for both prompt \jpsi mesons and \jpsi mesons from $b$-hadron
decays. The production cross-sections integrated over the kinematic coverage are $15.03\pm 0.03\pm 0.94\mub$ for
prompt \jpsi and $2.25\pm 0.01\pm 0.14\mub$ for \jpsi from $b$-hadron decays, assuming zero polarization of the \jpsi meson. The first uncertainties are statistical
and the second systematic. The cross-section reported for \jpsi mesons from $b$-hadron decays is used to extrapolate to a total $b\bar{b}$ cross-section.
The ratios of the cross-sections with respect to $\sqrt{s}=8\tev$ are also determined.
\end{abstract}

\vspace*{1.0cm}
\begin{center}
    Published in JHEP 10 (2016) 172 
\end{center}

\vspace{\fill}

{\footnotesize 
\centerline{\copyright~CERN on behalf of the \lhcb collaboration, licence \href{http://creativecommons.org/licenses/by/4.0/}{CC-BY-4.0}.}}
\vspace*{2mm}

\end{titlepage}


\newpage
\setcounter{page}{2}
\mbox{~}

\cleardoublepage

\renewcommand{\thefootnote}{\arabic{footnote}}
\setcounter{footnote}{0}
\pagestyle{plain} 
\setcounter{page}{1}
\pagenumbering{arabic}

\section{Introduction}
\label{sec:introduction}
The study of heavy quarkonium production in $pp$ collisions provides important information on both the perturbative and
the non-perturbative regimes of quantum chromodynamics (QCD).
Heavy quarkonium production can be described in two stages. The first is the short-distance production
of a heavy quark pair, $Q\overline{Q}$, which can be described perturbatively and the second is the non-perturbative hadronisation of the heavy quark pair into
quarkonium state, such as the $\jpsi$ meson.
The non-perturbative part cannot yet be determined reliably, and must be determined using experimental results.
After almost forty years of theoretical and experimental efforts, the hadronic production of quarkonia is still not fully understood.
In the colour singlet model (CSM)~\cite{Carlson:1976cd,Donnachie:1976ue,Ellis:1976fj,Fritzsch:1977ay,Gluck:1977zm,Chang:1979nn,Baier:1981uk},
the intermediate $Q\overline{Q}$ state is assumed to be colourless, and has the same $J^{PC}$ quantum numbers as the final-state quarkonium. 
In the non-relativistic QCD (NRQCD) approach~\cite{Bodwin:1994jh,Cho:1995vh,Cho:1995ce},
all viable colour and spin-parity quantum numbers are allowed for the intermediate $Q\overline{Q}$ state, and each
configuration is assigned a probability to transform into the specific quarkonium state.
The transition probabilities are described by universal long-distance matrix elements (LDME) determined from experimental data.

In $pp$ collisions, \jpsi mesons can be produced directly from hard collisions of partons, through the feed-down
of excited charmonium states, or via decays of $b$-flavoured hadrons. The first two sources are collectively referred to as prompt \jpsi
production, 
while the third process is referred to in the following as ``\jpsi-from-$b$''.

The $\jpsi$ differential production cross-section has been measured with \lhc data in $pp$ collisions
at centre-of-mass energies of $2.76\tev$~\cite{Abelev:2012kr,LHCb-PAPER-2012-039},
$7\tev$~\cite{LHCb-PAPER-2011-003,Khachatryan:2010yr,Aad:2011sp,Aamodt:2011gj,Chatrchyan:2011kc},
and $8\tev$~\cite{LHCb-PAPER-2013-016}.
Next-to-leading order CSM calculations~\cite{Campbell:2007ws,Lansberg:2011hi} give a better description of the experimental data than leading order (LO) calculations at high \jpsi transverse momentum ($\pt\!/c>M(\jpsi)$).
However, CSM calculations still underestimate the prompt \jpsi production cross-section. 
On the other hand, NRQCD calculations with LDME determined from CDF data~\cite{Abe:1992ww} describe well the \pt
dependence of the prompt \jpsi cross-section at both the Tevatron~\cite{Cacciari:1995yt,Braaten:1994vv} and the \lhc
experiments~\cite{Ma:2010jj,Gong:2008ft,Butenschoen:2010rq}.
The measured \jpsi-from-$b$ production cross-section 
and its dependence on \pt at the \lhc are in good agreement with predictions from fixed order plus next-to-leading logarithms (FONLL)~\cite{Cacciari:1998it} calculations.

The \jpsi polarisation was measured by the \alice~\cite{Abelev:2011md},
\cms~\cite{Chatrchyan:2013cla} and \lhcb~\cite{LHCb-PAPER-2013-008} collaborations in $pp$ collisions at $\sqrt{s}=7\tev$.
The measurements by \alice were performed for inclusive \jpsi meson production,
while \cms and \lhcb disentangled prompt \jpsi and \jpsi-from-$b$ mesons. Next-to-leading order CSM calculations predict
a large longitudinal polarisation of the \jpsi meson~\cite{Gong:2008sn}, while NRQCD calculations predict a large \jpsi
transverse polarisation at high \pt~\cite{Beneke:1995yb,Chao:2012iv,Gong:2012ug,Butenschoen:2012px,
    Campbell:2007ws,Lansberg:2008gk}. Neither prediction
is supported by experimental results~\cite{Abulencia:2007us,Abazov:2008aa,Abelev:2011md,Chatrchyan:2013cla,Chatrchyan:2012woa,LHCb-PAPER-2013-008,LHCb-PAPER-2013-067}.

This paper reports \jpsi cross-section measurements in $pp$ collisions at $\sqrt{s} = 13\tev$ in the \jpsi kinematic
range $\pt<14\gevc$ and $2.0<y<4.5$, using \jpsi decaying to $\mup\mun$ final states. Under the assumption of zero \jpsi
polarisation (see Sec.~\ref{sec:systematics} for detailed discussion), the following quantities are measured: the double
differential cross-sections as a function of \pt and $y$; the integrated production cross-sections for prompt \jpsi and
\jpsi-from-$b$; the $\bbbar$ production cross-section; and the ratio of cross-sections with respect to the \jpsi
cross-sections in $pp$ collisions at $\sqrt{s}=8~\tev$ previously measured by \lhcb~\cite{LHCb-PAPER-2013-016}.

\section{The \lhcb detector and data set}
\label{sec:Detector}
The data used in this analysis come from $pp$ collisions at $\sqrt{s} = 13\tev$, collected by the \lhcb detector in July
2015, with an average of 1.1 visible interactions per bunch crossing and correspond to an
integrated luminosity of $3.05\pm 0.12\invpb$. 
The \lhcb detector~\cite{Alves:2008zz,LHCb-DP-2014-002} is a single-arm forward
spectrometer covering the \mbox{pseudorapidity} range $2<\eta <5$,
designed for the study of particles containing \bquark or \cquark
quarks. The detector includes a high-precision tracking system
consisting of a silicon-strip vertex detector surrounding the $pp$
interaction region, a large-area silicon-strip detector located
upstream of a dipole magnet with a bending power of about
$4{\rm\,Tm}$, and three stations of silicon-strip detectors and straw
drift tubes placed downstream of the magnet.
The tracking system provides a measurement of momentum, \ptot, of charged particles with
a relative uncertainty that varies from 0.5\% at low momentum to 1.0\% at 200\gevc.
Different types of charged hadrons are distinguished using information
from two ring-imaging Cherenkov detectors. 
Photons, electrons and hadrons are identified by a calorimeter system consisting of
scintillating-pad and preshower detectors, an electromagnetic
calorimeter and a hadronic calorimeter. Muons are identified by a
system composed of alternating layers of iron and multiwire
proportional chambers~\cite{LHCb-DP-2012-002}.

The online event selection
consists of a hardware stage, based on information from the calorimeter and muon
systems, followed by a software stage, which performs \jpsi candidate
reconstruction. 
The hardware trigger selects events with at least one muon candidate with transverse momentum $\pt>0.9\gevc$. 
In the first stage of the software trigger, two muon tracks with $\pt>500\mevc$ are
required to form a \jpsi candidate with invariant mass $M(\mumu)>2.7\gevcc$. 
In the second stage, 
\jpsi candidates with good vertex fit quality and invariant mass within $150\mevcc$ of the known
value~\cite{PDG2014} are selected. 

This analysis benefits from a new scheme for the \lhcb software trigger introduced for \lhc Run 2.
Alignment and calibration is performed in near real-time ~\cite{Dujany:2017839} and updated constants are made available for the trigger.
The same alignment and calibration information is propagated to the offline reconstruction, to ensure consistent and high-quality
particle identification information for the trigger and offline. 
The larger timing budget available in the trigger with respect to \lhc Run 1 also results in the convergence of the online
and offline track reconstruction, such that offline performance is achieved in the trigger.
The identical performance of the online and offline 
reconstruction achieved in this way offers the opportunity to perform physics analyses directly using candidates
reconstructed in the trigger~\cite{LHCb-DP-2012-004}. The storage of only the triggered candidates enables a reduction
in the event size by an order of magnitude. The analysis described in this paper uses the online reconstruction for the
first time in \lhcb, and is checked the against the standard offline reconstruction chain. 

Simulated samples are used to evaluate the \jpsi detection efficiency. In the simulation, $pp$ collisions are generated using
\pythia 6~\cite{pythia6.4} with a specific \lhcb
configuration~\cite{LHCb-PROC-2010-056}. Decays of hadronic particles
are described by \evtgen~\cite{Lange:2001uf}, in which final-state
radiation is generated using \photos~\cite{Golonka:2005pn}. The prompt charmonium production is simulated in \pythia with contributions from both the leading order colour-singlet and colour-octet contributions~\cite{LHCb-PROC-2010-056,Bargiotti:2007zz}, and the charmonium is generated unpolarised.
The interaction of the generated particles with the detector, and its response,
are simulated using the \geant toolkit~\cite{Allison:2006ve, *Agostinelli:2002hh} as described in
Ref.~\cite{LHCb-PROC-2011-006}.  

\section{Selection of \jpsi candidates}
\label{sec:selection}
The \jpsi candidates are selected in the second step of the software trigger. Each event is required to have at
least one primary vertex (PV) reconstructed from at least four tracks found by the vertex detector. For events with
multiple PVs, the PV which has the smallest $\chi^2_\mathrm{IP}$ with respect to the $\jpsi$ candidate is chosen. The
$\chi^2_\mathrm{IP}$ is defined as the change of the primary vertex fit quality when the $\jpsi$ meson is 
excluded from the PV fit.
Each identified muon track is required to have $\pt>0.7\gevc$,
 $p>3\gevc$, and to have a good quality track fit. 
The muon tracks of the \jpsi candidate must form a good quality two-track vertex.
Duplicate tracks created by the reconstruction are suppressed to the level of  $0.5 \times 10^{-3}$.

The reconstructed vertex of the \jpsi mesons originating from $b$-hadron decays tends to be separated from the PVs,
and thus these can be distinguished from prompt \jpsi mesons by exploiting the pseudo decay time defined as
\begin{equation}
t_z=\frac{\left(z_{\jpsi}-z_\mathrm{PV}\right)\times M_{\jpsi}}{p_z},
\end{equation}
where $z_\jpsi-z_\mathrm{PV}$\/ is the distance along the beam axis between the \jpsi decay vertex and the PV, $p_z$\/ is the $z$-component of the \jpsi\ momentum, 
and $M_\jpsi$\/ the known \jpsi\ mass~\cite{PDG2014}. 

The \jpsi candidates with $\left|t_z\right|<10\ps$,
corresponding to less than 7 times the $b$-hadron lifetime, are selected for the fits to the $t_z$ distribution. To further
select good \jpsi candidates, the uncertainty on $t_z$, which is propagated from the uncertainties provided by the track reconstruction, 
is required to be less than $0.3\ps$. 

\section{Cross-section determination}
\label{sec:fit}
The double differential \jpsi\ production cross-section 
in each kinematic bin of $\pt$\ and $y$\ 
is defined as 
\begin{equation}
  \frac{\deriv^2\sigma}{\deriv y\deriv \pt} 
  = \frac{N(\jpsi\to\mup\mun)} {L_{\text{int}}\times\etot\times\BR(\jpsi\to\mumu)\times\Delta y \times \Delta
     \pt},\label{eqn:CrossSection}
\end{equation}
where $N(\jpsi\to\mumu)$ 
is the yield of prompt \jpsi\ or \jpsi-from-$b$ signal mesons,
$\etot$ is the total detection efficiency in the given kinematic bin, $L_{\text{int}}$ is the integrated luminosity, 
$\BR(\jpsi\to\mumu)=(5.961\pm0.033)\%$~\cite{PDG2014} is the branching ratio of the decay $\jpsi\to\mumu$ and
$\Delta\pt=1\gevc$ and $\Delta y=0.5$ are the bin widths. The measurements are restricted to $\pt<14\gevc$ due to limited data at higher transverse momenta.

The absolute luminosity is determined from the beam profiles and beam currents. The beam profiles and their overlap
integral are measured using a beam-gas imaging method~\cite{Barschel:1693671,LHCb-PAPER-2014-047}, where neon is injected into the beam vacuum around the
interaction point. The beam currents are measured by LHC instruments, which determine the bunch population fractions and
the total beam intensity. Furthermore, information on beam-gas interactions not originating from nominally filled
bunch slots is used to determine the charge fraction not participating in bunch collisions. The integrated luminosity for
this analysis is calibrated using the number of visible $pp$ interactions measured during the beam-gas imaging runs and
during the runs used for this analysis. From this procedure, the integrated luminosity is measured to be $3.05 \pm 0.12 \invpb$.

The efficiency $\etot$ is determined as the product of the reconstruction and selection efficiencies, muon identification efficiency and trigger efficiency, and is calculated using simulated samples in each $(\pt,y)$ bin, independently
for prompt \jpsi and \jpsi-from-$b$. The track reconstruction and the muon identification efficiency
are corrected using data-driven techniques,
while the trigger efficiencies are also validated using data, as explained in Sec.~\ref{sec:systematics}.

The yield of \jpsi signal events, both from prompt \jpsi and \jpsi-from-$b$, is determined from a two-dimensional 
unbinned maximum likelihood fit to the invariant mass and pseudo decay time of the candidates, performed independently
for each $(\pt,y)$ bin. 
The invariant mass distribution of the signal is described by the sum of two Crystal Ball (CB) functions~\cite{Skwarnicki:1986xj} with a common mean value and different widths.
The parameters of the power law tails, the relative fractions and the difference between the
widths of the two CB functions are fixed to values obtained from the simulation, leaving the mean and one width of the CB function as free parameters. 
The combinatorial background is described by an exponential distribution.

The fraction of \jpsi-from-$b$ candidates, $F_b$, is determined from the fit to the $t_z$ distribution. The $t_z$ distribution of
prompt \jpsi is described by a Dirac $\delta$ function at $t_z=0$, and that of \jpsi-from-$b$\ 
by an exponential decay function, which are both convolved with a double-Gaussian resolution function.
A \jpsi candidate can also be associated to a wrong PV,
 resulting in a long tail component in the $t_z$ distribution. The fraction of the tail is below 0.5\% of all events. Its shape is modelled from data by calculating $t_z$ with 
the \jpsi candidate from a given event and the closest PV in the next event of the sample. 
The background $t_z$
distribution is parametrised with an empirical function based on the observed shape of the $t_z$ distribution in the
\jpsi mass sidebands. The background comes from muons of semileptonic $b$- and $c$-hadron decays and from pions and kaons decaying in-flight. It is parametrised as the sum of a Dirac $\delta$ function and five
exponential functions, three for positive $t_z$ and two for negative $t_z$. 
The function is convolved with a double-Gaussian resolution function similar to that used for the signal, but with
different parameters. All parameters of the background $t_z$ distribution are fixed to values determined from 
the \jpsi mass sidebands independently in each $(\pt,y)$ bin.

\begin{figure}[!tbp]
\centering
\begin{minipage}[t]{0.49\textwidth}
\centering
\includegraphics[width=1.0\textwidth]{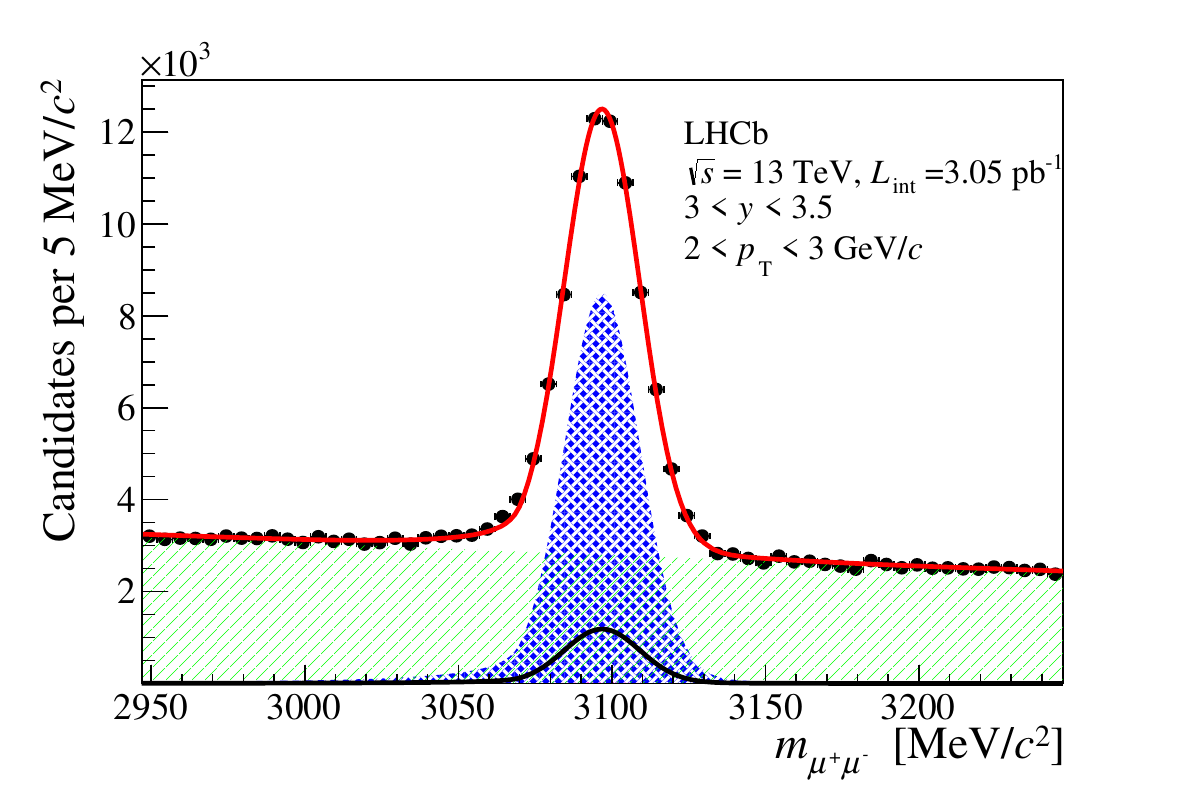}
\end{minipage}
\begin{minipage}[t]{0.49\textwidth}
\centering
\includegraphics[width=1.0\textwidth]{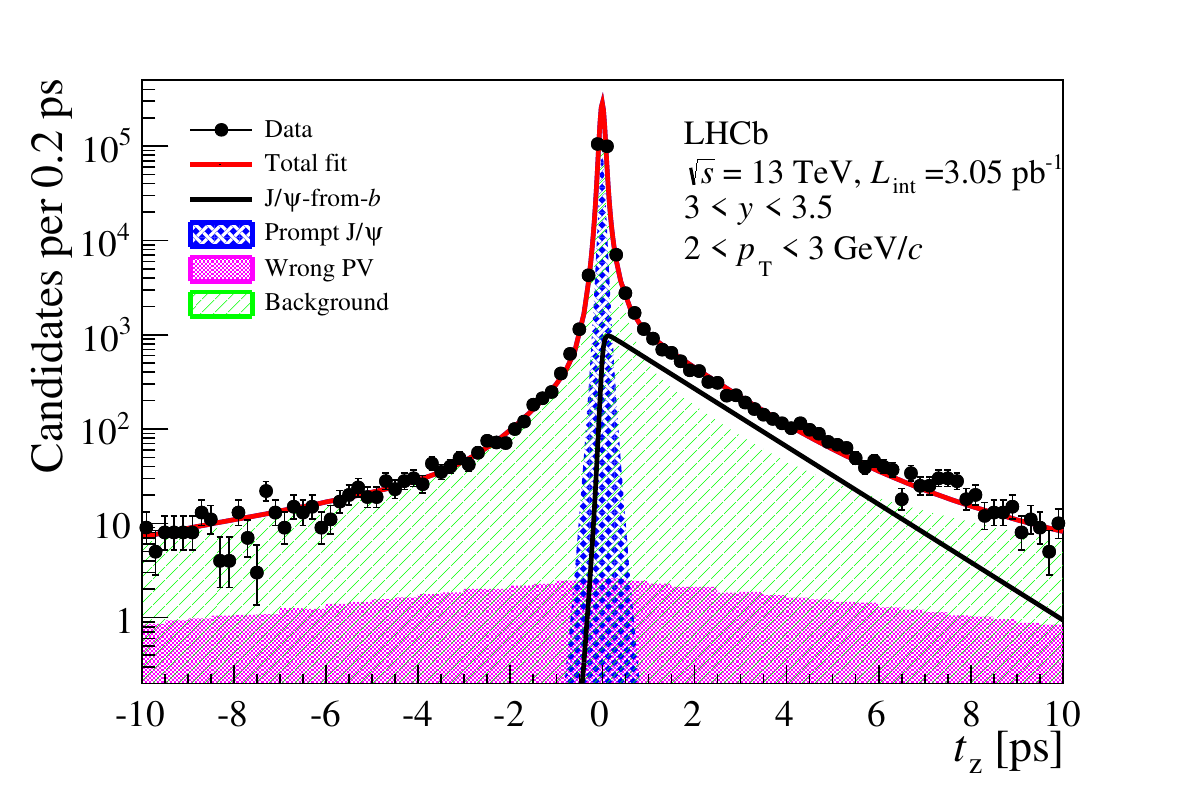}
\end{minipage}
\caption{Invariant mass (left) and pseudo decay time (right) distributions for the kinematic bin $2<\pt<3\gevc$,
   $3.0<y<3.5$, with fit results superimposed. The solid (red) line is the total fit function, the
      shaded (green) area corresponds to the background component. The prompt \jpsi contribution is shown in
      cross-hatched area (blue), \jpsi-from-$b$ in a solid (black) line and the tail contribution due to the association
      of \jpsi with the wrong PV is shown in full filled (magenta) area. The tail contribution is not visible in the
      invariant mass plot.}
\label{fig:FitResult}
\end{figure}

The total \jpsi signal yield determined from the fit is about one million events.
An example for one $(\pt,y)$ bin of the invariant mass and the pseudo decay time distributions is shown in
Fig.~\ref{fig:FitResult} with the one-dimensional projections of the fit result superimposed. 

\section{Systematic uncertainties}
\label{sec:systematics}
Systematic uncertainties, most of which apply to both prompt \jpsi and \jpsifromb mesons, are summarised in Table~\ref{tab:SystematicSummary} and described below. 
\begin{table}[!bp]
\caption{Relative systematic uncertainties (in \%) on the \jpsi cross-section measurements. The uncertainty from the $t_z$ fit
   only affects $\jpsi$-from-$b$ mesons. Most of the uncertainties are fully correlated between bins, with the exception of the \pt, $y$ spectrum dependence and the simulation statistics, which are considered uncorrelated.}
\centering
\begin{tabular}{c|c}
\hline
Source & Systematic uncertainty (\%)\\
\hline
Luminosity &  3.9 \\
Hardware trigger   &  $0.1-5.9$\\
Software trigger   &  1.5 \\
Muon ID &  1.8  \\
Tracking & $1.9-8.2$  \\
Radiative tail& 1.0    \\
$\jpsi$ vertex fit& 0.4 \\
Signal mass shape & 1.0\\
$\mathcal{B}(\jpsi\to\mup\mun)$& 0.6 \\
\pt, $y$ spectrum &  $0.1 -6.5$ \\
Simulation statistics&  $0.5 -10.0$\\
$t_z$ fit ($\jpsi$-from-$b$ only) &  0.1 \\
\hline
\end{tabular}\label{tab:SystematicSummary}
\end{table}

The uncertainty related to the modelling of the signal mass shape is studied by replacing the nominal model with a Hypatia function~\cite{Santos:2013gra}, which takes into account the mass uncertainty distribution. 
The relative difference of the signal yield is about $1.0\%$, which is taken as a fully correlated systematic uncertainty in each bin.

Due to the presence of bremsstrahlung in the \jpsimumu decay, a fraction of \jpsi events fall outside the mass window used in this analysis.
The efficiency of the mass window selection is determined from simulation, and based on a detailed comparison between
the radiative tails in simulation and data, a value of $1.0\%$ of the yield is assigned as the systematic uncertainty. 

To calibrate the muon identification efficiency determined from simulation, the single-track muon identification efficiency is measured with a \jpsimumu data sample using a tag-and-probe method.
In this method, the \jpsi candidates are reconstructed with only one track identified as a muon (``tag''). 
The single muon identification efficiency is measured as the probability of the other track (``probe'') to be identified
as a muon, in bins of momentum, $p_\mu$, and pseudorapidity, $\eta_\mu$ of the probe track.
The single-track muon identification efficiency obtained in data is weighted with the $(p_\mu,\eta_\mu)$ distribution of the muons from \jpsi mesons in simulation. 
The resulting efficiency is divided by that determined directly from simulation, giving $1.050\pm 0.017$, which is used to correct the muon identification efficiency in each $(\pt,y)$ bin of the \jpsi meson.
The uncertainty on the correction factor, limited by the calibration sample size in data, is used to obtain the systematic uncertainty.
The choice of binning scheme in $(p_\mu,\eta_\mu)$ is another source of systematic uncertainty and is evaluated by choosing alternative binning schemes.
In total the systematic uncertainty on the cross-sections due to the muon identification is $1.8\%$.

The tracking efficiency is studied with a data-driven tag-and-probe approach with \jpsi decays using partially
reconstructed tracks, where one muon track is fully reconstructed as the tag track, and the probe track is reconstructed
using only specific sub-detectors~\cite{LHCb-DP-2013-002}. 
The simulated sample is weighted to agree in event multiplicity with the data sample. 
A systematic uncertainty of the efficiency of $0.8\%$ per muon track is assigned to account for a different event multiplicity between data and simulation.
The tracking efficiency is determined to be the fraction of \jpsimumu decays where the probe track can be matched to a fully reconstructed track.
The ratio of the resulting tracking efficiency between data and simulation is used to weight the simulation sample according to pseudorapidity and momentum of muons to obtain an efficiency correction in each $(\pt,y)$ bin of the \jpsi meson.
In total the correction factor ranges from $0.91$ to $0.95$, depending on the \jpsi $(\pt,y)$ bin.
The uncertainty on the muon track efficiency correction factor is limited by the size of the calibration data sample, and is
propagated into the systematic uncertainty on the cross-section measurements.
In total, the systematic uncertainty on the cross-sections related to the tracking efficiency is in the range of $1.9-8.2\%$, depending on the \jpsi $(\pt,y)$ bin.

The \jpsi vertex fit quality requirement leads to an uncertainty of $0.4\%$ for the selection efficiency, which is determined by comparing the distributions of the vertex fit quality between data and simulation. 

The trigger efficiency obtained from simulation is cross-checked using data-driven methods using a fully reconstructed \jpsi
sample. For the hardware trigger, a tag-and-probe method is used to evaluate the single track muon trigger efficiency in
bins of muon transverse momentum, $\pt_\mu$, and pseudorapidity, $\eta_\mu$, for both simulation and data.
In this procedure, the \jpsi candidate is required to pass both steps of the software trigger. 
The \jpsi trigger efficiency is calculated by weighting $\pt_\mu$ and $\eta_\mu$ of the muons with the single-muon track
trigger efficiency obtained from data and simulation, and their relative difference is quoted as a systematic uncertainty. 
For most of the \jpsi $(\pt,y)$ bins, the systematic effect on the cross-section is found to be below $1.0\%$, but in two
bins an uncertainty of 5.9\% is found.
The software trigger efficiency is determined using a subset of events that would  have triggered if the \jpsi signals were excluded~\cite{LHCb-DP-2012-004}.
The efficiency is computed in each $(\pt,y)$ bin for data and simulation, and the relative difference between data and simulation of about $1.5\%$ is taken as a systematic uncertainty. 

The possible discrepancy between the \pt and $y$ distributions of \jpsi mesons in data and simulation for each bin is studied by reweighting the distribution in simulation to that in data.
The relative difference between the efficiency after reweighting and the nominal efficiency is taken as a systematic
uncertainty and found to be in the range $0.1\%-6.5\%$, depending on the $(\pt,y)$ bin of the \jpsi meson, where the
largest value corresponds to the bins at the \lhcb acceptance boundary.

The uncertainty associated with the luminosity determination is $3.9\%$, and the branching fraction uncertainty of the
$\jpsi\to\mumu$ decay is $0.6\%$. 
The limited size of the simulated sample in each bin leads to an uncertainty between $0.5\%$ and $10\%$, which is less
than or comparable to the data statistical uncertainty in each bin.

The detection efficiency is dependent on the polarisation of the \jpsi meson. 
Previous measurements by \cdf~\cite{Abulencia:2007us} in $p\overline{p}$ collisions at $1.96\tev$, 
\alice~\cite{Abelev:2011md}, \cms~\cite{Chatrchyan:2013cla} and \lhcb~\cite{LHCb-PAPER-2013-008} in $pp$ collisions at $7\tev$, showed that the prompt \jpsi polarisation in hadron collisions is small.
The \lhcb experiment studied \jpsi polarisation in the helicity frame~\cite{Jacob:1959at} and measured the longitudinal
polarisation parameter $\lambda_{\theta}$~\cite{LHCb-PAPER-2013-008} to be on average $-0.145\pm 0.027$ in the range
$2<\pt<14\gevc$\ and $2.0<y<4.5$.
If the longitudinal polarisation is assumed to be $-20$\%, the measured \jpsi cross-section would decrease by values
between $0.7\%$ and $6.2\%$ depending on the \jpsi $(\pt,y)$ bin, with an average value of $2.5\%$, which is smaller than the total systematic uncertainty. 
The efficiency changes in different \pt and $y$\ bins obtained under this assumption are discussed in the appendix.
Therefore, since no polarisation measurement has yet been made for data collected at $\sqrt{s}=13 \tev$, the polarisation is assumed to be zero, and no corresponding systematic uncertainty is quoted on the cross-section related to this effect.

There are sources of systematic uncertainties that are related to the $t_z$ fit, that affect only \jpsi-from-$b$\ decays and are negligible for prompt \jpsi production.
The modelling of the $t_z$ resolution is studied by adding a third Gaussian to the nominal resolution model.
The variation in $F_b$ is found to be negligible. 
Background modelling is tested using the \sPlot~\cite{Pivk:2004ty} method to extract the background $t_z$ distribution and the relative variation in $F_b$ is taken as a systematic uncertainty.
The impact of the choice of $t_z$ parametrisation for the long tail component is studied using an exponential
function with equal magnitude for positive and negative slopes 
and the relative difference of $F_b$ is taken as a systematic uncertainty.
The total relative systematic uncertainty on the \jpsifromb cross-section related to the $t_z$ fit is 0.1\%.

\section{Results}\label{sec:results}
\begin{figure}[!tbp]
\centering
\begin{minipage}[t]{0.7\textwidth}
\centering
\includegraphics[width=1.0\textwidth]{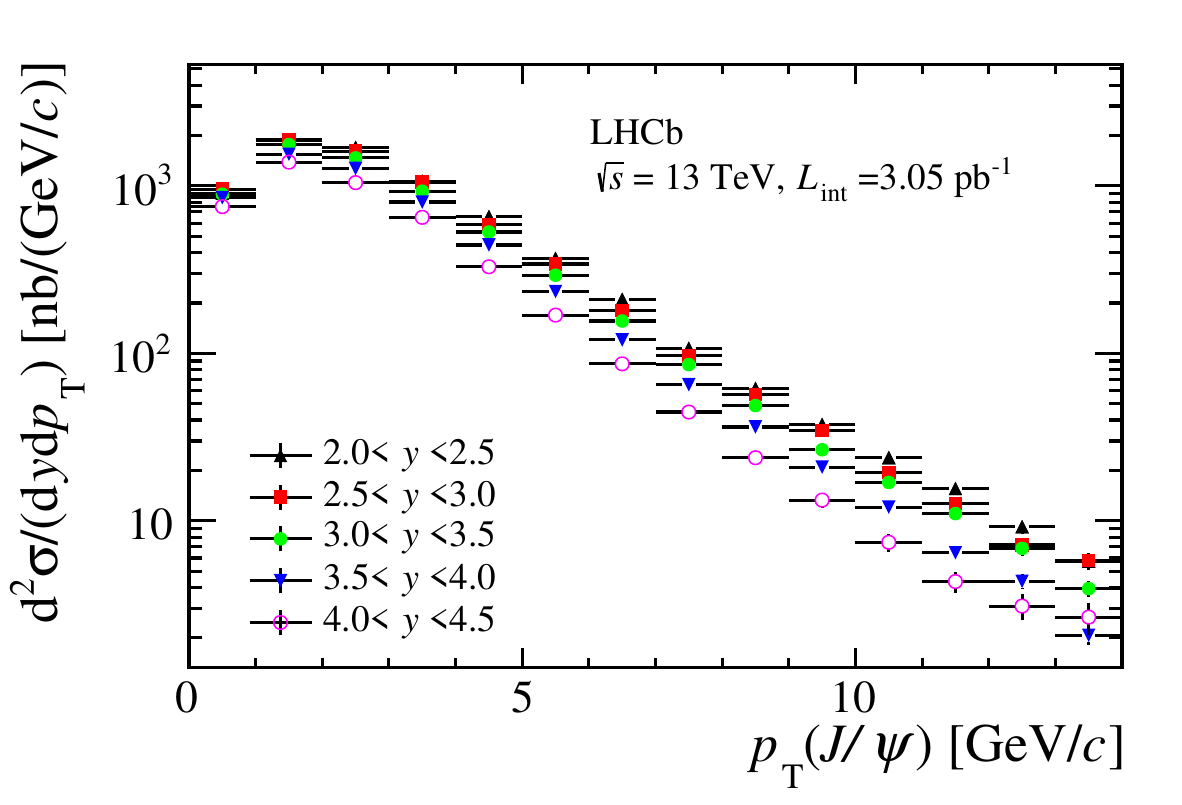}
\end{minipage}
\caption{Double differential cross-section for prompt \jpsi mesons as a function of \pt in bins of $y$. Statistical and systematic uncertainties are added in quadrature.} 
\label{fig:ResultPromptJpsi}
\end{figure}

\begin{figure}[!tbp]
\centering
\begin{minipage}[t]{0.7\textwidth}
\centering
\includegraphics[width=1.0\textwidth]{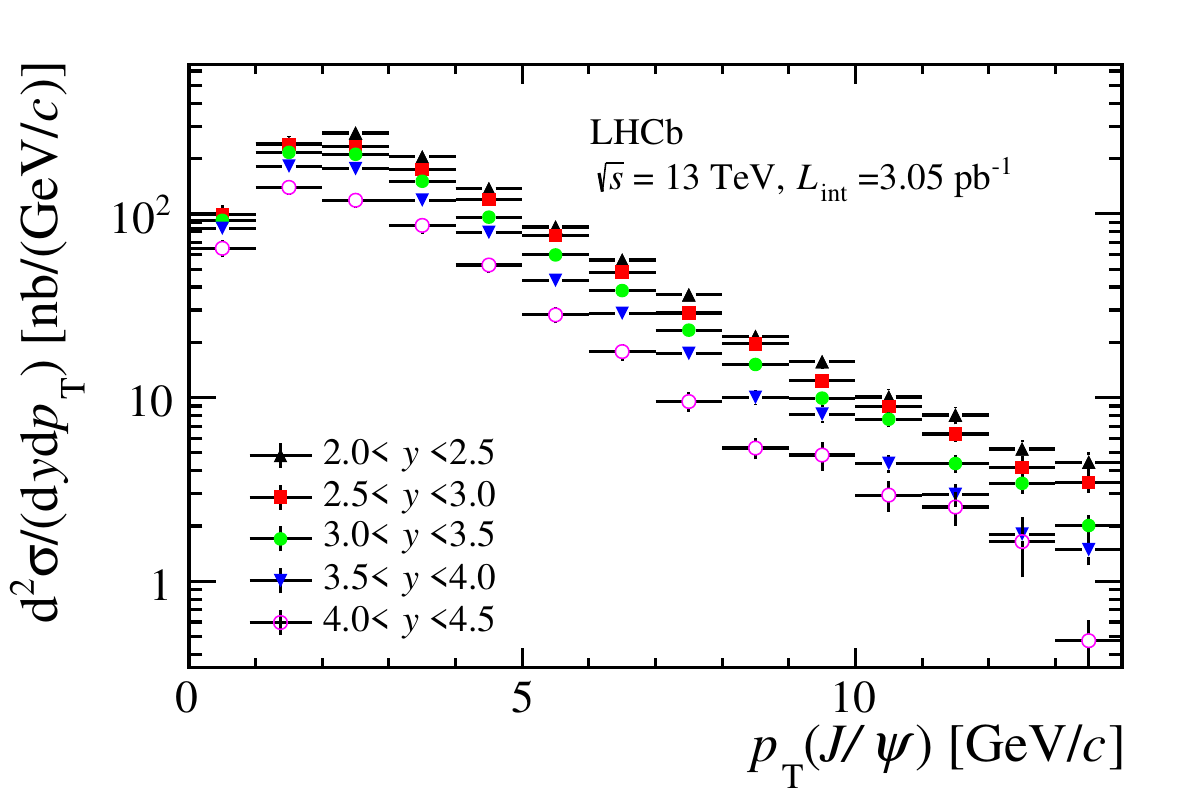}
\end{minipage}
\caption{Double differential cross-section for \jpsi-from-$b$ mesons as a function of \pt in bins of $y$. Statistical and systematic
   uncertainties are added in quadrature.} 
\label{fig:ResultJpsiFromB}
\end{figure}

The measured double differential cross-sections for prompt \jpsi and \jpsi-from-$b$ mesons,
assuming no polarisation, are shown in Figs.~\ref{fig:ResultPromptJpsi} and~\ref{fig:ResultJpsiFromB}, and given in Tables~\ref{tab:ResultPromptJpsi} and \ref{tab:ResultJpsiFromB}. The cross-sections for prompt \jpsi and \jpsi-from-$b$ mesons in the acceptance \mbox{$\pt<14\gevc$} and $2.0<y<4.5$,
integrated over all $(\pt,y)$ bins, are:
\begin{eqnarray}
\sigma(\mathrm{prompt}~ \jpsi, \pt<14\gevc,2.0<y<4.5) &=& 15.03\pm 0.03\pm 0.94\mub,\nonumber\\
\sigma(\jpsi\text{-from-}b, \pt<14\gevc,2.0<y<4.5) &=&  \phantom0 2.25\pm 0.01\pm 0.14\mub,\nonumber
\end{eqnarray}
where the first uncertainties are statistical and the second systematic.
\renewcommand{\arraystretch}{1.35}
\begin{table}[!htbp]
\begin{center}
\caption{\small \label{tab:ResultPromptJpsi} Double differential production cross-section in
      $\mathrm{nb}/(\!\gevc)$
   for prompt \jpsi mesons in bins of ($\pt,y$). The first uncertainties are statistical, the second are the correlated
      systematic uncertainties shared between bins and the last are the uncorrelated systematic uncertainties.}
\vskip 0.5cm
\scalebox{0.9}{
\begin{tabular}{@{}r@{}c@{}lrr@{}c@{}r@{}c@{}r@{}c@{}rr@{}c@{}r@{}c@{}r@{}c@{}rr@{}c@{}r@{}c@{}r@{}c@{}r@{}}
\\
\toprule
\multicolumn{4}{c}{$\pt\,[\!\gevc]$} & \multicolumn{7}{c}{$2.0<y<2.5$} & \multicolumn{7}{c}{$2.5<y<3.0$} & \multicolumn{7}{c}{$3.0<y<3.5$} \\
\midrule
$0$&$-$&$1$&&$906\,\,\,\tpm  14\,\,\,\tpm 44\,\,\,\tpm 24\,\,\,$& $955\,\,\,\tpm   9\,\,\,\tpm 40\,\,\,\tpm 12\,\,\,$& $892\,\,\,\tpm   8\,\,\,\tpm 41\,\,\,\tpm 10\,\,\,$\\
 $1$&$-$&$2$&&$1880\,\,\,\tpm  20\,\,\,\tpm 88\,\,\,\tpm 46\,\,\,$& $1876\,\,\,\tpm  12\,\,\,\tpm 77\,\,\,\tpm 17\,\,\,$& $1764\,\,\,\tpm  11\,\,\,\tpm 81\,\,\,\tpm 14\,\,\,$\\
 $2$&$-$&$3$&&$1697\,\,\,\tpm  16\,\,\,\tpm 75\,\,\,\tpm 41$& $1612\,\,\,\tpm  10\,\,\,\tpm 66\,\,\,\tpm 15\,\,\,$& $1470\,\,\,\tpm   9\,\,\,\tpm 66\,\,\,\tpm 12\,\,\,$\\
 $3$&$-$&$4$&&$1069\,\,\,\tpm  11\,\,\,\tpm 46\,\,\,\tpm 20\,\,\,$& $1055\,\,\,\tpm   7\,\,\,\tpm 43\,\,\,\tpm 12\,\,\,$& $930\,\,\,\tpm   6\,\,\,\tpm 39\,\,\,\tpm  9\,\,\,$\\
 $4$&$-$&$5$&&$656\,\,\,\tpm   7\,\,\,\tpm 28\,\,\,\tpm 14\,\,\,$& $586\,\,\,\tpm   5\,\,\,\tpm 24\,\,\,\tpm  7\,\,\,$& $531\,\,\,\tpm   4\,\,\,\tpm 22\,\,\,\tpm  6\,\,\,$\\
 $5$&$-$&$6$&&$369\,\,\,\tpm   5\,\,\,\tpm 15\,\,\,\tpm  9\,\,\,$& $342\,\,\,\tpm   3\,\,\,\tpm 14\,\,\,\tpm  4\,\,\,$& $293\,\,\,\tpm   3\,\,\,\tpm 12\,\,\,\tpm  4\,\,\,$\\
 $6$&$-$&$7$&&$210.3\,\tpm 3.3\,\tpm8.6\,\tpm5.2$& $180.3\,\tpm 2.1\,\tpm7.3\,\tpm2.8$& $156.1\,\tpm 1.9\,\tpm6.3\,\tpm2.4$\\
 $7$&$-$&$8$&&$107.3\,\tpm 2.1\,\tpm4.4\,\tpm3.3$& $96.7\,\tpm 1.5\,\tpm3.9\,\tpm1.8$& $85.8\,\tpm 1.4\,\tpm3.5\,\tpm1.7$\\
 $8$&$-$&$9$&&$61.7\,\tpm 1.5\,\tpm2.5\,\tpm2.1$& $56.8\,\tpm 1.1\,\tpm2.3\,\tpm1.4$& $48.8\,\tpm 1.0\,\tpm2.0\,\tpm1.3$\\
 $9$&$-$&$10$&&$37.6\,\tpm 1.1\,\tpm1.5\,\tpm1.5$& $34.6\,\tpm 0.9\,\tpm1.4\,\tpm1.0$& $26.6\,\tpm 0.7\,\tpm1.1\,\tpm0.8$\\
 $10$&$-$&$11$&&$23.9\,\tpm 0.9\,\tpm1.0\,\tpm1.3$& $19.5\,\tpm 0.6\,\tpm0.8\,\tpm0.7$& $17.0\,\tpm 0.6\,\tpm0.7\,\tpm0.7$\\
 $11$&$-$&$12$&&$15.6\,\tpm 0.7\,\tpm0.6\,\tpm1.0$& $12.7\,\tpm 0.5\,\tpm0.5\,\tpm0.6$& $11.0\,\tpm 0.5\,\tpm0.4\,\tpm0.5$\\
 $12$&$-$&$13$&&$9.2\,\tpm 0.5\,\tpm0.4\,\tpm0.6$& $7.2\,\tpm 0.4\,\tpm0.3\,\tpm0.4$& $6.8\,\tpm 0.4\,\tpm0.3\,\tpm0.4$\\
 $13$&$-$&$14$&&$5.8\,\tpm 0.4\,\tpm0.2\,\tpm0.5$& $5.8\,\tpm 0.4\,\tpm0.2\,\tpm0.4$& $3.9\,\tpm 0.3\,\tpm0.2\,\tpm0.3$\\
 \bottomrule
&&&& \multicolumn{7}{c}{$3.5<y<4.0$} & \multicolumn{7}{c}{$4.0<y<4.5$}\\ \midrule
$0$&$-$&$1$&&$850\,\,\,\tpm   8\,\,\,\tpm 48\,\,\,\tpm 11\,\,\,$& $752\,\,\,\tpm   9\,\,\,\tpm 50\,\,\,\tpm 16\,\,\,$&\\
 $1$&$-$&$2$&&$1545\,\,\,\tpm  10\,\,\,\tpm 90\,\,\,\tpm 14\,\,\,$& $1387\,\,\,\tpm  12\,\,\,\tpm 99\,\,\,\tpm 23\,\,\,$&\\
 $2$&$-$&$3$&&$1272\,\,\,\tpm   8\,\,\,\tpm 71\,\,\,\tpm 13\,\,\,$& $1046\,\,\,\tpm  10\,\,\,\tpm 76\,\,\,\tpm 24\,\,\,$&\\
 $3$&$-$&$4$&&$801\,\,\,\tpm   6\,\,\,\tpm 42\,\,\,\tpm  9\,\,\,$& $649\,\,\,\tpm   8\,\,\,\tpm 44\,\,\,\tpm 19\,\,\,$&\\
 $4$&$-$&$5$&&$444\,\,\,\tpm   4\,\,\,\tpm 21\,\,\,\tpm  6\,\,\,$& $329\,\,\,\tpm   5\,\,\,\tpm 19\,\,\,\tpm  8\,\,\,$&\\
 $5$&$-$&$6$&&$234\,\,\,\tpm   3\,\,\,\tpm 11\,\,\,\tpm  4\,\,\,$& $169\,\,\,\tpm   3\,\,\,\tpm  9\,\,\,\tpm  5\,\,\,$&\\
 $6$&$-$&$7$&&$119.6\,\tpm 1.7\,\tpm5.2\,\tpm0.8$& $87.3\,\tpm 2.1\,\tpm4.5\,\tpm1.0$&\\
 $7$&$-$&$8$&&$65.0\,\tpm 1.3\,\tpm2.8\,\tpm1.5$& $44.6\,\tpm 1.4\,\tpm2.2\,\tpm2.0$&\\
 $8$&$-$&$9$&&$36.4\,\tpm 0.9\,\tpm1.5\,\tpm1.0$& $23.8\,\tpm 1.0\,\tpm1.1\,\tpm1.2$&\\
 $9$&$-$&$10$&&$20.9\,\tpm 0.7\,\tpm0.9\,\tpm0.8$& $13.3\,\tpm 0.7\,\tpm0.6\,\tpm1.0$&\\
 $10$&$-$&$11$&&$12.1\,\tpm 0.5\,\tpm0.5\,\tpm0.6$& $7.4\,\tpm 0.5\,\tpm0.3\,\tpm0.6$&\\
 $11$&$-$&$12$&&$6.4\,\tpm 0.3\,\tpm0.3\,\tpm0.3$& $4.3\,\tpm 0.4\,\tpm0.2\,\tpm0.4$&\\
 $12$&$-$&$13$&&$4.4\,\tpm 0.3\,\tpm0.2\,\tpm0.3$& $3.1\,\tpm 0.3\,\tpm0.2\,\tpm0.4$&\\
 $13$&$-$&$14$&&$2.1\,\tpm 0.2\,\tpm0.1\,\tpm0.2$& $2.7\,\tpm 0.3\,\tpm0.1\,\tpm0.5$&\\
  
\bottomrule
\end{tabular}
}
\end{center}
\end{table}
\renewcommand{\arraystretch}{1.35}
\begin{table}[!htbp]
\begin{center}
\caption{\small \label{tab:ResultJpsiFromB} Double differential production cross-section in $\mathrm{nb}/(\!\gevc)$
   for $\jpsi$-from-$b$ mesons in bins of ($\pt,y$). The first uncertainties are statistical, the second are the correlated
      systematic uncertainties shared between bins and the last are the uncorrelated systematic uncertainties.}
\vskip 0.5cm
\scalebox{0.9}{
\begin{tabular}{@{}r@{}c@{}lrr@{}c@{}r@{}c@{}r@{}c@{}rr@{}c@{}r@{}c@{}r@{}c@{}rr@{}c@{}r@{}c@{}r@{}c@{}r@{}}
\\
\toprule
\multicolumn{4}{c}{$\pt\,[\!\gevc]$} & \multicolumn{7}{c}{$2.0<y<2.5$} & \multicolumn{7}{c}{$2.5<y<3.0$} & \multicolumn{7}{c}{$3.0<y<3.5$} \\
\midrule
$0$&$-$&$1$&&$99.3\,\tpm 4.8\,\tpm9.6\,\tpm5.1$& $98.8\,\tpm 2.8\,\tpm6.1\,\tpm2.6$& $92.3\,\tpm 2.6\,\tpm5.4\,\tpm2.2$\\
 $1$&$-$&$2$&&$242.3\,\tpm 6.1\,\tpm20.6\,\tpm8.7$& $238.1\,\tpm 3.8\,\tpm14.2\,\tpm4.2$& $216.4\,\tpm 3.4\,\tpm12.6\,\tpm3.6$\\
 $2$&$-$&$3$&&$275.6\,\tpm 5.8\,\tpm19.9\,\tpm9.5$& $233.4\,\tpm 3.4\,\tpm13.4\,\tpm3.9$& $211.3\,\tpm 3.1\,\tpm12.0\,\tpm3.5$\\
 $3$&$-$&$4$&&$204.3\,\tpm 4.6\,\tpm13.1\,\tpm6.3$& $174.8\,\tpm 2.7\,\tpm9.6\,\tpm3.2$& $150.6\,\tpm 2.3\,\tpm8.2\,\tpm2.7$\\
 $4$&$-$&$5$&&$137.3\,\tpm 3.3\,\tpm8.1\,\tpm4.7$& $120.0\,\tpm 2.0\,\tpm6.4\,\tpm2.3$& $96.0\,\tpm 1.7\,\tpm5.1\,\tpm1.9$\\
 $5$&$-$&$6$&&$84.9\,\tpm 2.3\,\tpm4.7\,\tpm2.9$& $76.3\,\tpm 1.5\,\tpm4.0\,\tpm1.6$& $59.8\,\tpm 1.3\,\tpm3.1\,\tpm1.3$\\
 $6$&$-$&$7$&&$56.2\,\tpm 1.7\,\tpm3.0\,\tpm2.1$& $48.2\,\tpm 1.1\,\tpm2.5\,\tpm1.1$& $38.3\,\tpm 1.0\,\tpm2.0\,\tpm1.0$\\
 $7$&$-$&$8$&&$36.3\,\tpm 1.3\,\tpm1.9\,\tpm1.6$& $28.9\,\tpm 0.8\,\tpm1.5\,\tpm0.8$& $23.3\,\tpm 0.7\,\tpm1.2\,\tpm0.7$\\
 $8$&$-$&$9$&&$21.5\,\tpm 0.9\,\tpm1.1\,\tpm1.0$& $19.6\,\tpm 0.7\,\tpm1.0\,\tpm0.6$& $15.2\,\tpm 0.6\,\tpm0.8\,\tpm0.6$\\
 $9$&$-$&$10$&&$15.7\,\tpm 0.7\,\tpm0.8\,\tpm0.8$& $12.3\,\tpm 0.5\,\tpm0.6\,\tpm0.5$& $9.9\,\tpm 0.5\,\tpm0.5\,\tpm0.5$\\
 $10$&$-$&$11$&&$10.1\,\tpm 0.6\,\tpm0.5\,\tpm0.6$& $9.0\,\tpm 0.5\,\tpm0.5\,\tpm0.4$& $7.6\,\tpm 0.4\,\tpm0.4\,\tpm0.5$\\
 $11$&$-$&$12$&&$8.0\,\tpm 0.5\,\tpm0.4\,\tpm0.6$& $6.3\,\tpm 0.4\,\tpm0.3\,\tpm0.3$& $4.4\,\tpm 0.3\,\tpm0.2\,\tpm0.3$\\
 $12$&$-$&$13$&&$5.3\,\tpm 0.4\,\tpm0.3\,\tpm0.4$& $4.2\,\tpm 0.3\,\tpm0.2\,\tpm0.3$& $3.4\,\tpm 0.3\,\tpm0.2\,\tpm0.3$\\
 $13$&$-$&$14$&&$4.5\,\tpm 0.4\,\tpm0.2\,\tpm0.4$& $3.5\,\tpm 0.3\,\tpm0.2\,\tpm0.3$& $2.0\,\tpm 0.2\,\tpm0.1\,\tpm0.2$\\
 \bottomrule
&&&& \multicolumn{7}{c}{$3.5<y<4.0$} & \multicolumn{7}{c}{$4.0<y<4.5$}\\ \midrule
$0$&$-$&$1$&&$83.5\,\tpm 2.7\,\tpm5.5\,\tpm2.5$& $65.0\,\tpm 3.8\,\tpm4.8\,\tpm3.5$&\\
 $1$&$-$&$2$&&$182.1\,\tpm 3.4\,\tpm12.2\,\tpm3.8$& $139.5\,\tpm 4.6\,\tpm11.0\,\tpm5.3$&\\
 $2$&$-$&$3$&&$176.3\,\tpm 3.0\,\tpm11.5\,\tpm3.8$& $118.7\,\tpm 3.6\,\tpm9.5\,\tpm5.0$&\\
 $3$&$-$&$4$&&$118.9\,\tpm 2.3\,\tpm7.3\,\tpm2.7$& $86.6\,\tpm 3.0\,\tpm6.6\,\tpm4.4$&\\
 $4$&$-$&$5$&&$79.4\,\tpm 1.7\,\tpm4.6\,\tpm2.0$& $52.7\,\tpm 2.1\,\tpm3.7\,\tpm2.7$&\\
 $5$&$-$&$6$&&$43.5\,\tpm 1.2\,\tpm2.5\,\tpm1.2$& $28.2\,\tpm 1.4\,\tpm1.9\,\tpm1.6$&\\
 $6$&$-$&$7$&&$28.8\,\tpm 0.9\,\tpm1.6\,\tpm1.0$& $17.8\,\tpm 1.0\,\tpm1.1\,\tpm1.1$&\\
 $7$&$-$&$8$&&$17.4\,\tpm 0.7\,\tpm1.0\,\tpm0.7$& $9.5\,\tpm 0.7\,\tpm0.6\,\tpm0.7$&\\
 $8$&$-$&$9$&&$10.0\,\tpm 0.5\,\tpm0.6\,\tpm0.5$& $5.3\,\tpm 0.5\,\tpm0.3\,\tpm0.4$&\\
 $9$&$-$&$10$&&$8.1\,\tpm 0.5\,\tpm0.5\,\tpm0.5$& $4.9\,\tpm 0.5\,\tpm0.3\,\tpm0.6$&\\
 $10$&$-$&$11$&&$4.4\,\tpm 0.3\,\tpm0.3\,\tpm0.3$& $2.9\,\tpm 0.3\,\tpm0.2\,\tpm0.4$&\\
 $11$&$-$&$12$&&$3.0\,\tpm 0.3\,\tpm0.2\,\tpm0.3$& $2.5\,\tpm 0.3\,\tpm0.2\,\tpm0.4$&\\
 $12$&$-$&$13$&&$1.8\,\tpm 0.2\,\tpm0.1\,\tpm0.2$& $1.6\,\tpm 0.3\,\tpm0.1\,\tpm0.5$&\\
 $13$&$-$&$14$&&$1.5\,\tpm 0.2\,\tpm0.1\,\tpm0.2$& $0.5\,\tpm 0.1\,\tpm0.0\,\tpm0.1$&\\
  
\bottomrule
\end{tabular}
}
\end{center}
\end{table}

\subsection{Fraction of \jpsi-from-$b$ mesons}
\begin{figure}[!tbp]
\centering
\begin{minipage}[t]{0.7\textwidth}
\centering
\includegraphics[width=1.0\textwidth]{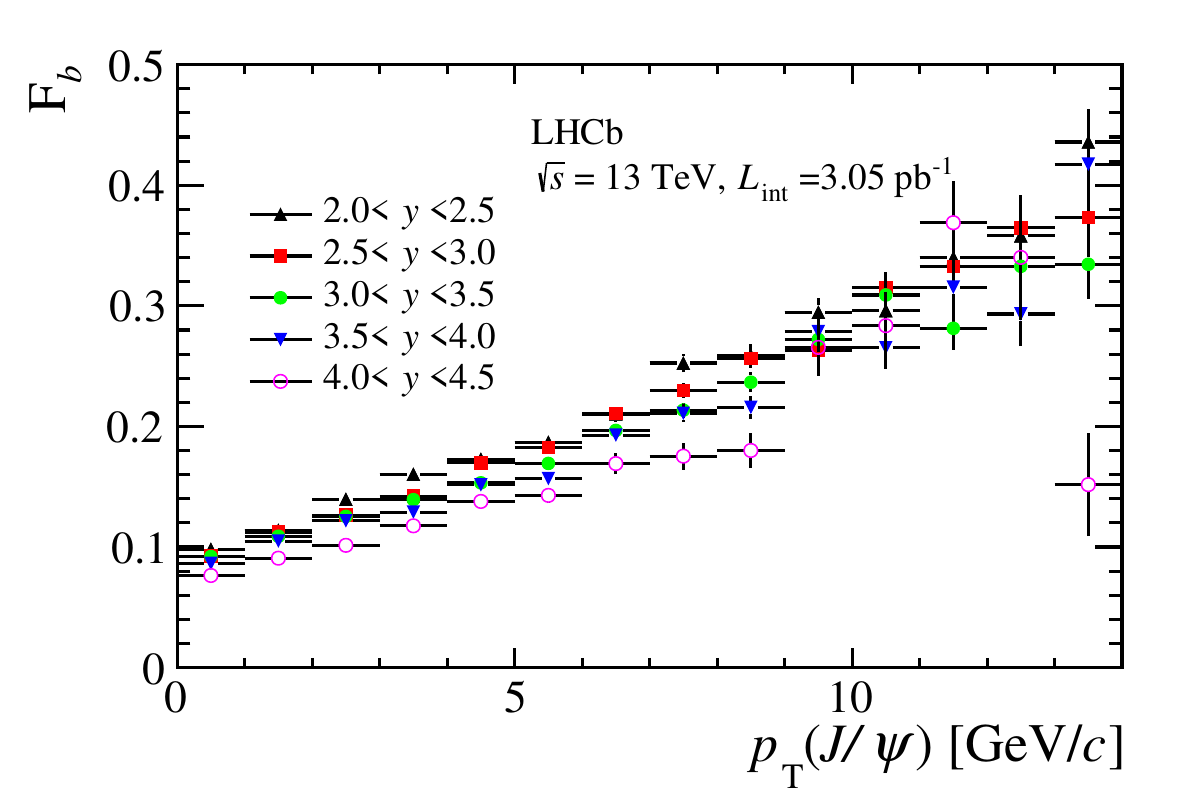}
\end{minipage}
\caption{Fractions of $\jpsi$-from-$b$ mesons in bins of \jpsi \pt and $y$. Statistical and systematic uncertainties are added in quadrature.} 
\label{fig:BFraction}
\end{figure}

The fractions of \jpsi-from-$b$ mesons in different kinematic bins are given in Fig.~\ref{fig:BFraction} and Table~\ref{tab:BFraction}. 
The fraction increases as a function of \pt, and 
tends to decrease with increasing rapidity.
These trends are consistent with the measurements at $\sqrt{s}=7\tev$ and $\sqrt{s}=8\tev$~\cite{LHCb-PAPER-2011-003, LHCb-PAPER-2013-016}.
A measurement of the non-prompt \jpsi production fraction at $\sqrt{s}=$13\tev has also been performed by the \atlas collaboration \cite{ATLAS-CONF-2015-030}.

\subsection{Extrapolation to the total $\bquark \bquarkbar$\ cross-section}\label{extrapolation}
The total $\bquark \bquarkbar$ production cross-section is calculated
using:
\begin{eqnarray}
\sigma(pp \rightarrow \bquark \bquarkbar X) &= \alpha_{4 \pi}
\frac{\displaystyle\sigma\left(J/\psi \, \text{-from-} b,\,  p_{T} < 14 \gevc, \, 2.0 < y < 4.5 \right)}{\displaystyle2 \mathcal{B}\left(b \rightarrow J/\psi
      X\right)},
 \end{eqnarray}
where $\alpha_{4 \pi}$ is the extrapolation factor to the full kinematic region and $\mathcal{B}(b\rightarrow J/\psi X) = 1.16 \pm 0.10 \%$~\cite{PDG2014} is the inclusive $b \rightarrow J/\psi X$ branching fraction.
Using the \lhcb tuning of \pythia 6~\cite{pythia6.4}, $\alpha_{4 \pi}$ is found to be 5.2. 
The extrapolation predictions given by \pythia 8  and FONLL~\cite{Cacciari:1998it} are $\alpha_{4 \pi  } = 5.1$ and $\alpha_{4\pi  } = 5.0$ respectively at $\sqrt{s}=13 \tev$.
Their predictions at $\sqrt{s}=7\tev$ are compatible with the estimate using LHC \jpsi cross-section measurements in
different rapidity ranges~\cite{LHCb-PAPER-2013-008, Chatrchyan:2011kc}.
Using the extrapolation factor from \pythia 6, the total $\bquark \bquarkbar$ production cross-section is found to be
$\sigma(pp\rightarrow\bquark\bquarkbar X) = 495\pm 2\pm 52 \mub$, where the first uncertainty is statistical and the second systematic. 
No uncertainty on $\alpha_{4 \pi }$ is included in this estimate. 

\subsection{Comparison with lower energy results}
\begin{figure}[!tbp]
\centering
\begin{minipage}[t]{0.45\textwidth}
\centering
\includegraphics[width=1.0\textwidth]{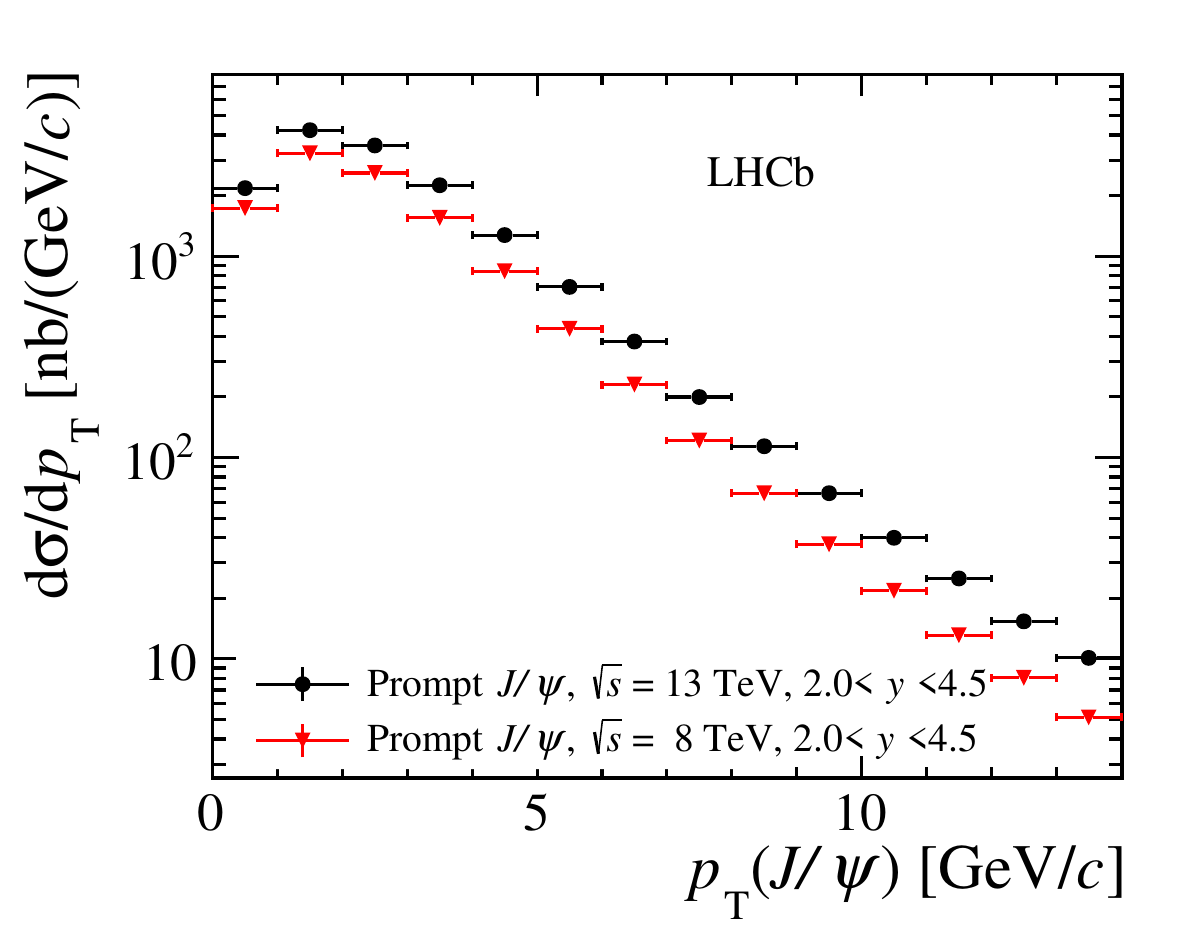}
\end{minipage}
\begin{minipage}[t]{0.45\textwidth}
\centering
\includegraphics[width=1.0\textwidth]{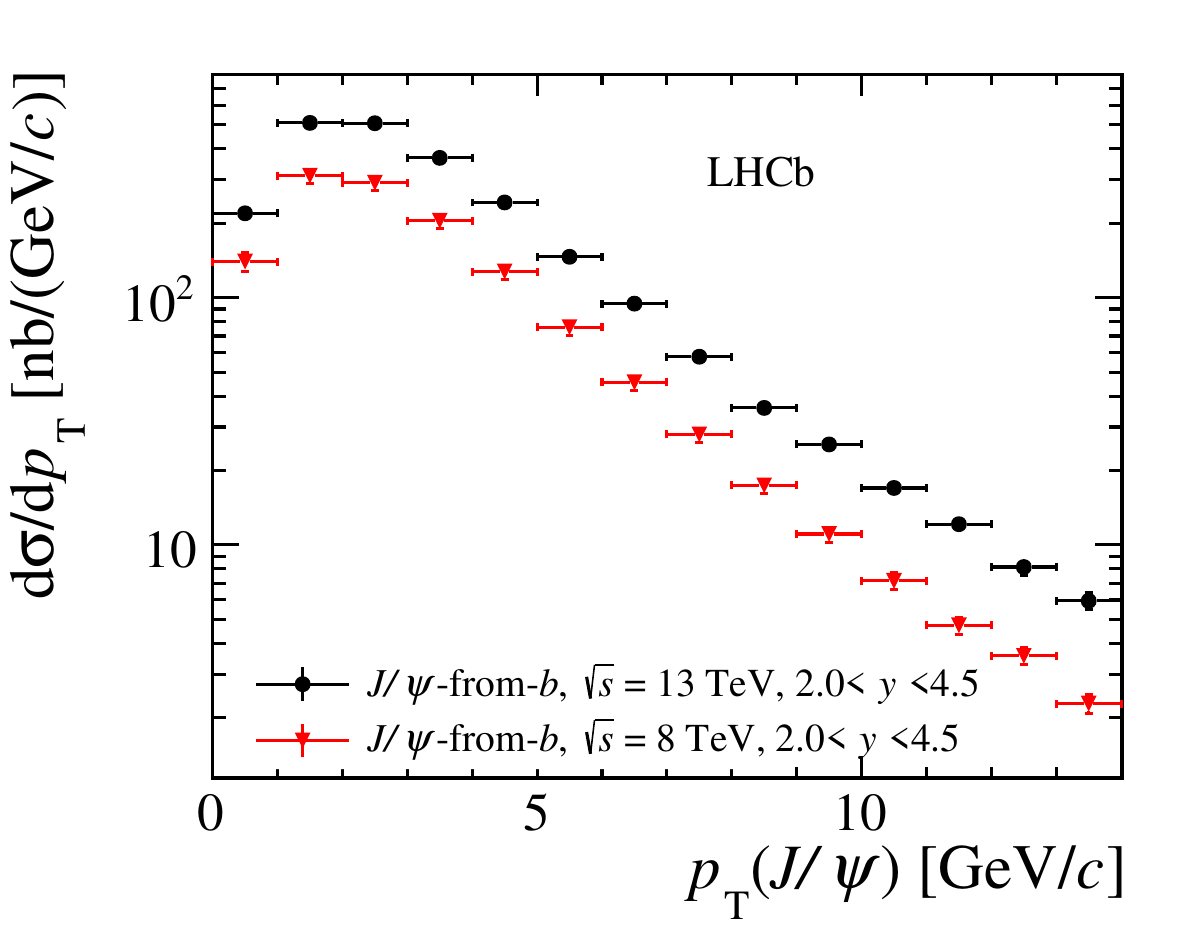}
\end{minipage}
\caption{Differential cross-sections as a function of $\pt$ integrated over $y$ for (left) prompt \jpsi and (right)
   \jpsi-from-$b$ mesons.} 
\label{fig:PTXsection}
\end{figure}

\begin{figure}[!tbp]
\centering
\begin{minipage}[t]{0.45\textwidth}
\centering
\includegraphics[width=1.0\textwidth]{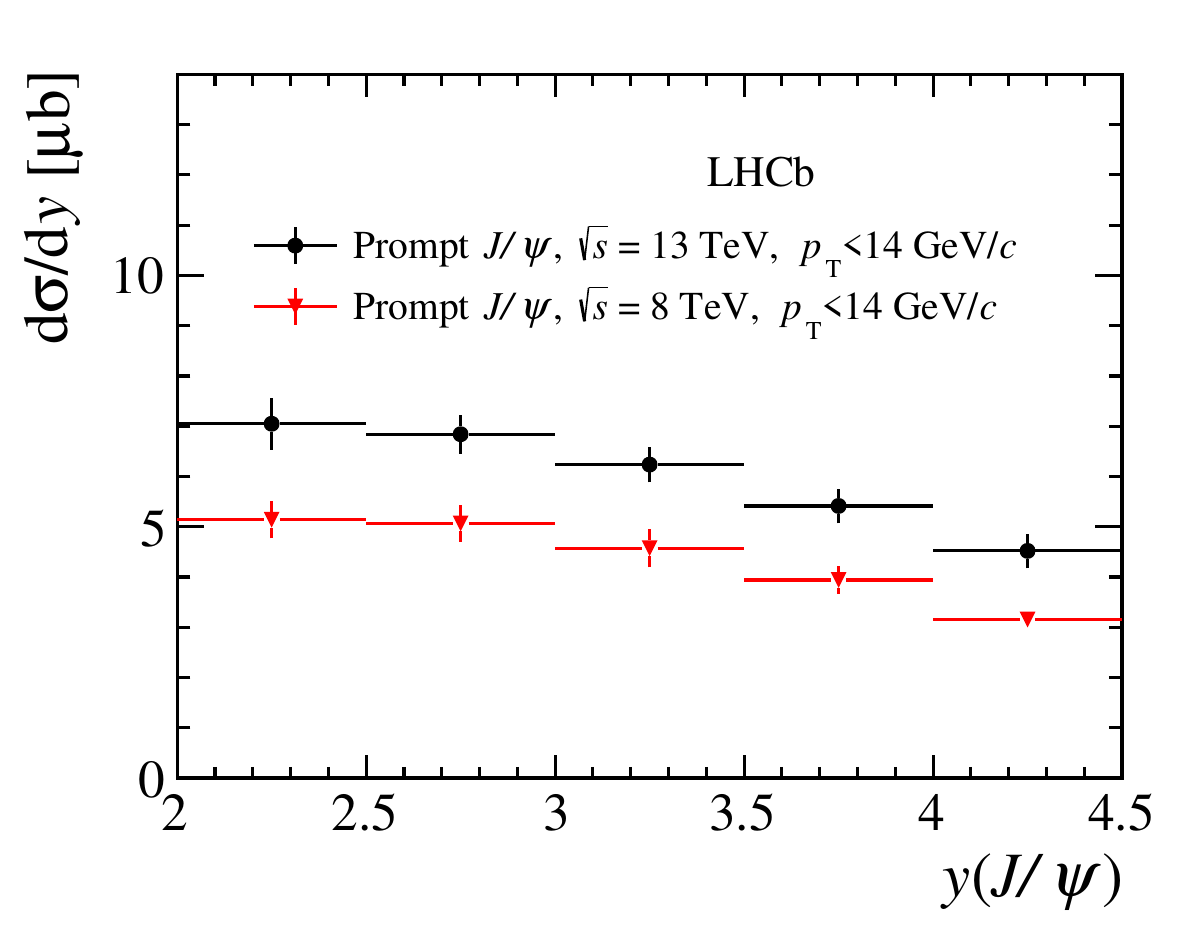}
\end{minipage}
\begin{minipage}[t]{0.45\textwidth}
\centering
\includegraphics[width=1.0\textwidth]{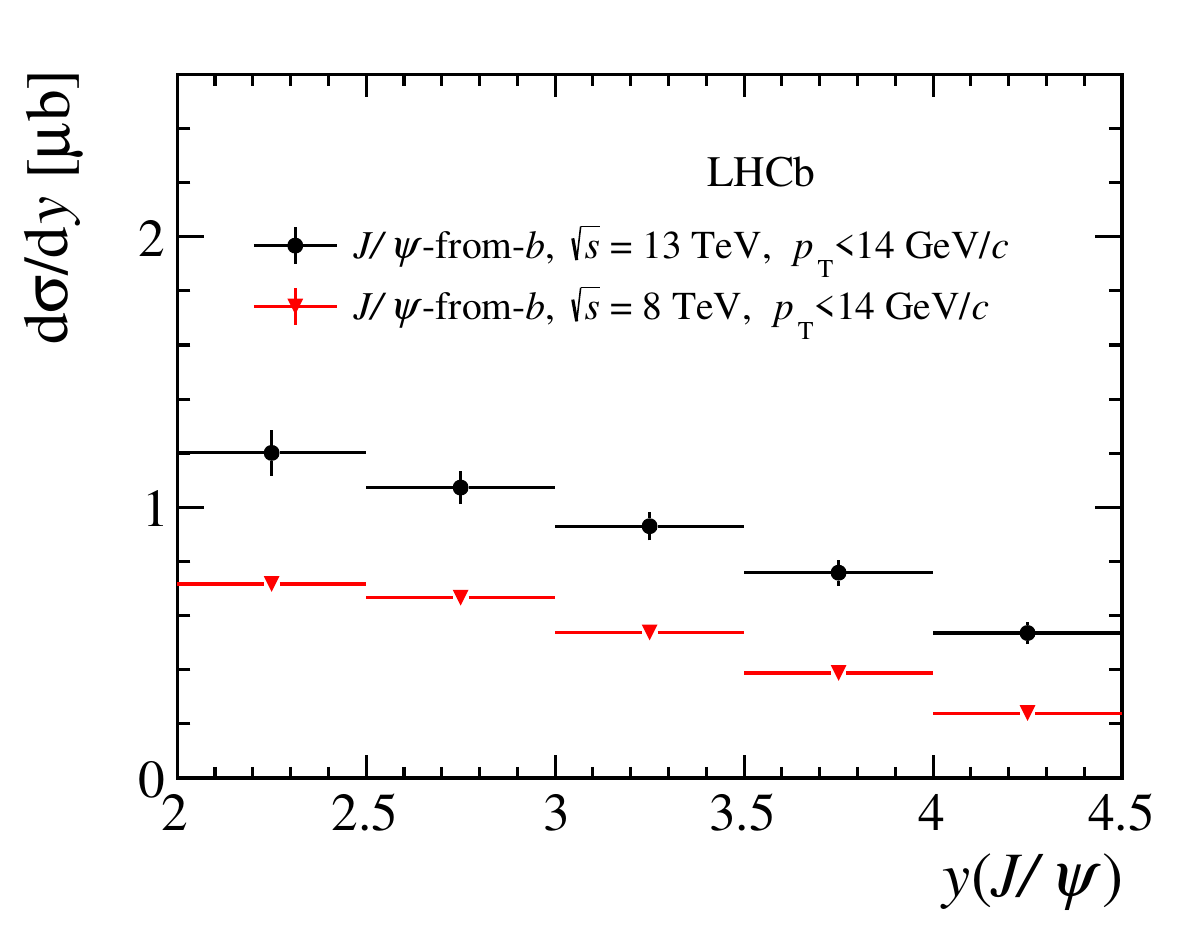}
\end{minipage}
\caption{Differential cross-sections as a  function of $y$ integrated over \pt for (left) prompt \jpsi and (right)
   \jpsi-from-$b$ mesons.} 
\label{fig:YXsection}
\end{figure}

The \jpsi cross-sections measured at $\sqrt{s}=13\tev$ are compared to previous \lhcb measurements~\cite{LHCb-PAPER-2013-016, LHCb-PAPER-2012-039, LHCb-PAPER-2013-008}.
In all previous \lhcb measurements of the \jpsi production cross-section, the branching fraction from Ref.~\cite{PDG2012}, $\mathcal{B}(\jpsimumu)=(5.94\pm0.06)\%$, was used. 
When the measurements at $13\tev$ are compared with those at lower energy, the previous results are updated with the improved branching fraction value, $\mathcal{B}(\jpsimumu)=(5.961\pm0.033)\%$~\cite{PDG2014}. 
The corresponding systematic uncertainty is totally correlated among the measurements.
The differential cross-section as a function of \pt integrated over $y$ is shown in Fig.~\ref{fig:PTXsection}, including all uncertainties, compared to measurements with $pp$ collisions at $\sqrt{s}=8\tev$, for prompt \jpsi and \jpsi-from-$b$ mesons. 
In Fig.~\ref{fig:YXsection}, the differential cross-section as a function of $y$ integrated over $\pt$ is shown, compared to measurements with $pp$ collisions at $\sqrt{s}=8\tev$.
Tables~\ref{tab:CrossSectionWithPT} and \ref{tab:CrossSectionWithY} show the differential cross-sections integrated over
$y$ and $\pt$ for prompt \jpsi and \jpsi-from-$b$ mesons. 


\begin{figure}[!tbp]
\centering
\begin{minipage}[t]{0.45\textwidth}
\centering
\includegraphics[width=1.0\textwidth]{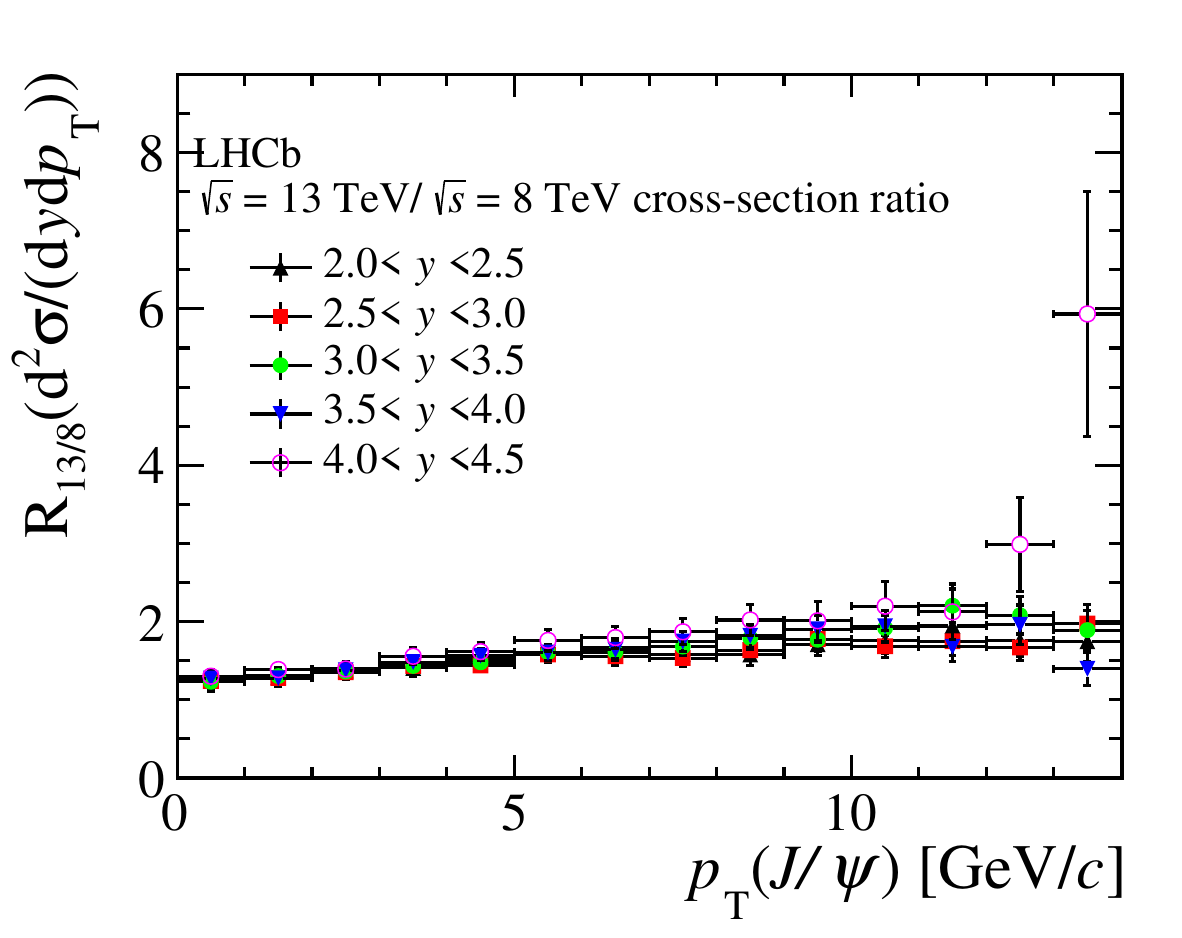}
\end{minipage}
\begin{minipage}[t]{0.45\textwidth}
\centering
\includegraphics[width=1.0\textwidth]{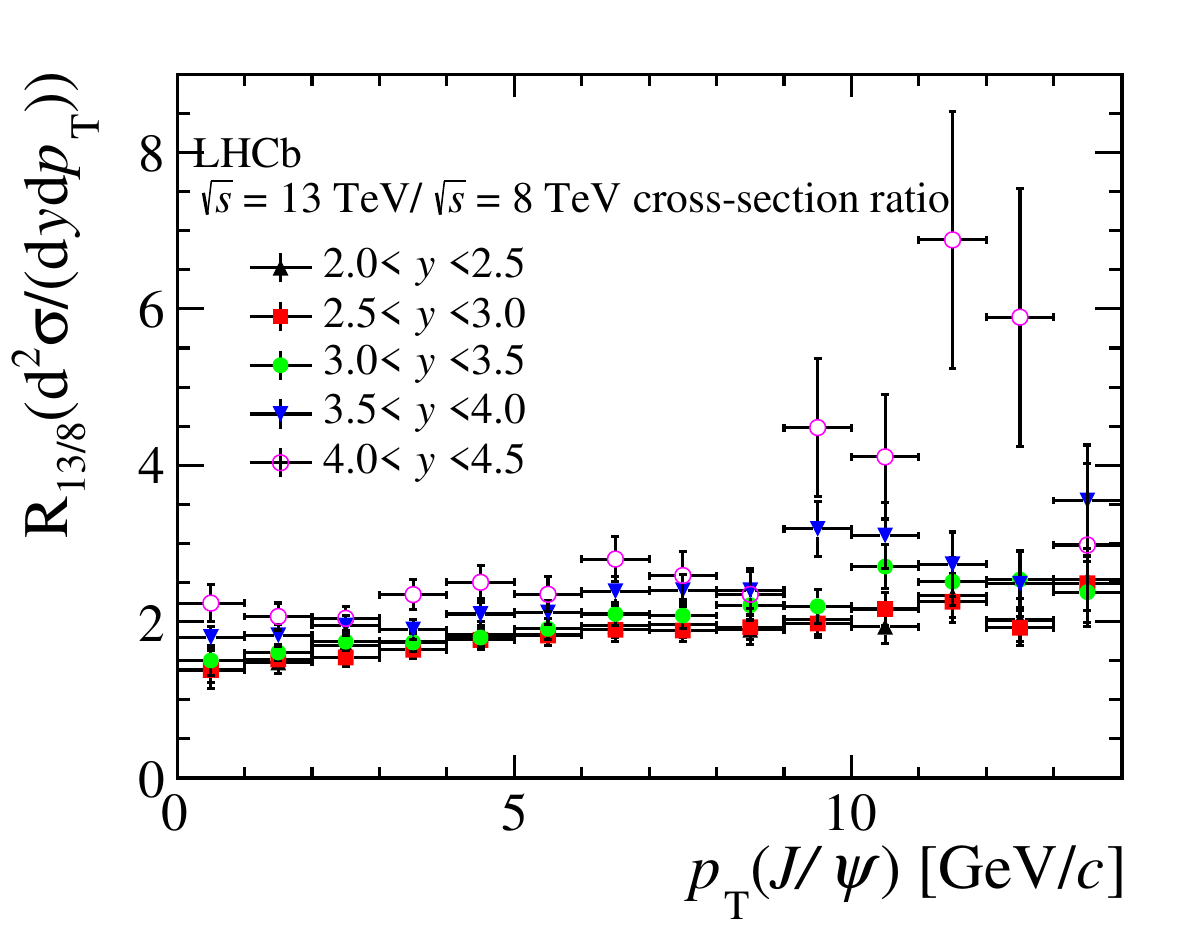}
\end{minipage}
\caption{Ratios of differential cross-sections between measurements at $\sqrt{s}=13\protect\tev$ and $\sqrt{s}=8\protect\tev$ as a function
   of \pt in bins of $y$ for (left) prompt \jpsi mesons and (right) \jpsi-from-$b$ mesons.} 
\label{fig:RatioAsDifferential}
\end{figure}

\begin{figure}[!tbp]
\centering
\begin{minipage}[t]{0.45\textwidth}
\centering
\includegraphics[width=1.0\textwidth]{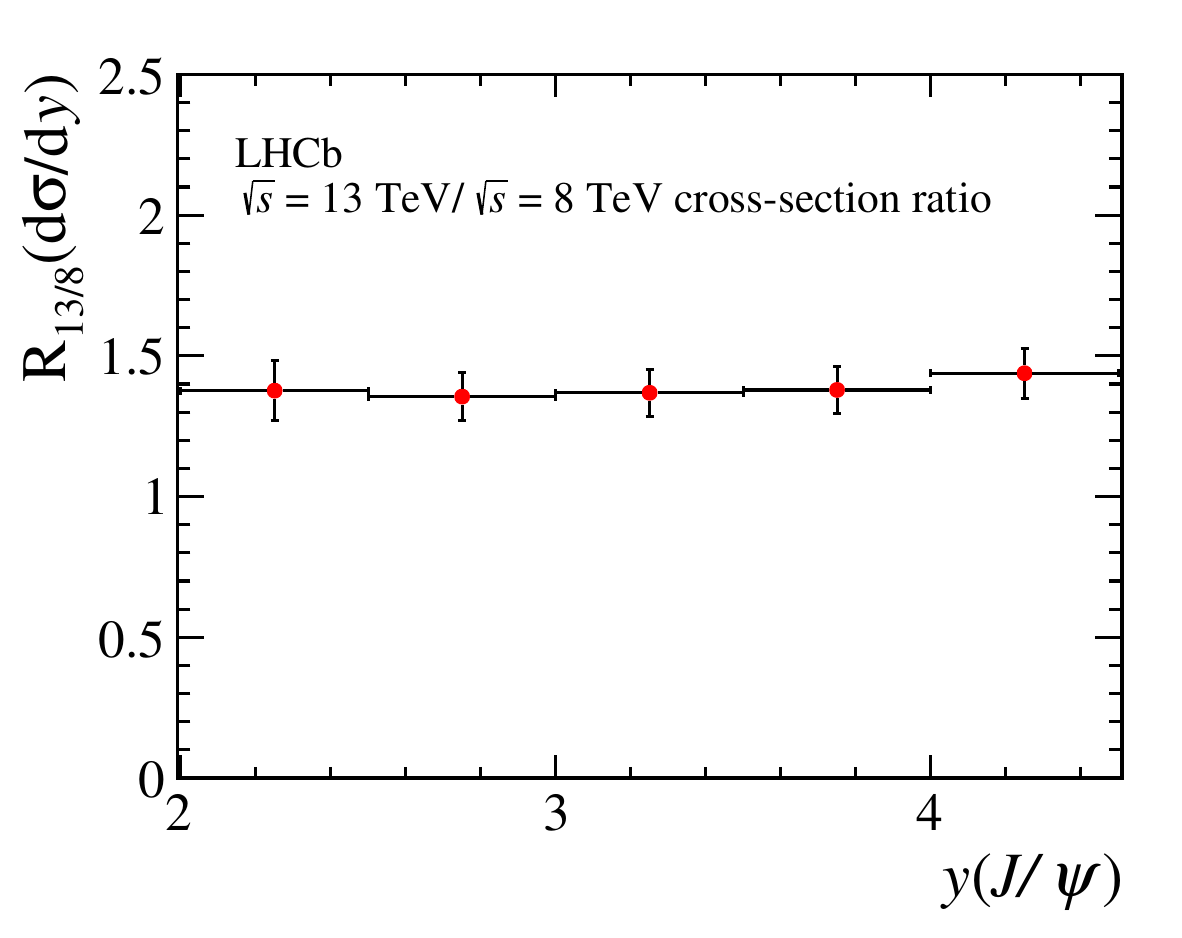}
\end{minipage}
\begin{minipage}[t]{0.45\textwidth}
\centering
\includegraphics[width=1.0\textwidth]{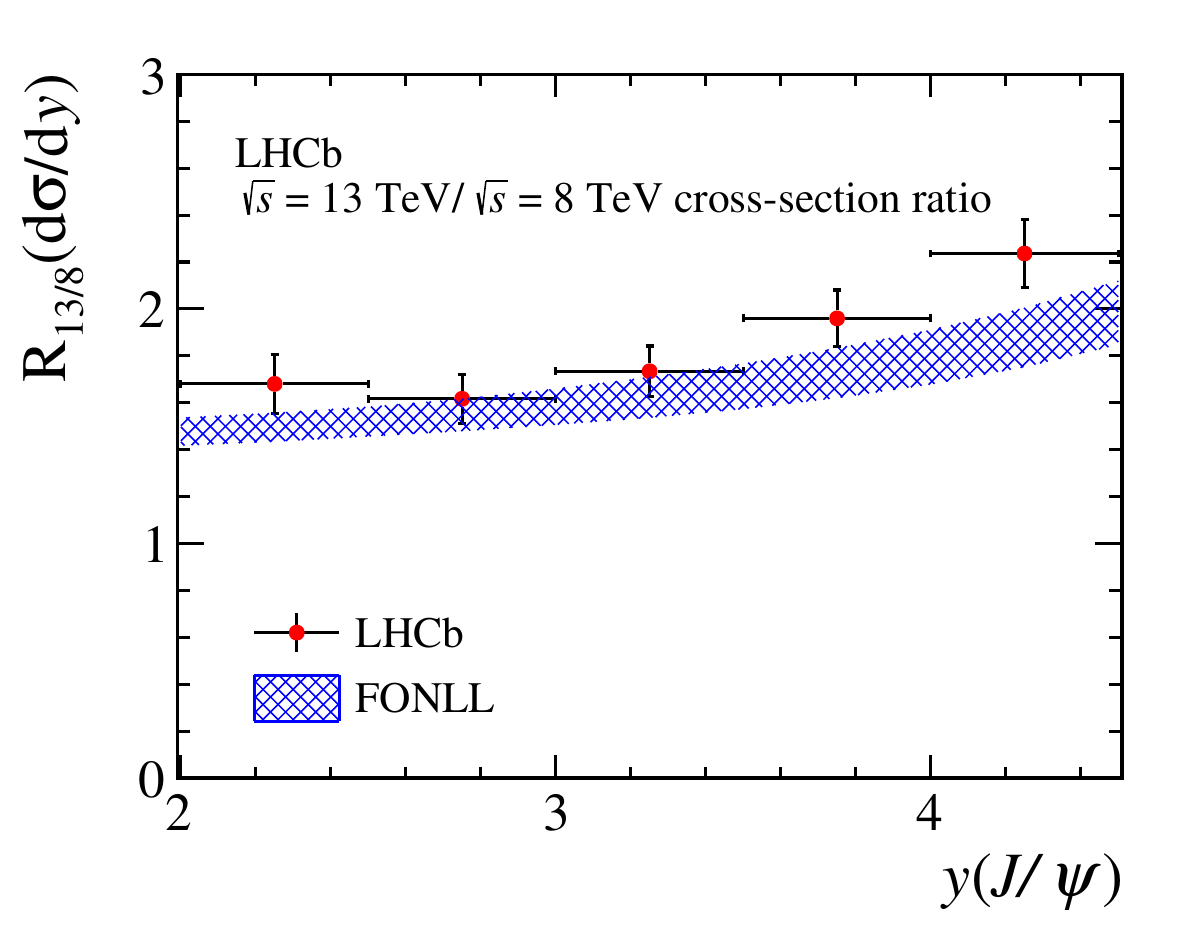}
\end{minipage}
\caption{Ratios of differential cross-sections between measurements at $\sqrt{s}=13\protect\tev$ and $\sqrt{s}=8\protect\tev$ as a function
   of $y$ integrated over \pt for (left) prompt \jpsi and (right) \jpsi-from-$b$ mesons. 
   The FONLL calculation~\cite{Cacciari:2015fta} is compared to the measured \jpsi-from-$b$ production ratio.
      } 
\label{fig:RatioAsY}
\end{figure}

\begin{figure}[!tbp]
\centering
\begin{minipage}[t]{0.45\textwidth}
\centering
\includegraphics[width=1.0\textwidth]{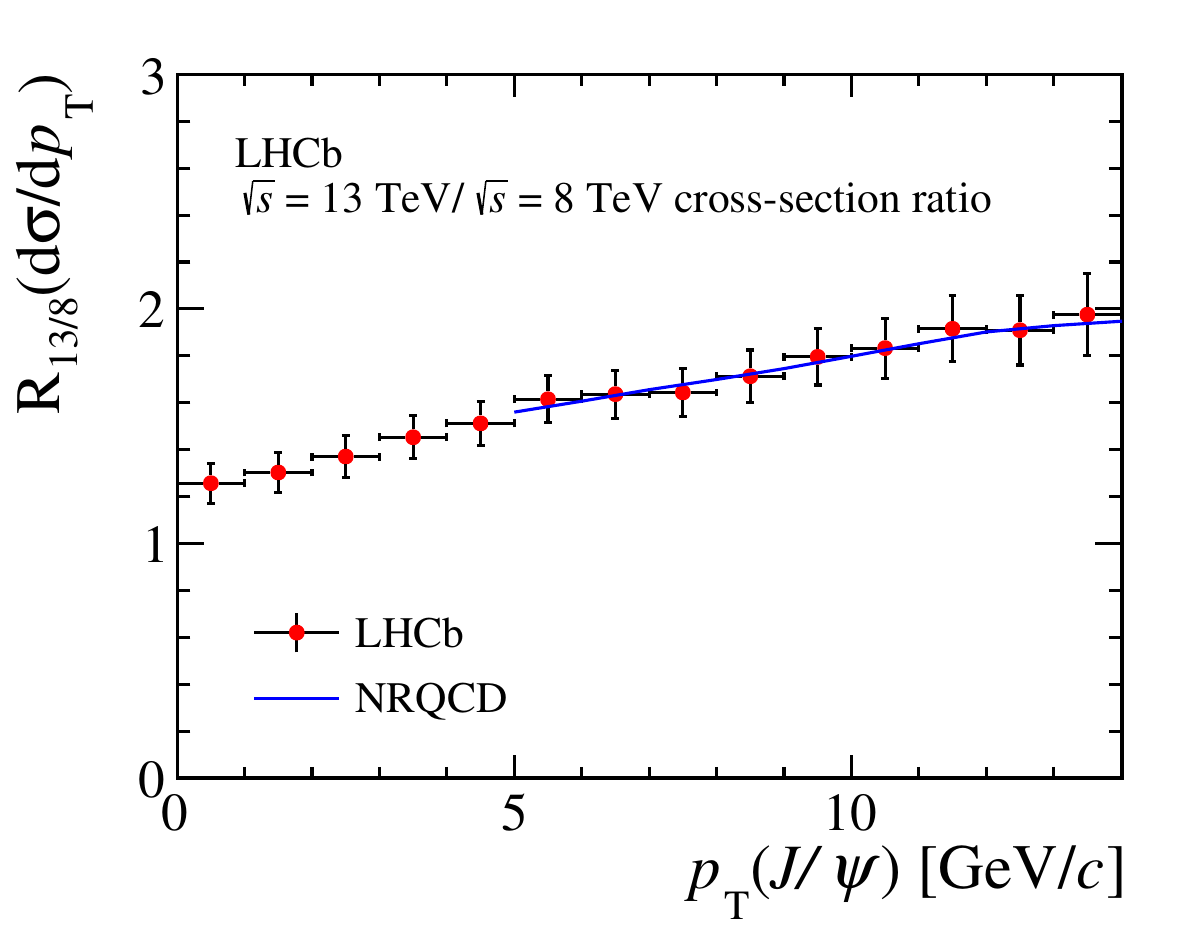}
\end{minipage}
\begin{minipage}[t]{0.45\textwidth}
\centering
\includegraphics[width=1.0\textwidth]{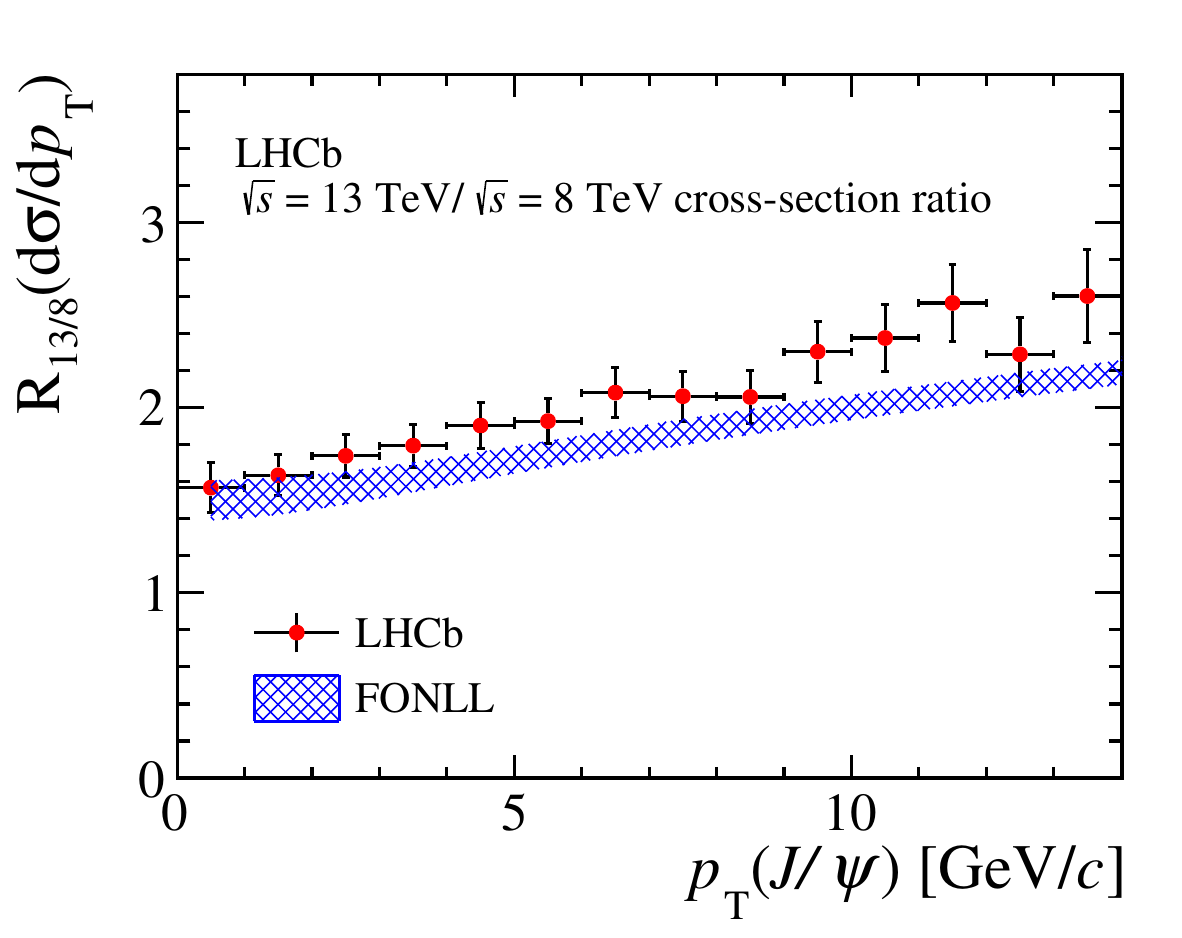}
\end{minipage}
\caption{Ratios of differential cross-sections between measurements at $\sqrt{s}=13\protect\tev$ and $\sqrt{s}=8\protect\tev$ as a function
   of $\pt$ integrated over $y$ for (left) prompt \jpsi mesons and (right) \jpsi-from-$b$ mesons. Calculations of
      NRQCD~\cite{Shao:2014yta} and FONLL~\cite{Cacciari:2015fta} are compared to prompt \jpsi mesons and \jpsi-from-$b$
      mesons, respectively.} 
\label{fig:RatioAsPT}
\end{figure}

In Fig.~\ref{fig:RatioAsDifferential}, the ratios $R_{13/8}$ of the double differential cross-sections in $pp$ collisions at
$\sqrt{s}=13 \tev$ and at $\sqrt{s}=8 \tev$ are given for prompt \jpsi and \jpsi-from-$b$ mesons, taking into account
the correlations of various systematic uncertainties. 
The ratios of the cross-sections in bins of $y$ integrated over \pt are shown in Fig.~\ref{fig:RatioAsY}, while those in bins of \pt integrated over $y$ are in Fig.~\ref{fig:RatioAsPT}.  
The cross-section ratios are summarised in Table~\ref{tab:XsectionRatio2D} and~\ref{tab:XsectionRatio2DFromB}
for prompt \jpsi and \jpsi-from-$b$ respectively.  

In the cross-section ratios, many of the systematic uncertainties cancel because of correlations between the two
measurements. 
The uncertainty of the luminosity determination, which is the dominating systematic uncertainty, is determined to be $50\%$ correlated~\cite{LHCb-PAPER-2014-047}, yielding a total uncertainty in the ratio of $4.6\%$.
The uncertainties due to the signal mass shape, vertex fit quality requirement, radiative tail, muon identification, tracking efficiency, $\jpsi\to\mumu$ branching fraction and trigger are also totally or partially correlated.
The remaining systematic uncertainties of the cross-section ratio are summarised in Table~\ref{tab:SystematicRatio}.


\begin{table}[bp]
\caption{Relative systematic uncertainty (in \%) on the ratio of the cross-section in $pp$ collisions at $\sqrt{s}=13
   \protect\tev$ relative to that at $\sqrt{s}=8\protect\tev$. The systematic uncertainty from $t_z$ fits only affects
      \jpsifromb.}
\centering
\renewcommand{\arraystretch}{1.}
\begin{tabular}{c|c}
\hline
Source  & Systematic uncertainty (\%) \\
\hline
Luminosity &  4.6 \\
Trigger   &  1.5 \\
Muon ID   &  2.2  \\
Tracking  &   2.0   \\
Signal mass shape & 2.0  \\
\pt, $y$\ spectrum, simulation statistics ($t_z$ fits)&   $1.1-18.9$ \\
\hline
\end{tabular}
\label{tab:SystematicRatio}
\end{table}


\begin{figure}[!tbp]
\centering
\begin{minipage}[t]{0.45\textwidth}
\centering
\includegraphics[width=1.0\textwidth]{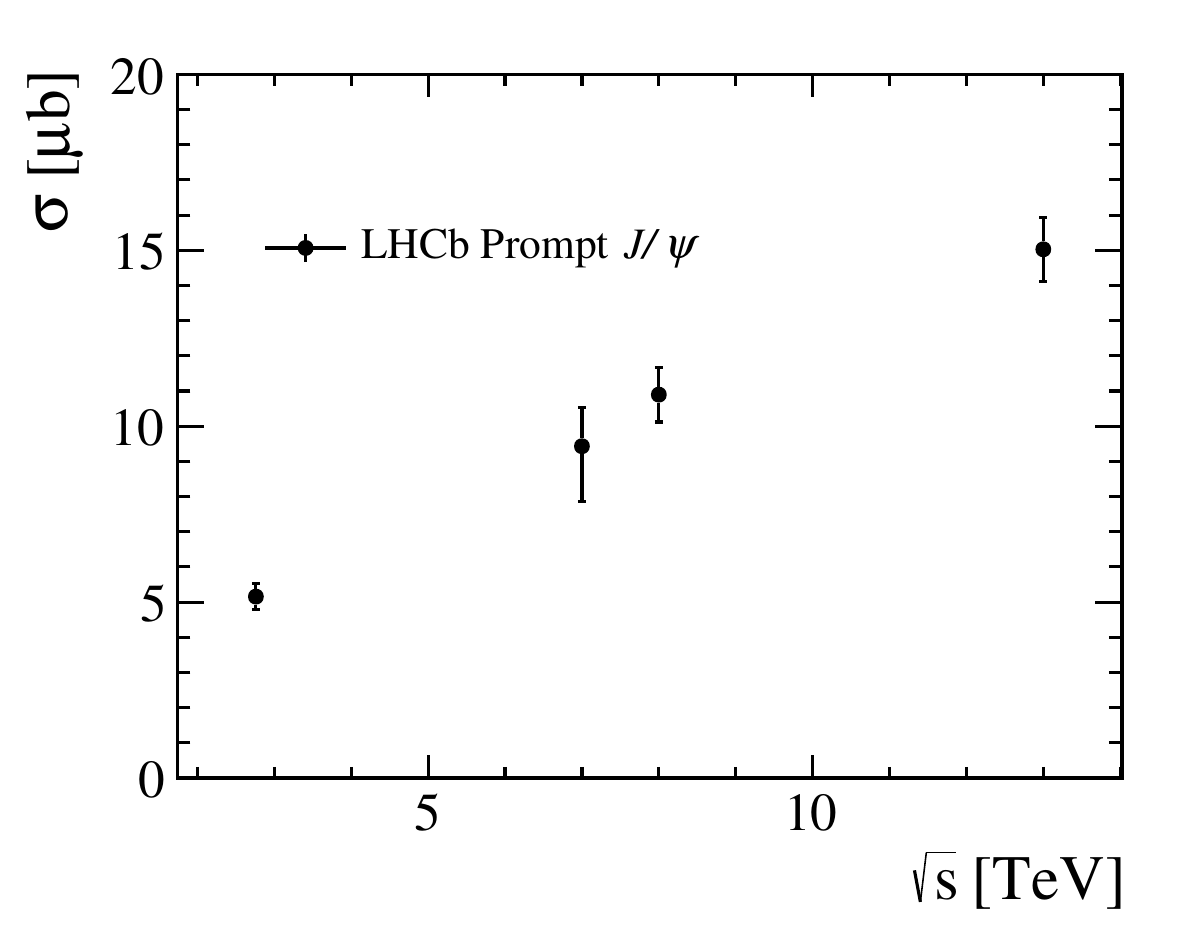}
\end{minipage}
\begin{minipage}[t]{0.45\textwidth}
\centering
\includegraphics[width=1.0\textwidth]{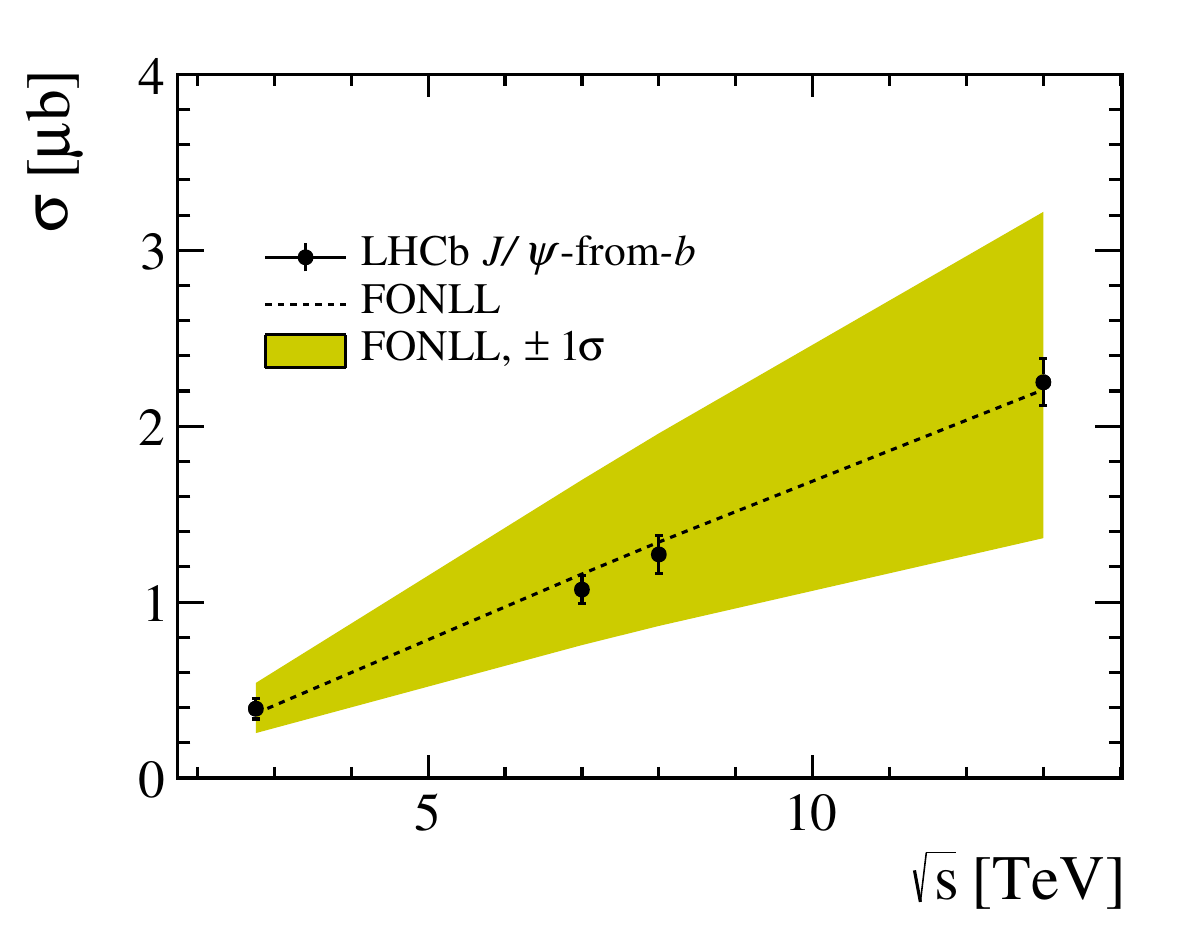}
\end{minipage}
\caption{The \jpsi production cross-section for (left) prompt \jpsi and (right) \jpsi-from-$b$\ mesons as a function of
   $pp$ collision energy in the \lhcb fiducial region compared to the FONLL calculation~\cite{Cacciari:1998it}. In
      general, the correlated and uncorrelated systematic uncertainties among different measurements are of comparable magnitude.} 
\label{fig:XsectionWithEnergy}
\end{figure}

In Fig.~\ref{fig:XsectionWithEnergy} the total cross-section in the fiducial region $\pt<14\gevc$ and $2.0<y<4.5$, as a function of $pp$ centre-of-mass energy, is shown for prompt \jpsi and \jpsi-from-$b$ mesons. 
The larger cross-section for \jpsi meson production at  $\sqrt{s}=13\protect\tev$ compared to  $\sqrt{s}=8\protect\tev$
is mostly due to the increased $pp$\ collision energy, but is also partly due to the increased boost of the produced
$b$-hadron into the fiducial region. 
In Table~\ref{tab:IntegratedCrossSectionWithEnergy}, the cross-sections of prompt \jpsi and \jpsi-from-$b$ mesons
integrated over the kinematic range  $2.0<y<4.5$, $\pt<14\gevc$ ($\pt<12\gevc$ for the analysis in $pp$ at
$\sqrt{s}=2.76\tev$) are given for $pp$ collisions in different centre-of-mass energies. 
The cross-section values for $pp$ collisions at $\sqrt{s}=2.76\tev$, $\sqrt{s}=7\tev$ and $\sqrt{s}=8\tev$ are taken from Refs.~\cite{LHCb-PAPER-2012-039,LHCb-PAPER-2013-008} and \cite{LHCb-PAPER-2013-016}. 
The uncertainties are split into parts that are correlated and uncorrelated between the measurements at different $pp$ centre-of-mass energies.

\begin{table}[!tbp]
\caption{\small  Production cross-sections of prompt \jpsi and \jpsi-from-$b$ mesons, integrated over the \lhcb fiducial
region, in $pp$ collisions at various centre-of-mass energies~\cite{LHCb-PAPER-2012-039,LHCb-PAPER-2013-008,LHCb-PAPER-2013-016}. The first uncertainty is the uncorrelated component, and the second the correlated one.}
\centering
\small
\begin{tabular}{ccccc}
\hline
$\sigma_\mathrm{tot}$ ($\mathrm{\mub}$)    & $\sqrt{s}=2.76\tev$   &$\sqrt{s}=7\tev$   &$\sqrt{s}=8\tev$    &$\sqrt{s}=13\tev$\\
\hline
Prompt \jpsi&$ 5.2   \pm0.3  \pm0.3$&$ 9.4   \pm0.5  ^{+0.7}_{-1.0}$&$10.9   \pm0.5  \pm0.6$&$15.0   \pm0.6  \pm0.7$\\
\jpsi-from-$b$&$0.39   \pm0.04  \pm0.04$&$1.07   \pm0.05  \pm0.06$&$1.27   \pm0.06  \pm0.09$&$2.25   \pm0.09  \pm0.10$\\
\hline
\end{tabular}\label{tab:IntegratedCrossSectionWithEnergy}
\end{table}

\subsection{Comparison with theoretical models}\label{sec:theorycomparison}

\begin{figure}[!tbp]
\centering
\begin{minipage}[t]{0.45\textwidth}
\centering
\includegraphics[width=1.0\textwidth]{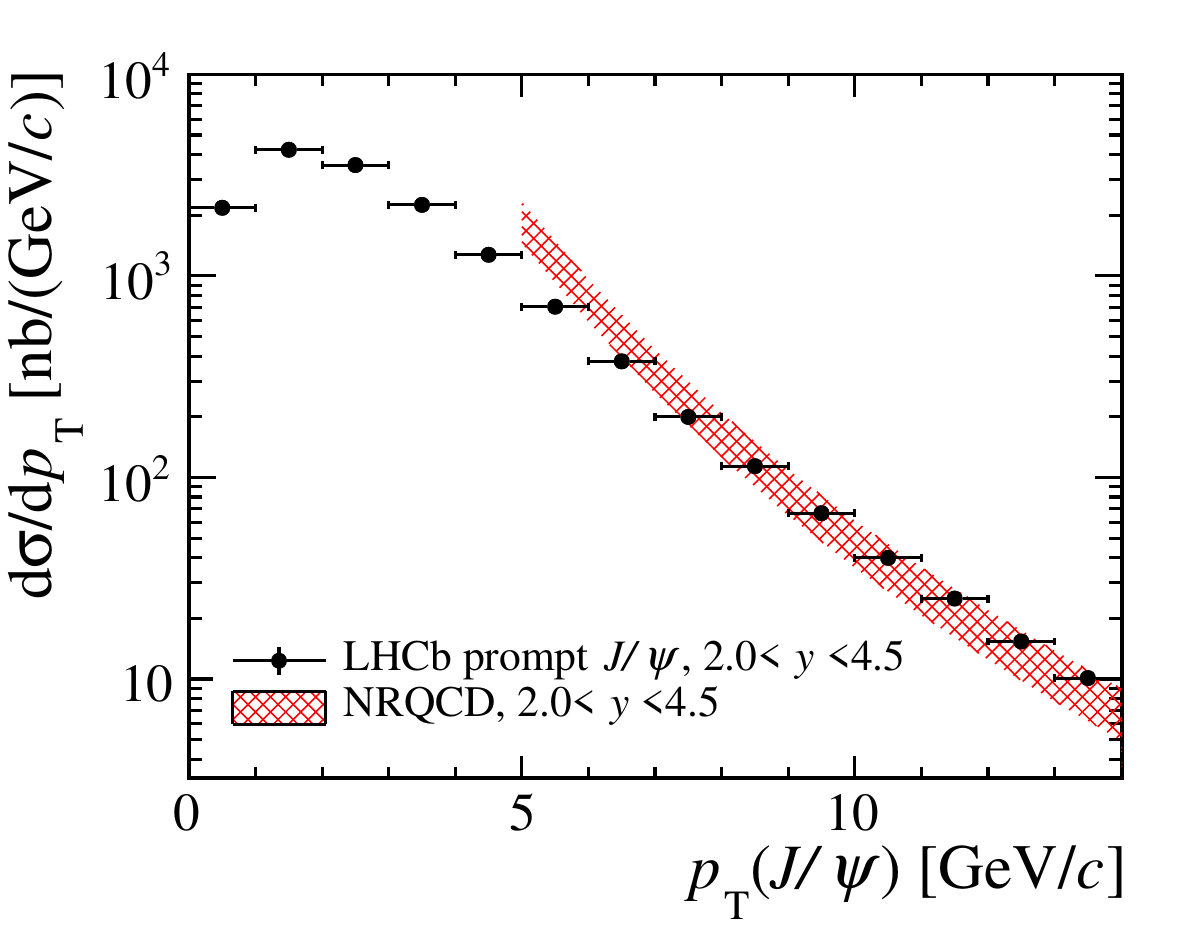}
\end{minipage}
\begin{minipage}[t]{0.45\textwidth}
\centering
\includegraphics[width=1.0\textwidth]{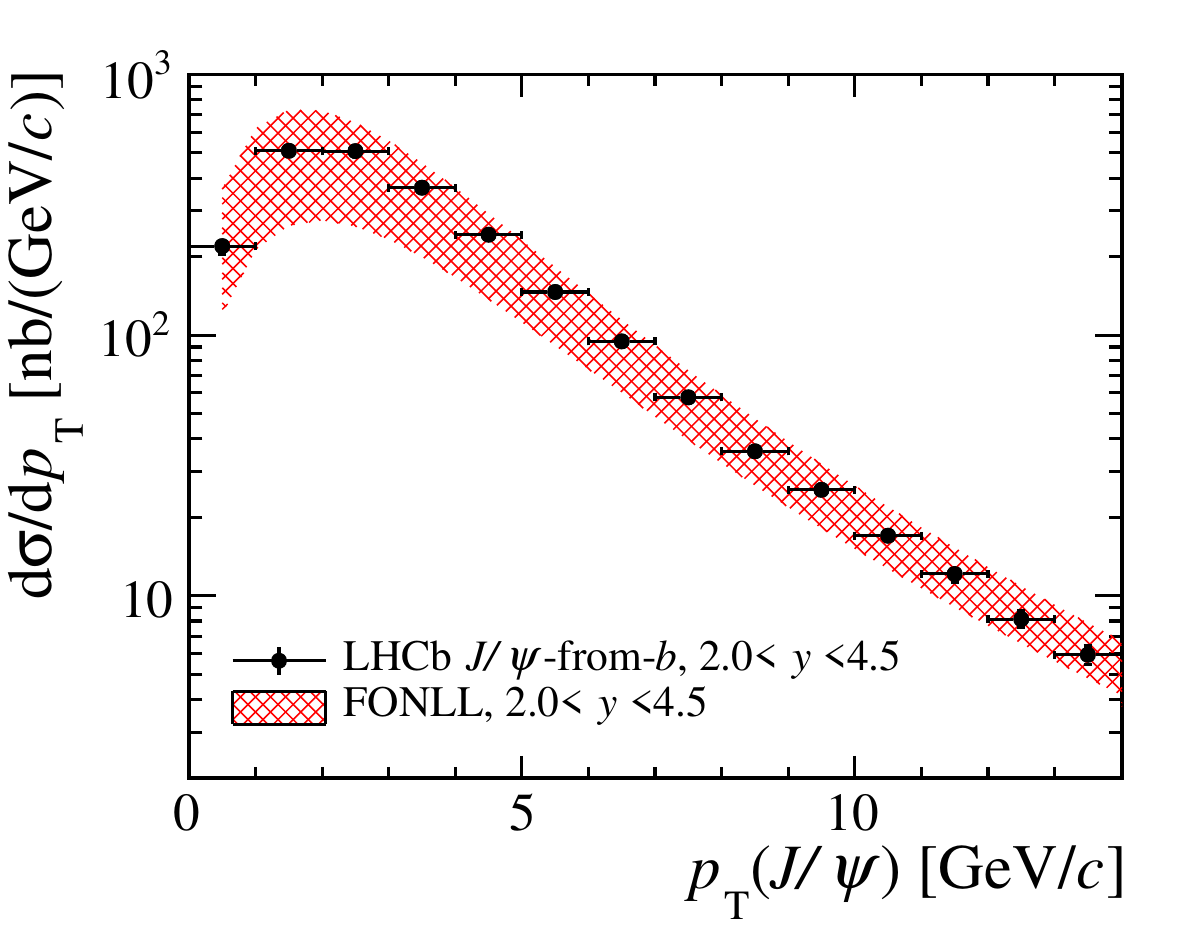}
\end{minipage}
\caption{Differential cross-sections as a function of \pt integrated over $y$ in the range $2.0<y<4.5$, (left) compared with the NRQCD calculation~\cite{Shao:2014yta} for prompt \jpsi and  (right) compared with the FONLL calculation~\cite{Cacciari:1998it} for \jpsi-from-$b$\ mesons.} 
\label{fig:TheoryPTComparison}
\end{figure}

The measured \jpsi cross-sections are compared to the calculations of NRQCD and FONLL for prompt \jpsi and for
\jpsi-from-$b$.  Figure~\ref{fig:TheoryPTComparison} (left) shows the comparison between the NRQCD
calculation~\cite{Shao:2014yta} and the measured prompt \jpsi cross-section as a function of transverse momentum,
   integrated over $y$ in the range $2.0<y<4.5$. 
In the NRQCD calculation, only uncertainties associated with LDME are considered since these are the dominating
uncertainties for the absolute production cross-section prediction. 
The FONLL calculation~\cite{Cacciari:1998it} is compared to the measurements of the \jpsi-from-$b$ cross-section as a
function of transverse momentum integrated over $y$ in the range $2.0<y<4.5$ in Fig.~\ref{fig:TheoryPTComparison}
(right). 
The FONLL calculation includes the uncertainties due to the $b$-quark mass and the renormalisation and factorisation scales for the prediction of the absolute production cross-section. 
Good agreement is found between the measurements and the theoretical calculations.

Fig.~\ref{fig:RatioAsY}
(right) shows the ratio of the cross-sections as a function of $y$ integrated over \pt in the range $\pt<14\gevc$ compared
with the FONLL calculation based on Ref.~\cite{Cacciari:2015fta} for \jpsi-from-$b$. 
The ratio of the cross-sections as a function of \pt integrated over $y$ in the range $2.0<y<4.5$ is compared with the
NRQCD calculation~\cite{Shao:2014yta} for prompt \jpsi mesons, and with predictions by FONLL based
on Ref.~\cite{Cacciari:2015fta} for \jpsifromb, shown in Fig.~\ref{fig:RatioAsPT}. 
The uncertainty of the NRQCD prediction, considering only that from LDME, almost cancels in the cross-section ratio
between the $13\tev$ and $8\tev$ measurements, so no uncertainty is given for the calculations in
Fig.~\ref{fig:RatioAsPT} (left). 
Besides those due to the $b$-quark mass and the scales, the FONLL calculation for the cross-section ratio also takes into account the gluon PDF uncertainty.
The NRQCD prediction agrees remarkably well with the experimental data for the prompt \jpsi production cross-section ratio,
    while the FONLL prediction also provides a reasonably good agreement with our measurements for the \jpsi-from-$b$
cross-section ratio. 

\begin{table}
      \caption{\small The fraction of \jpsi-from-$b$ mesons (in \%) in bins of the \jpsi transverse momentum and
         rapidity. The uncertainties are statistical only. The systematic uncertainties are negligible.}
      \centering
      \begin{tabular}{c|ccccc}
      \hline
      $\pt[\!\gevc]$& $2<y<2.5$& $2.5<y<3$& $3<y<3.5$& $3.5<y<4$& $4<y<4.5$\\
      \hline

$ 0- 1$&$\,\,\,9.8\pm 0.5$& $\,\,\,9.2\pm 0.3$& $\,\,\,9.2\pm 0.3$& $\,\,\,8.6\pm 0.3$& $\,\,\,7.6\pm 0.4$\\
 $ 1- 2$&$11.4\pm 0.3$& $11.2\pm 0.2$& $10.9\pm 0.2$& $10.5\pm 0.2$& $\,\,\,9.1\pm 0.3$\\
 $ 2- 3$&$14.0\pm 0.3$& $12.6\pm 0.2$& $12.5\pm 0.2$& $12.2\pm 0.2$& $10.1\pm 0.3$\\
 $ 3- 4$&$16.0\pm 0.3$& $14.2\pm 0.2$& $13.9\pm 0.2$& $12.9\pm 0.2$& $11.8\pm 0.4$\\
 $ 4- 5$&$17.3\pm 0.4$& $17.0\pm 0.3$& $15.3\pm 0.3$& $15.2\pm 0.3$& $13.8\pm 0.5$\\
 $ 5- 6$&$18.7\pm 0.5$& $18.2\pm 0.3$& $16.9\pm 0.3$& $15.7\pm 0.4$& $14.3\pm 0.6$\\
 $ 6- 7$&$21.0\pm 0.6$& $21.1\pm 0.4$& $19.7\pm 0.5$& $19.3\pm 0.6$& $16.9\pm 0.9$\\
 $ 7- 8$&$25.3\pm 0.8$& $23.0\pm 0.6$& $21.3\pm 0.6$& $21.1\pm 0.7$& $17.5\pm 1.1$\\
 $ 8- 9$&$25.9\pm 1.0$& $25.6\pm 0.8$& $23.7\pm 0.8$& $21.6\pm 1.0$& $18.0\pm 1.5$\\
 $\,\,\, 9-10$&$29.5\pm 1.2$& $26.3\pm 1.0$& $27.2\pm 1.1$& $27.9\pm 1.3$& $26.6\pm 2.4$\\
 $10-11$&$29.6\pm 1.5$& $31.5\pm 1.3$& $30.9\pm 1.4$& $26.5\pm 1.7$& $28.4\pm 2.8$\\
 $11-12$&$34.0\pm 1.9$& $33.3\pm 1.6$& $28.1\pm 1.8$& $31.5\pm 2.3$& $36.9\pm 3.5$\\
 $12-13$&$35.8\pm 2.3$& $36.5\pm 2.1$& $33.3\pm 2.2$& $29.3\pm 2.7$& $34.0\pm 5.2$\\
 $13-14$&$43.6\pm 2.7$& $37.3\pm 2.3$& $33.4\pm 2.8$& $41.7\pm 3.9$& $15.2\pm 4.3$\\
 \hline
      \end{tabular}\label{tab:BFraction}
      \end{table}

\begin{table}[!tbp]
\caption{\small Differential cross-sections $d\sigma/dp_\mathrm{T}$ (in  $\mathrm{nb}/(\!\gevc)$) for prompt \jpsi and \jpsi-from-$b$ mesons, integrated over $y$. The
first uncertainties are statistical and the second (third) are uncorrelated (correlated) systematic uncertainties
amongst bins.}
\centering
\small
\begin{tabular}{@{}r@{}c@{}lrr@{}c@{}r@{}c@{}r@{}c@{}rr@{}c@{}r@{}c@{}r@{}c@{}rr@{}}
\hline
\multicolumn{4}{c}{$\pt\,[\!\gevc]$} & \multicolumn{7}{c}{Prompt \jpsi} & \multicolumn{7}{c}{\jpsi-from-$b$} \\ 
\hline
$0$&$-$&$1$&&$  \,2177\, \tpm \, 10 \, \tpm \, 17 \, \tpm \, 146$&$ 219.4 \, \tpm \, 3.9 \, \tpm \, 1.8 \, \tpm \, 14.8$\\
 $1$&$-$&$2$&&$  4226 \, \tpm \, 14 \, \tpm \, 29 \, \tpm \, 278$&$ 509.2 \, \tpm \, 4.9 \, \tpm \, 3.6 \, \tpm \, 33.5$\\
 $2$&$-$&$3$&&$  3548 \, \tpm \, 12 \, \tpm \, 26 \, \tpm \, 223$&$ 507.6 \, \tpm \, 4.4 \, \tpm \, 4.0 \, \tpm \, 31.9$\\
$3$&$-$&$4$&&$  2251 \, \tpm \,  9 \, \tpm \, 16 \, \tpm \, 134$&$ 367.6 \, \tpm \, 3.5 \, \tpm \, 2.7 \, \tpm \, 21.9$\\
 $4$&$-$&$5$&&$  1273 \, \tpm \,  5 \, \tpm \,  9 \, \tpm \, 72$&$ 242.7 \, \tpm \, 2.5 \, \tpm \, 1.9 \, \tpm \, 13.8$\\
$5$&$-$&$6$&&$ 703.7 \, \tpm \, 3.8 \, \tpm \, 6.0 \, \tpm \, 38.9$&$ 146.3 \, \tpm \, 1.8 \, \tpm \, 1.3 \, \tpm \, 8.1$\\
$6$&$-$&$7$&&$ 376.8 \, \tpm \, 2.6 \, \tpm \, 3.7 \, \tpm \, 20.5$&$  94.6 \, \tpm \, 1.3 \, \tpm \, 0.9 \, \tpm \, 5.1$\\
$7$&$-$&$8$&&$ 199.7 \, \tpm \, 1.7 \, \tpm \, 2.4 \, \tpm \, 10.8$&$  57.7 \, \tpm \, 1.0 \, \tpm \, 0.7 \, \tpm \, 3.1$\\
$8$&$-$&$9$&&$ 113.8 \, \tpm \, 1.2 \, \tpm \, 1.6 \, \tpm \, 6.1$&$  35.8 \, \tpm \, 0.7 \, \tpm \, 0.5 \, \tpm \, 1.9$\\
$9$&$-$&$10$&& $  66.5 \, \tpm \, 0.9 \, \tpm \, 1.2 \, \tpm \, 3.6$&$  25.5 \, \tpm \, 0.6 \, \tpm \, 0.5 \, \tpm \, 1.4$\\
$10$&$-$&$11$&&$  39.9 \, \tpm \, 0.7 \, \tpm \, 0.9 \, \tpm \, 2.1$&$  17.0 \, \tpm \, 0.5 \, \tpm \, 0.4 \, \tpm \, 0.9$\\
$11$&$-$&$12$&&$  25.1 \, \tpm \, 0.6 \, \tpm \, 0.7 \, \tpm \, 1.3$&$  12.1 \, \tpm \, 0.4 \, \tpm \, 0.3 \, \tpm \, 0.6$\\
$12$&$-$&$13$&&$  15.4 \, \tpm \, 0.4 \, \tpm \, 0.5 \, \tpm \, 0.8$&$   8.1 \, \tpm \, 0.3 \, \tpm \, 0.3 \, \tpm \, 0.4$\\
$13$&$-$&$14$&&$  10.1 \, \tpm \, 0.3 \, \tpm \, 0.4 \, \tpm \, 0.5$&$   5.9 \, \tpm \, 0.3 \, \tpm \, 0.2 \, \tpm \, 0.3$\\
\hline
\end{tabular}\label{tab:CrossSectionWithPT}
\end{table}

\begin{table}[!tbp]
\caption{\small Differential cross-sections $d\sigma/dy$ (in \mub) for prompt \jpsi and \jpsi-from-$b$ mesons, integrated over \pt. The
first uncertainties are statistical and the second (third) are the uncorrelated (correlated) systematic uncertainties.}
\centering
\small
\begin{tabular}{c|cc}
\hline
$y$      &Prompt \jpsi   &\jpsi-from-$b$ \\
\hline
$2.0-2.5$&$ 7.049 \pm 0.033 \pm 0.072 \pm 0.516$&$ 1.201 \pm 0.012 \pm 0.011 \pm 0.083$\\
$2.5-3.0$&$ 6.840 \pm 0.021 \pm 0.029 \pm 0.390$&$ 1.073 \pm 0.007 \pm 0.004 \pm 0.060$\\
$3.0-3.5$&$ 6.236 \pm 0.018 \pm 0.024 \pm 0.350$&$ 0.930 \pm 0.006 \pm 0.003 \pm 0.052$\\
$3.5-4.0$&$ 5.413 \pm 0.017 \pm 0.025 \pm 0.344$&$ 0.759 \pm 0.006 \pm 0.003 \pm 0.048$\\
$4.0-4.5$&$ 4.519 \pm 0.020 \pm 0.043 \pm 0.343$&$ 0.536 \pm 0.008 \pm 0.005 \pm 0.040$\\
\hline
\end{tabular}\label{tab:CrossSectionWithY}
\end{table}

\begin{table}[!tbp]
\caption{\small The ratio of cross-sections between measurements at 13 TeV and 8 TeV in different bins of \pt and $y$
for prompt \jpsi mesons.  The systematic errors are negligible.}
\centering
\scalebox{0.82}{
\begin{tabular}{c|ccccc|c}
\hline
$\pt\,[\!\gevc]$& $2<y<2.5$& $2.5<y<3$& $3<y<3.5$& $3.5<y<4$& $4<y<4.5$& $2<y<4.5$\\
\hline
$ 0- 1$&$1.25\pm 0.14  $&$1.24\pm 0.09  $&$1.23\pm 0.08  $&$1.27\pm 0.08  $&$1.30\pm 0.09  $&$1.26\pm 0.09  $\\
$ 1- 2$&$1.29\pm 0.12  $&$1.28\pm 0.09  $&$1.30\pm 0.08  $&$1.28\pm 0.08  $&$1.39\pm 0.09  $&$1.30\pm 0.09  $\\
$ 2- 3$&$1.38\pm 0.11  $&$1.35\pm 0.09  $&$1.36\pm 0.08  $&$1.38\pm 0.08  $&$1.39\pm 0.09  $&$1.37\pm 0.09  $\\
$ 3- 4$&$1.41\pm 0.11  $&$1.43\pm 0.09  $&$1.43\pm 0.09  $&$1.48\pm 0.09  $&$1.56\pm 0.11  $&$1.45\pm 0.09  $\\
$ 4- 5$&$1.52\pm 0.13  $&$1.44\pm 0.09  $&$1.48\pm 0.09  $&$1.56\pm 0.10  $&$1.62\pm 0.11  $&$1.51\pm 0.10  $\\
$ 5- 6$&$1.60\pm 0.11  $&$1.58\pm 0.10  $&$1.61\pm 0.10  $&$1.60\pm 0.10  $&$1.76\pm 0.14  $&$1.61\pm 0.10  $\\
$ 6- 7$&$1.67\pm 0.12  $&$1.56\pm 0.11  $&$1.61\pm 0.11  $&$1.64\pm 0.11  $&$1.80\pm 0.14  $&$1.64\pm 0.10  $\\
$ 7- 8$&$1.58\pm 0.12  $&$1.53\pm 0.10  $&$1.68\pm 0.12  $&$1.75\pm 0.13  $&$1.87\pm 0.17  $&$1.64\pm 0.10  $\\
$ 8- 9$&$1.58\pm 0.14  $&$1.64\pm 0.12  $&$1.79\pm 0.14  $&$1.82\pm 0.15  $&$2.02\pm 0.20  $&$1.71\pm 0.11  $\\
$\,\,\, 9-10$&$1.71\pm 0.15  $&$1.78\pm 0.14  $&$1.77\pm 0.15  $&$1.90\pm 0.17  $&$2.01\pm 0.25  $&$1.80\pm 0.12  $\\
$10-11$&$1.76\pm 0.17  $&$1.69\pm 0.14  $&$1.92\pm 0.17  $&$1.94\pm 0.20  $&$2.20\pm 0.32  $&$1.83\pm 0.13  $\\
$11-12$&$1.94\pm 0.21  $&$1.75\pm 0.18  $&$2.20\pm 0.21  $&$1.68\pm 0.19  $&$2.12\pm 0.36  $&$1.92\pm 0.14  $\\
$12-13$&$1.76\pm 0.21  $&$1.67\pm 0.17  $&$2.08\pm 0.24  $&$1.96\pm 0.25  $&$2.99\pm 0.60  $&$1.91\pm 0.15  $\\
$13-14$&$1.75\pm 0.25  $&$1.98\pm 0.25  $&$1.89\pm 0.26  $&$1.40\pm 0.21  $&$5.94\pm 1.57  $&$1.98\pm 0.17  $\\
\hline
$0-14$& $1.38\pm 0.11  $&$1.36\pm 0.09  $&$1.37\pm 0.08  $&$1.38\pm 0.08  $&$1.44\pm 0.09  $&---\\
\hline
\end{tabular}
}
\label{tab:XsectionRatio2D}
\end{table}

\begin{table}[!tbp]
\caption{\small The ratio of cross-sections between measurements at 13 TeV and 8 TeV in different bins of \pt and $y$
for \jpsi-from-$b$ mesons.  The systematic uncertainties are negligible.}
\centering
\scalebox{0.82}{
\begin{tabular}{c|ccccc|c}
\hline
$\pt\,[\!\gevc]$& $2<y<2.5$& $2.5<y<3$& $3<y<3.5$& $3.5<y<4$& $4<y<4.5$& $2<y<4.5$\\
\hline
$ 0- 1$&$1.39\pm 0.24  $&$1.38\pm 0.16  $&$1.50\pm 0.19  $&$1.80\pm 0.14  $&$2.24\pm 0.24  $&$1.57\pm 0.13  $\\
$ 1- 2$&$1.48\pm 0.14  $&$1.52\pm 0.11  $&$1.61\pm 0.11  $&$1.83\pm 0.14  $&$2.07\pm 0.17  $&$1.63\pm 0.11  $\\
$ 2- 3$&$1.70\pm 0.15  $&$1.54\pm 0.11  $&$1.74\pm 0.11  $&$1.95\pm 0.13  $&$2.04\pm 0.15  $&$1.74\pm 0.11  $\\
$ 3- 4$&$1.75\pm 0.14  $&$1.64\pm 0.11  $&$1.73\pm 0.11  $&$1.90\pm 0.13  $&$2.35\pm 0.19  $&$1.80\pm 0.12  $\\
$ 4- 5$&$1.84\pm 0.16  $&$1.77\pm 0.12  $&$1.80\pm 0.12  $&$2.10\pm 0.14  $&$2.50\pm 0.21  $&$1.90\pm 0.12  $\\
$ 5- 6$&$1.84\pm 0.14  $&$1.82\pm 0.13  $&$1.91\pm 0.13  $&$2.12\pm 0.15  $&$2.35\pm 0.22  $&$1.93\pm 0.12  $\\
$ 6- 7$&$1.95\pm 0.15  $&$1.90\pm 0.15  $&$2.10\pm 0.15  $&$2.39\pm 0.18  $&$2.80\pm 0.29  $&$2.08\pm 0.13  $\\
$ 7- 8$&$1.96\pm 0.17  $&$1.89\pm 0.14  $&$2.08\pm 0.16  $&$2.40\pm 0.20  $&$2.59\pm 0.31  $&$2.06\pm 0.14  $\\
$ 8- 9$&$1.90\pm 0.19  $&$1.93\pm 0.16  $&$2.21\pm 0.20  $&$2.40\pm 0.24  $&$2.35\pm 0.33  $&$2.06\pm 0.14  $\\
$\,\,\, 9-10$&$2.03\pm 0.20  $&$1.98\pm 0.18  $&$2.20\pm 0.22  $&$3.19\pm 0.35  $&$4.48\pm 0.89  $&$2.30\pm 0.16  $\\
$10-11$&$1.93\pm 0.22  $&$2.16\pm 0.21  $&$2.70\pm 0.28  $&$3.10\pm 0.42  $&$4.11\pm 0.79  $&$2.38\pm 0.18  $\\
$11-12$&$2.33\pm 0.28  $&$2.25\pm 0.26  $&$2.51\pm 0.31  $&$2.73\pm 0.41  $&$6.88\pm 1.64  $&$2.57\pm 0.21  $\\
$12-13$&$2.02\pm 0.28  $&$1.92\pm 0.23  $&$2.54\pm 0.36  $&$2.48\pm 0.42  $&$5.89\pm 1.65  $&$2.29\pm 0.20  $\\
$13-14$&$2.54\pm 0.40  $&$2.49\pm 0.35  $&$2.38\pm 0.39  $&$3.55\pm 0.71  $&$2.98\pm 1.04  $&$2.60\pm 0.25  $\\
\hline
$0-14$ &$1.68\pm 0.13  $&$1.62\pm 0.10  $&$1.73\pm 0.11  $&$1.96\pm 0.12  $&$2.24\pm 0.15  $&---\\
\hline
\end{tabular}}\label{tab:XsectionRatio2DFromB}
\end{table}

\FloatBarrier

\section{Conclusions}\label{sec:conclusion}
The differential \jpsi production cross-section in $pp$ collisions at $\sqrt{s}=13\tev$ is measured as a function of the
\jpsi transverse momentum and rapidity in the range of $\pt<14\gev$ and $2.0<y<4.5$. 
The analysis is based on a data sample corresponding to an integrated luminosity of $3.05\pm 0.12 \invpb$, collected with the \lhcb detector in July 2015. 
The production cross-sections of prompt \jpsi and \jpsi-from-$b$\ mesons are measured separately. 
The ratios of the \jpsi cross-sections in $pp$ collisions at a centre-of-mass energy of $13\tev$ relative to those at
$8\tev$ are also determined.

The \pt distribution of \jpsi mesons produced in $\sqrt{s}=13\tev$ $pp$ collisions is harder than at $\sqrt{s}=8\tev$.
The measured prompt \jpsi meson production cross-section as a function of transverse momentum is in good
agreement with theoretical calculations in the NRQCD framework.
Theoretical predictions based on FONLL calculations describe well the measured cross-section for \jpsi-from-$b$ mesons
and its dependence on the centre-of-mass energy of $pp$ collisions. The predicted ratio between the
cross-section at $\sqrt{s}=13\tev$\ and $\sqrt{s}=8\tev$ is also consistent with data.

\section*{Acknowledgements}
 
\noindent 
The authors would like to thank K.-T. Chao, H. Han and H.-S. Shao for providing the NRQCD calculations, and M. Cacciari, M. L.
Mangano and P. Nason for the FONLL predictions that are compared with the measurements discussed in the paper.
We express our gratitude to our colleagues in the CERN
accelerator departments for the excellent performance of the LHC. We
thank the technical and administrative staff at the LHCb
institutes. We acknowledge support from CERN and from the national
agencies: CAPES, CNPq, FAPERJ and FINEP (Brazil); NSFC (China);
CNRS/IN2P3 (France); BMBF, DFG, HGF and MPG (Germany); INFN (Italy); 
FOM and NWO (The Netherlands); MNiSW and NCN (Poland); MEN/IFA (Romania); 
MinES and FANO (Russia); MinECo (Spain); SNSF and SER (Switzerland); 
NASU (Ukraine); STFC (United Kingdom); NSF (USA).
The Tier1 computing centres are supported by IN2P3 (France), KIT and BMBF 
(Germany), INFN (Italy), NWO and SURF (The Netherlands), PIC (Spain), GridPP 
(United Kingdom).
We are indebted to the communities behind the multiple open 
source software packages on which we depend. We are also thankful for the 
computing resources and the access to software R\&D tools provided by Yandex LLC (Russia).
Individual groups or members have received support from 
EPLANET, Marie Sk\l{}odowska-Curie Actions and ERC (European Union), 
Conseil g\'{e}n\'{e}ral de Haute-Savoie, Labex ENIGMASS and OCEVU, 
R\'{e}gion Auvergne (France), RFBR (Russia), XuntaGal and GENCAT (Spain), Royal Society and Royal
Commission for the Exhibition of 1851 (United Kingdom).

\addcontentsline{toc}{section}{References}
\setboolean{inbibliography}{true}
\bibliographystyle{LHCb}
\bibliography{main,local,LHCb-PAPER,LHCb-CONF,LHCb-DP,LHCb-TDR}

\ifx\mcitethebibliography\mciteundefinedmacro
\PackageError{LHCb.bst}{mciteplus.sty has not been loaded}
{This bibstyle requires the use of the mciteplus package.}\fi
\providecommand{\href}[2]{#2}
\begin{mcitethebibliography}{10}
\mciteSetBstSublistMode{n}
\mciteSetBstMaxWidthForm{subitem}{\alph{mcitesubitemcount})}
\mciteSetBstSublistLabelBeginEnd{\mcitemaxwidthsubitemform\space}
{\relax}{\relax}

\bibitem{Carlson:1976cd}
C.~E. Carlson and R.~Suaya,
  \ifthenelse{\boolean{articletitles}}{\emph{{Hadronic production of the
  $\psi/J$ meson}}, }{}\href{http://dx.doi.org/10.1103/PhysRevD.14.3115}{Phys.\
  Rev.\  \textbf{D14} (1976) 3115}\relax
\mciteBstWouldAddEndPuncttrue
\mciteSetBstMidEndSepPunct{\mcitedefaultmidpunct}
{\mcitedefaultendpunct}{\mcitedefaultseppunct}\relax
\EndOfBibitem
\bibitem{Donnachie:1976ue}
A.~Donnachie and P.~V. Landshoff,
  \ifthenelse{\boolean{articletitles}}{\emph{{Production of lepton pairs,
  $J/\psi$ and charm with hadron beams}},
  }{}\href{http://dx.doi.org/10.1016/0550-3213(76)90532-0}{Nucl.\ Phys.\
  \textbf{B112} (1976) 233}\relax
\mciteBstWouldAddEndPuncttrue
\mciteSetBstMidEndSepPunct{\mcitedefaultmidpunct}
{\mcitedefaultendpunct}{\mcitedefaultseppunct}\relax
\EndOfBibitem
\bibitem{Ellis:1976fj}
S.~D. Ellis, M.~B. Einhorn, and C.~Quigg,
  \ifthenelse{\boolean{articletitles}}{\emph{{Comment on hadronic production of
  psions}}, }{}\href{http://dx.doi.org/10.1103/PhysRevLett.36.1263}{Phys.\
  Rev.\ Lett.\  \textbf{36} (1976) 1263}\relax
\mciteBstWouldAddEndPuncttrue
\mciteSetBstMidEndSepPunct{\mcitedefaultmidpunct}
{\mcitedefaultendpunct}{\mcitedefaultseppunct}\relax
\EndOfBibitem
\bibitem{Fritzsch:1977ay}
H.~Fritzsch, \ifthenelse{\boolean{articletitles}}{\emph{{Producing heavy quark
  flavors in hadronic collisions: A test of quantum chromodynamics}},
  }{}\href{http://dx.doi.org/10.1016/0370-2693(77)90108-3}{Phys.\ Lett.\
  \textbf{B67} (1977) 217}\relax
\mciteBstWouldAddEndPuncttrue
\mciteSetBstMidEndSepPunct{\mcitedefaultmidpunct}
{\mcitedefaultendpunct}{\mcitedefaultseppunct}\relax
\EndOfBibitem
\bibitem{Gluck:1977zm}
M.~Gluck, J.~F. Owens, and E.~Reya,
  \ifthenelse{\boolean{articletitles}}{\emph{{Gluon contribution to hadronic
  $J/\psi$ production}},
  }{}\href{http://dx.doi.org/10.1103/PhysRevD.17.2324}{Phys.\ Rev.\
  \textbf{D17} (1978) 2324}\relax
\mciteBstWouldAddEndPuncttrue
\mciteSetBstMidEndSepPunct{\mcitedefaultmidpunct}
{\mcitedefaultendpunct}{\mcitedefaultseppunct}\relax
\EndOfBibitem
\bibitem{Chang:1979nn}
C.-H. Chang, \ifthenelse{\boolean{articletitles}}{\emph{{Hadronic production of
  $J/\psi$ associated with a gluon}},
  }{}\href{http://dx.doi.org/10.1016/0550-3213(80)90175-3}{Nucl.\ Phys.\
  \textbf{B172} (1980) 425}\relax
\mciteBstWouldAddEndPuncttrue
\mciteSetBstMidEndSepPunct{\mcitedefaultmidpunct}
{\mcitedefaultendpunct}{\mcitedefaultseppunct}\relax
\EndOfBibitem
\bibitem{Baier:1981uk}
{R.\ Baier and R.\ R\"{u}ckl},
  \ifthenelse{\boolean{articletitles}}{\emph{{Hadronic production of $J/\psi$
  and $\Upsilon$: Transverse momentum distributions}},
  }{}\href{http://dx.doi.org/10.1016/0370-2693(81)90636-5}{Phys.\ Lett.\
  \textbf{B102} (1981) 364}\relax
\mciteBstWouldAddEndPuncttrue
\mciteSetBstMidEndSepPunct{\mcitedefaultmidpunct}
{\mcitedefaultendpunct}{\mcitedefaultseppunct}\relax
\EndOfBibitem
\bibitem{Bodwin:1994jh}
G.~T. Bodwin, E.~Braaten, and G.~P. Lepage,
  \ifthenelse{\boolean{articletitles}}{\emph{{Rigorous QCD analysis of
  inclusive annihilation and production of heavy quarkonium}},
  }{}\href{http://dx.doi.org/10.1103/PhysRevD.51.1125}{Phys.\ Rev.\
  \textbf{D51} (1995) 1125}, \href{http://arxiv.org/abs/hep-ph/9407339}{{\tt
  arXiv:hep-ph/9407339}}\relax
\mciteBstWouldAddEndPuncttrue
\mciteSetBstMidEndSepPunct{\mcitedefaultmidpunct}
{\mcitedefaultendpunct}{\mcitedefaultseppunct}\relax
\EndOfBibitem
\bibitem{Cho:1995vh}
P.~L. Cho and A.~K. Leibovich,
  \ifthenelse{\boolean{articletitles}}{\emph{{Color-octet quarkonia
  production}}, }{}\href{http://dx.doi.org/10.1103/PhysRevD.53.150}{Phys.\
  Rev.\  \textbf{D53} (1996) 150},
  \href{http://arxiv.org/abs/hep-ph/9505329}{{\tt arXiv:hep-ph/9505329}}\relax
\mciteBstWouldAddEndPuncttrue
\mciteSetBstMidEndSepPunct{\mcitedefaultmidpunct}
{\mcitedefaultendpunct}{\mcitedefaultseppunct}\relax
\EndOfBibitem
\bibitem{Cho:1995ce}
P.~L. Cho and A.~K. Leibovich,
  \ifthenelse{\boolean{articletitles}}{\emph{{Color-octet quarkonia production.
  II.}}, }{}\href{http://dx.doi.org/10.1103/PhysRevD.53.6203}{Phys.\ Rev.\
  \textbf{D53} (1996) 6203}, \href{http://arxiv.org/abs/hep-ph/9511315}{{\tt
  arXiv:hep-ph/9511315}}\relax
\mciteBstWouldAddEndPuncttrue
\mciteSetBstMidEndSepPunct{\mcitedefaultmidpunct}
{\mcitedefaultendpunct}{\mcitedefaultseppunct}\relax
\EndOfBibitem
\bibitem{Abelev:2012kr}
ALICE collaboration, B.~Abelev {\em et~al.},
  \ifthenelse{\boolean{articletitles}}{\emph{{Inclusive $J/\psi$ production in
  $pp$ collisions at $\sqrt{s} = 2.76$ TeV}},
  }{}\href{http://dx.doi.org/10.1016/j.physletb.2012.10.078}{Phys.\ Lett.\
  \textbf{B718} (2012) 295}, \href{http://arxiv.org/abs/1203.3641}{{\tt
  arXiv:1203.3641}}\relax
\mciteBstWouldAddEndPuncttrue
\mciteSetBstMidEndSepPunct{\mcitedefaultmidpunct}
{\mcitedefaultendpunct}{\mcitedefaultseppunct}\relax
\EndOfBibitem
\bibitem{LHCb-PAPER-2012-039}
LHCb collaboration, R.~Aaij {\em et~al.},
  \ifthenelse{\boolean{articletitles}}{\emph{{Measurement of $J/\psi$
  production in $pp$ collisions at $\sqrt{s}=2.76$ TeV}},
  }{}\href{http://dx.doi.org/10.1007/JHEP02(2013)041}{JHEP \textbf{02} (2013)
  041}, \href{http://arxiv.org/abs/1212.1045}{{\tt arXiv:1212.1045}}\relax
\mciteBstWouldAddEndPuncttrue
\mciteSetBstMidEndSepPunct{\mcitedefaultmidpunct}
{\mcitedefaultendpunct}{\mcitedefaultseppunct}\relax
\EndOfBibitem
\bibitem{LHCb-PAPER-2011-003}
LHCb collaboration, R.~Aaij {\em et~al.},
  \ifthenelse{\boolean{articletitles}}{\emph{{Measurement of $J/\psi$
  production in $pp$ collisions at $\sqrt{s}=7$ TeV}},
  }{}\href{http://dx.doi.org/10.1140/epjc/s10052-011-1645-y}{Eur.\ Phys.\ J.\
  \textbf{C71} (2011) 1645}, \href{http://arxiv.org/abs/1103.0423}{{\tt
  arXiv:1103.0423}}\relax
\mciteBstWouldAddEndPuncttrue
\mciteSetBstMidEndSepPunct{\mcitedefaultmidpunct}
{\mcitedefaultendpunct}{\mcitedefaultseppunct}\relax
\EndOfBibitem
\bibitem{Khachatryan:2010yr}
CMS collaboration, V.~Khachatryan {\em et~al.},
  \ifthenelse{\boolean{articletitles}}{\emph{{Prompt and non-prompt $J/\psi$
  production in $pp$ collisions at $\sqrt{s}=7$ TeV}},
  }{}\href{http://dx.doi.org/10.1140/epjc/s10052-011-1575-8}{Eur.\ Phys.\ J.\
  \textbf{C71} (2011) 1575}, \href{http://arxiv.org/abs/1011.4193}{{\tt
  arXiv:1011.4193}}\relax
\mciteBstWouldAddEndPuncttrue
\mciteSetBstMidEndSepPunct{\mcitedefaultmidpunct}
{\mcitedefaultendpunct}{\mcitedefaultseppunct}\relax
\EndOfBibitem
\bibitem{Aad:2011sp}
ATLAS collaboration, G.~Aad {\em et~al.},
  \ifthenelse{\boolean{articletitles}}{\emph{{Measurement of the differential
  cross-sections of inclusive, prompt and non-prompt $J/\psi$ production in
  proton-proton collisions at $\sqrt{s}=7$ TeV}},
  }{}\href{http://dx.doi.org/10.1016/j.nuclphysb.2011.05.015}{Nucl.\ Phys.\
  \textbf{B850} (2011) 387}, \href{http://arxiv.org/abs/1104.3038}{{\tt
  arXiv:1104.3038}}\relax
\mciteBstWouldAddEndPuncttrue
\mciteSetBstMidEndSepPunct{\mcitedefaultmidpunct}
{\mcitedefaultendpunct}{\mcitedefaultseppunct}\relax
\EndOfBibitem
\bibitem{Aamodt:2011gj}
ALICE collaboration, K.~Aamodt {\em et~al.},
  \ifthenelse{\boolean{articletitles}}{\emph{{Rapidity and transverse momentum
  dependence of inclusive $J/\psi$ production in $pp$ collisions at $\sqrt{s} =
  7$ TeV}}, }{}\href{http://dx.doi.org/10.1016/j.physletb.2011.09.054}{Phys.\
  Lett.\  \textbf{B704} (2011) 442}, \href{http://arxiv.org/abs/1105.0380}{{\tt
  arXiv:1105.0380}}\relax
\mciteBstWouldAddEndPuncttrue
\mciteSetBstMidEndSepPunct{\mcitedefaultmidpunct}
{\mcitedefaultendpunct}{\mcitedefaultseppunct}\relax
\EndOfBibitem
\bibitem{Chatrchyan:2011kc}
CMS collaboration, S.~Chatrchyan {\em et~al.},
  \ifthenelse{\boolean{articletitles}}{\emph{{$J/\psi$ and $\psi(2S)$
  production in $pp$ collisions at $\sqrt{s}=7$ TeV}},
  }{}\href{http://dx.doi.org/10.1007/JHEP02(2012)011}{JHEP \textbf{02} (2012)
  011}, \href{http://arxiv.org/abs/1111.1557}{{\tt arXiv:1111.1557}}\relax
\mciteBstWouldAddEndPuncttrue
\mciteSetBstMidEndSepPunct{\mcitedefaultmidpunct}
{\mcitedefaultendpunct}{\mcitedefaultseppunct}\relax
\EndOfBibitem
\bibitem{LHCb-PAPER-2013-016}
LHCb collaboration, R.~Aaij {\em et~al.},
  \ifthenelse{\boolean{articletitles}}{\emph{{Production of $J/\psi$ and
  $\Upsilon$ mesons in $pp$ collisions at $\sqrt{s}=8$ TeV}},
  }{}\href{http://dx.doi.org/10.1007/JHEP06(2013)064}{JHEP \textbf{06} (2013)
  064}, \href{http://arxiv.org/abs/1304.6977}{{\tt arXiv:1304.6977}}\relax
\mciteBstWouldAddEndPuncttrue
\mciteSetBstMidEndSepPunct{\mcitedefaultmidpunct}
{\mcitedefaultendpunct}{\mcitedefaultseppunct}\relax
\EndOfBibitem
\bibitem{Campbell:2007ws}
J.~M. Campbell, F.~Maltoni, and F.~Tramontano,
  \ifthenelse{\boolean{articletitles}}{\emph{{QCD corrections to $J/\psi$ and
  $\Upsilon$ production at hadron colliders}},
  }{}\href{http://dx.doi.org/10.1103/PhysRevLett.98.252002}{Phys.\ Rev.\ Lett.\
   \textbf{98} (2007) 252002}, \href{http://arxiv.org/abs/hep-ph/0703113}{{\tt
  arXiv:hep-ph/0703113}}\relax
\mciteBstWouldAddEndPuncttrue
\mciteSetBstMidEndSepPunct{\mcitedefaultmidpunct}
{\mcitedefaultendpunct}{\mcitedefaultseppunct}\relax
\EndOfBibitem
\bibitem{Lansberg:2011hi}
J.~P. Lansberg, \ifthenelse{\boolean{articletitles}}{\emph{{$J/\psi$ production
  at $\sqrt s$=1.96 and 7 TeV: Color-Singlet Model, NNLO* and polarisation}},
  }{}\href{http://dx.doi.org/10.1088/0954-3899/38/12/124110}{J.\ Phys.\
  \textbf{G38} (2011) 124110}, \href{http://arxiv.org/abs/1107.0292}{{\tt
  arXiv:1107.0292}}\relax
\mciteBstWouldAddEndPuncttrue
\mciteSetBstMidEndSepPunct{\mcitedefaultmidpunct}
{\mcitedefaultendpunct}{\mcitedefaultseppunct}\relax
\EndOfBibitem
\bibitem{Abe:1992ww}
CDF collaboration, F.~Abe {\em et~al.},
  \ifthenelse{\boolean{articletitles}}{\emph{{Inclusive $J/\psi$, $\psi(2S)$
  and $b$ quark production in $\bar{p}p$ collisions at $\sqrt{s} = 1.8$ TeV}},
  }{}\href{http://dx.doi.org/10.1103/PhysRevLett.69.3704}{Phys.\ Rev.\ Lett.\
  \textbf{69} (1992) 3704}\relax
\mciteBstWouldAddEndPuncttrue
\mciteSetBstMidEndSepPunct{\mcitedefaultmidpunct}
{\mcitedefaultendpunct}{\mcitedefaultseppunct}\relax
\EndOfBibitem
\bibitem{Cacciari:1995yt}
M.~Cacciari, M.~Greco, M.~L. Mangano, and A.~Petrelli,
  \ifthenelse{\boolean{articletitles}}{\emph{{Charmonium production at the
  Tevatron}}, }{}\href{http://dx.doi.org/10.1016/0370-2693(95)00868-L}{Phys.\
  Lett.\  \textbf{B356} (1995) 553},
  \href{http://arxiv.org/abs/hep-ph/9505379}{{\tt arXiv:hep-ph/9505379}}\relax
\mciteBstWouldAddEndPuncttrue
\mciteSetBstMidEndSepPunct{\mcitedefaultmidpunct}
{\mcitedefaultendpunct}{\mcitedefaultseppunct}\relax
\EndOfBibitem
\bibitem{Braaten:1994vv}
E.~Braaten and S.~Fleming,
  \ifthenelse{\boolean{articletitles}}{\emph{{Color-octet fragmentation and the
  $\psi'$ surplus at the Fermilab Tevatron}},
  }{}\href{http://dx.doi.org/10.1103/PhysRevLett.74.3327}{Phys.\ Rev.\ Lett.\
  \textbf{74} (1995) 3327}, \href{http://arxiv.org/abs/hep-ph/9411365}{{\tt
  arXiv:hep-ph/9411365}}\relax
\mciteBstWouldAddEndPuncttrue
\mciteSetBstMidEndSepPunct{\mcitedefaultmidpunct}
{\mcitedefaultendpunct}{\mcitedefaultseppunct}\relax
\EndOfBibitem
\bibitem{Ma:2010jj}
Y.-Q. Ma, K.~Wang, and K.-T. Chao,
  \ifthenelse{\boolean{articletitles}}{\emph{{Complete next-to-leading order
  calculation of the $J/\psi$ and $\psi'$ production at hadron colliders}},
  }{}\href{http://dx.doi.org/10.1103/PhysRevD.84.114001}{Phys.\ Rev.\
  \textbf{D84} (2011) 114001}, \href{http://arxiv.org/abs/1012.1030}{{\tt
  arXiv:1012.1030}}\relax
\mciteBstWouldAddEndPuncttrue
\mciteSetBstMidEndSepPunct{\mcitedefaultmidpunct}
{\mcitedefaultendpunct}{\mcitedefaultseppunct}\relax
\EndOfBibitem
\bibitem{Gong:2008ft}
B.~Gong, X.~Q. Li, and J.-X. Wang,
  \ifthenelse{\boolean{articletitles}}{\emph{{QCD corrections to $J/\psi$
  production via color-octet states at Tevatron and LHC}},
  }{}\href{http://dx.doi.org/10.1016/j.physletb.2009.02.026}{Phys.\ Lett.\
  \textbf{B673} (2009) 197}, \href{http://arxiv.org/abs/0805.4751}{{\tt
  arXiv:0805.4751}}\relax
\mciteBstWouldAddEndPuncttrue
\mciteSetBstMidEndSepPunct{\mcitedefaultmidpunct}
{\mcitedefaultendpunct}{\mcitedefaultseppunct}\relax
\EndOfBibitem
\bibitem{Butenschoen:2010rq}
M.~Butenschoen and B.~A. Kniehl,
  \ifthenelse{\boolean{articletitles}}{\emph{{Reconciling $J/\psi$ production
  at HERA, RHIC, Tevatron, and LHC with nonrelativistic QCD factorization at
  next-to-leading order}},
  }{}\href{http://dx.doi.org/10.1103/PhysRevLett.106.022003}{Phys.\ Rev.\
  Lett.\  \textbf{106} (2011) 022003},
  \href{http://arxiv.org/abs/1009.5662}{{\tt arXiv:1009.5662}}\relax
\mciteBstWouldAddEndPuncttrue
\mciteSetBstMidEndSepPunct{\mcitedefaultmidpunct}
{\mcitedefaultendpunct}{\mcitedefaultseppunct}\relax
\EndOfBibitem
\bibitem{Cacciari:1998it}
M.~Cacciari, M.~Greco, and P.~Nason,
  \ifthenelse{\boolean{articletitles}}{\emph{{The $p_T$ spectrum in heavy
  flavor hadroproduction}},
  }{}\href{http://dx.doi.org/10.1088/1126-6708/1998/05/007}{JHEP \textbf{05}
  (1998) 007}, \href{http://arxiv.org/abs/hep-ph/9803400}{{\tt
  arXiv:hep-ph/9803400}}\relax
\mciteBstWouldAddEndPuncttrue
\mciteSetBstMidEndSepPunct{\mcitedefaultmidpunct}
{\mcitedefaultendpunct}{\mcitedefaultseppunct}\relax
\EndOfBibitem
\bibitem{Abelev:2011md}
ALICE collaboration, B.~Abelev {\em et~al.},
  \ifthenelse{\boolean{articletitles}}{\emph{{$J/\psi$ polarization in $pp$
  collisions at $\sqrt{s}=7$ TeV}},
  }{}\href{http://dx.doi.org/10.1103/PhysRevLett.108.082001}{Phys.\ Rev.\
  Lett.\  \textbf{108} (2012) 082001},
  \href{http://arxiv.org/abs/1111.1630}{{\tt arXiv:1111.1630}}\relax
\mciteBstWouldAddEndPuncttrue
\mciteSetBstMidEndSepPunct{\mcitedefaultmidpunct}
{\mcitedefaultendpunct}{\mcitedefaultseppunct}\relax
\EndOfBibitem
\bibitem{Chatrchyan:2013cla}
CMS collaboration, S.~Chatrchyan {\em et~al.},
  \ifthenelse{\boolean{articletitles}}{\emph{{Measurement of the prompt
  $J/\psi$ and $\psi$(2S) polarizations in $pp$ collisions at $\sqrt{s}$ = 7
  TeV}}, }{}\href{http://dx.doi.org/10.1016/j.physletb.2013.10.055}{Phys.\
  Lett.\  \textbf{B727} (2013) 381}, \href{http://arxiv.org/abs/1307.6070}{{\tt
  arXiv:1307.6070}}\relax
\mciteBstWouldAddEndPuncttrue
\mciteSetBstMidEndSepPunct{\mcitedefaultmidpunct}
{\mcitedefaultendpunct}{\mcitedefaultseppunct}\relax
\EndOfBibitem
\bibitem{LHCb-PAPER-2013-008}
LHCb collaboration, R.~Aaij {\em et~al.},
  \ifthenelse{\boolean{articletitles}}{\emph{{Measurement of $\jpsi$
  polarization in $pp$ collisions at $\sqrt{s}=7$ TeV}},
  }{}\href{http://dx.doi.org/10.1140/epjc/s10052-013-2631-3}{Eur.\ Phys.\ J.\
  \textbf{C73} (2013) 2631}, \href{http://arxiv.org/abs/1307.6379}{{\tt
  arXiv:1307.6379}}\relax
\mciteBstWouldAddEndPuncttrue
\mciteSetBstMidEndSepPunct{\mcitedefaultmidpunct}
{\mcitedefaultendpunct}{\mcitedefaultseppunct}\relax
\EndOfBibitem
\bibitem{Gong:2008sn}
B.~Gong and J.-X. Wang,
  \ifthenelse{\boolean{articletitles}}{\emph{{Next-to-leading-order QCD
  corrections to $J/\psi$ polarization at Tevatron and Large-Hadron-Collider
  energies}}, }{}\href{http://dx.doi.org/10.1103/PhysRevLett.100.232001}{Phys.\
  Rev.\ Lett.\  \textbf{100} (2008) 232001},
  \href{http://arxiv.org/abs/0802.3727}{{\tt arXiv:0802.3727}}\relax
\mciteBstWouldAddEndPuncttrue
\mciteSetBstMidEndSepPunct{\mcitedefaultmidpunct}
{\mcitedefaultendpunct}{\mcitedefaultseppunct}\relax
\EndOfBibitem
\bibitem{Beneke:1995yb}
M.~Beneke and I.~Z. Rothstein,
  \ifthenelse{\boolean{articletitles}}{\emph{{$\psi^{'}$ polarization as a test
  of colour octet quarkonium production}},
  }{}\href{http://dx.doi.org/10.1016/0370-2693(96)00030-5}{Phys.\ Lett.\
  \textbf{B372} (1996) 157}, \href{http://arxiv.org/abs/hep-ph/9509375}{{\tt
  arXiv:hep-ph/9509375}}\relax
\mciteBstWouldAddEndPuncttrue
\mciteSetBstMidEndSepPunct{\mcitedefaultmidpunct}
{\mcitedefaultendpunct}{\mcitedefaultseppunct}\relax
\EndOfBibitem
\bibitem{Chao:2012iv}
K.-T. Chao {\em et~al.}, \ifthenelse{\boolean{articletitles}}{\emph{{$J/\psi$
  polarization at hadron colliders in nonrelativistic QCD}},
  }{}\href{http://dx.doi.org/10.1103/PhysRevLett.108.242004}{Phys.\ Rev.\
  Lett.\  \textbf{108} (2012) 242004},
  \href{http://arxiv.org/abs/1201.2675}{{\tt arXiv:1201.2675}}\relax
\mciteBstWouldAddEndPuncttrue
\mciteSetBstMidEndSepPunct{\mcitedefaultmidpunct}
{\mcitedefaultendpunct}{\mcitedefaultseppunct}\relax
\EndOfBibitem
\bibitem{Gong:2012ug}
B.~Gong, L.-P. Wan, J.-X. Wang, and H.-F. Zhang,
  \ifthenelse{\boolean{articletitles}}{\emph{{Polarization for prompt $J/\psi$
  and $\psi(2S)$ production at the Tevatron and LHC}},
  }{}\href{http://dx.doi.org/10.1103/PhysRevLett.110.042002}{Phys.\ Rev.\
  Lett.\  \textbf{110} (2013) 042002},
  \href{http://arxiv.org/abs/1205.6682}{{\tt arXiv:1205.6682}}\relax
\mciteBstWouldAddEndPuncttrue
\mciteSetBstMidEndSepPunct{\mcitedefaultmidpunct}
{\mcitedefaultendpunct}{\mcitedefaultseppunct}\relax
\EndOfBibitem
\bibitem{Butenschoen:2012px}
M.~Butenschoen and B.~A. Kniehl,
  \ifthenelse{\boolean{articletitles}}{\emph{{$J/\psi$ polarization at Tevatron
  and LHC: Nonrelativistic-QCD factorization at the crossroads}},
  }{}\href{http://dx.doi.org/10.1103/PhysRevLett.108.172002}{Phys.\ Rev.\
  Lett.\  \textbf{108} (2012) 172002},
  \href{http://arxiv.org/abs/1201.1872}{{\tt arXiv:1201.1872}}\relax
\mciteBstWouldAddEndPuncttrue
\mciteSetBstMidEndSepPunct{\mcitedefaultmidpunct}
{\mcitedefaultendpunct}{\mcitedefaultseppunct}\relax
\EndOfBibitem
\bibitem{Lansberg:2008gk}
J.~P. Lansberg, \ifthenelse{\boolean{articletitles}}{\emph{{On the mechanisms
  of heavy-quarkonium hadroproduction}},
  }{}\href{http://dx.doi.org/10.1140/epjc/s10052-008-0826-9}{Eur.\ Phys.\ J.\
  \textbf{C61} (2009) 693}, \href{http://arxiv.org/abs/0811.4005}{{\tt
  arXiv:0811.4005}}\relax
\mciteBstWouldAddEndPuncttrue
\mciteSetBstMidEndSepPunct{\mcitedefaultmidpunct}
{\mcitedefaultendpunct}{\mcitedefaultseppunct}\relax
\EndOfBibitem
\bibitem{Abulencia:2007us}
CDF collaboration, A.~Abulencia {\em et~al.},
  \ifthenelse{\boolean{articletitles}}{\emph{{Polarization of $J/\psi$ and
  $\psi(2S)$ mesons produced in $p \bar{p}$ collisions at $\sqrt{s}$ = 1.96
  TeV}}, }{}\href{http://dx.doi.org/10.1103/PhysRevLett.99.132001}{Phys.\ Rev.\
  Lett.\  \textbf{99} (2007) 132001},
  \href{http://arxiv.org/abs/0704.0638}{{\tt arXiv:0704.0638}}\relax
\mciteBstWouldAddEndPuncttrue
\mciteSetBstMidEndSepPunct{\mcitedefaultmidpunct}
{\mcitedefaultendpunct}{\mcitedefaultseppunct}\relax
\EndOfBibitem
\bibitem{Abazov:2008aa}
D0 collaboration, V.~M. Abazov {\em et~al.},
  \ifthenelse{\boolean{articletitles}}{\emph{{Measurement of the polarization
  of the $\Upsilon(1S)$ and $\Upsilon(2S)$ states in $p \bar{p}$ collisions at
  $\sqrt{s}$ = 1.96 TeV}},
  }{}\href{http://dx.doi.org/10.1103/PhysRevLett.101.182004}{Phys.\ Rev.\
  Lett.\  \textbf{101} (2008) 182004},
  \href{http://arxiv.org/abs/0804.2799}{{\tt arXiv:0804.2799}}\relax
\mciteBstWouldAddEndPuncttrue
\mciteSetBstMidEndSepPunct{\mcitedefaultmidpunct}
{\mcitedefaultendpunct}{\mcitedefaultseppunct}\relax
\EndOfBibitem
\bibitem{Chatrchyan:2012woa}
CMS collaboration, S.~Chatrchyan {\em et~al.},
  \ifthenelse{\boolean{articletitles}}{\emph{{Measurement of the
  $\Upsilon(1S)$, $\Upsilon(2S)$ and $\Upsilon(3S)$ polarizations in $pp$
  collisions at $\sqrt{s}=7$ TeV}},
  }{}\href{http://dx.doi.org/10.1103/PhysRevLett.110.081802}{Phys.\ Rev.\
  Lett.\  \textbf{110} (2013) 081802},
  \href{http://arxiv.org/abs/1209.2922}{{\tt arXiv:1209.2922}}\relax
\mciteBstWouldAddEndPuncttrue
\mciteSetBstMidEndSepPunct{\mcitedefaultmidpunct}
{\mcitedefaultendpunct}{\mcitedefaultseppunct}\relax
\EndOfBibitem
\bibitem{LHCb-PAPER-2013-067}
LHCb collaboration, R.~Aaij {\em et~al.},
  \ifthenelse{\boolean{articletitles}}{\emph{{Measurement of $\psi(2S)$
  polarisation in $pp$ collisions at $\sqrt{s}=7$ TeV}},
  }{}\href{http://dx.doi.org/10.1140/epjc/s10052-014-2872-9}{Eur.\ Phys.\ J.\
  \textbf{C74} (2014) 2872}, \href{http://arxiv.org/abs/1403.1339}{{\tt
  arXiv:1403.1339}}\relax
\mciteBstWouldAddEndPuncttrue
\mciteSetBstMidEndSepPunct{\mcitedefaultmidpunct}
{\mcitedefaultendpunct}{\mcitedefaultseppunct}\relax
\EndOfBibitem
\bibitem{Alves:2008zz}
LHCb collaboration, A.~A. Alves~Jr.\ {\em et~al.},
  \ifthenelse{\boolean{articletitles}}{\emph{{The \lhcb detector at the LHC}},
  }{}\href{http://dx.doi.org/10.1088/1748-0221/3/08/S08005}{JINST \textbf{3}
  (2008) S08005}\relax
\mciteBstWouldAddEndPuncttrue
\mciteSetBstMidEndSepPunct{\mcitedefaultmidpunct}
{\mcitedefaultendpunct}{\mcitedefaultseppunct}\relax
\EndOfBibitem
\bibitem{LHCb-DP-2014-002}
LHCb collaboration, R.~Aaij {\em et~al.},
  \ifthenelse{\boolean{articletitles}}{\emph{{LHCb detector performance}},
  }{}\href{http://dx.doi.org/10.1142/S0217751X15300227}{Int.\ J.\ Mod.\ Phys.\
  \textbf{A30} (2015) 1530022}, \href{http://arxiv.org/abs/1412.6352}{{\tt
  arXiv:1412.6352}}\relax
\mciteBstWouldAddEndPuncttrue
\mciteSetBstMidEndSepPunct{\mcitedefaultmidpunct}
{\mcitedefaultendpunct}{\mcitedefaultseppunct}\relax
\EndOfBibitem
\bibitem{LHCb-DP-2012-002}
A.~A. Alves~Jr.\ {\em et~al.},
  \ifthenelse{\boolean{articletitles}}{\emph{{Performance of the LHCb muon
  system}}, }{}\href{http://dx.doi.org/10.1088/1748-0221/8/02/P02022}{JINST
  \textbf{8} (2013) P02022}, \href{http://arxiv.org/abs/1211.1346}{{\tt
  arXiv:1211.1346}}\relax
\mciteBstWouldAddEndPuncttrue
\mciteSetBstMidEndSepPunct{\mcitedefaultmidpunct}
{\mcitedefaultendpunct}{\mcitedefaultseppunct}\relax
\EndOfBibitem
\bibitem{PDG2014}
Particle Data Group, K.~A. Olive {\em et~al.},
  \ifthenelse{\boolean{articletitles}}{\emph{{\href{http://pdg.lbl.gov/}{Review
  of particle physics}}},
  }{}\href{http://dx.doi.org/10.1088/1674-1137/38/9/090001}{Chin.\ Phys.\
  \textbf{C38} (2014) 090001}\relax
\mciteBstWouldAddEndPuncttrue
\mciteSetBstMidEndSepPunct{\mcitedefaultmidpunct}
{\mcitedefaultendpunct}{\mcitedefaultseppunct}\relax
\EndOfBibitem
\bibitem{Dujany:2017839}
G.~Dujany and B.~Storaci, \ifthenelse{\boolean{articletitles}}{\emph{{Real-time
  alignment and calibration of the LHCb Detector in Run II}}, }{}
  \href{http://cdsweb.cern.ch/search?p=LHCb-PROC-2015-011&f=reportnumber&action_search=Search&c=LHCb+Conference+Proceedings}
  {LHCb-PROC-2015-011}\relax
\mciteBstWouldAddEndPuncttrue
\mciteSetBstMidEndSepPunct{\mcitedefaultmidpunct}
{\mcitedefaultendpunct}{\mcitedefaultseppunct}\relax
\EndOfBibitem
\bibitem{LHCb-DP-2012-004}
R.~Aaij {\em et~al.}, \ifthenelse{\boolean{articletitles}}{\emph{{The \lhcb
  trigger and its performance in 2011}},
  }{}\href{http://dx.doi.org/10.1088/1748-0221/8/04/P04022}{JINST \textbf{8}
  (2013) P04022}, \href{http://arxiv.org/abs/1211.3055}{{\tt
  arXiv:1211.3055}}\relax
\mciteBstWouldAddEndPuncttrue
\mciteSetBstMidEndSepPunct{\mcitedefaultmidpunct}
{\mcitedefaultendpunct}{\mcitedefaultseppunct}\relax
\EndOfBibitem
\bibitem{pythia6.4}
T.~Sj\"{o}strand, S.~Mrenna, and P.~Skands,
  \ifthenelse{\boolean{articletitles}}{\emph{{PYTHIA 6.4 physics and manual}},
  }{}\href{http://dx.doi.org/10.1088/1126-6708/2006/05/026}{JHEP \textbf{05}
  (2006) 026}, \href{http://arxiv.org/abs/hep-ph/0603175}{{\tt
  arXiv:hep-ph/0603175}}\relax
\mciteBstWouldAddEndPuncttrue
\mciteSetBstMidEndSepPunct{\mcitedefaultmidpunct}
{\mcitedefaultendpunct}{\mcitedefaultseppunct}\relax
\EndOfBibitem
\bibitem{LHCb-PROC-2010-056}
I.~Belyaev {\em et~al.}, \ifthenelse{\boolean{articletitles}}{\emph{{Handling
  of the generation of primary events in Gauss, the LHCb simulation
  framework}}, }{}\href{http://dx.doi.org/10.1088/1742-6596/331/3/032047}{{J.\
  Phys.\ Conf.\ Ser.\ } \textbf{331} (2011) 032047}\relax
\mciteBstWouldAddEndPuncttrue
\mciteSetBstMidEndSepPunct{\mcitedefaultmidpunct}
{\mcitedefaultendpunct}{\mcitedefaultseppunct}\relax
\EndOfBibitem
\bibitem{Lange:2001uf}
D.~J. Lange, \ifthenelse{\boolean{articletitles}}{\emph{{The EvtGen particle
  decay simulation package}},
  }{}\href{http://dx.doi.org/10.1016/S0168-9002(01)00089-4}{Nucl.\ Instrum.\
  Meth.\  \textbf{A462} (2001) 152}\relax
\mciteBstWouldAddEndPuncttrue
\mciteSetBstMidEndSepPunct{\mcitedefaultmidpunct}
{\mcitedefaultendpunct}{\mcitedefaultseppunct}\relax
\EndOfBibitem
\bibitem{Golonka:2005pn}
P.~Golonka and Z.~Was, \ifthenelse{\boolean{articletitles}}{\emph{{PHOTOS Monte
  Carlo: A precision tool for QED corrections in $Z$ and $W$ decays}},
  }{}\href{http://dx.doi.org/10.1140/epjc/s2005-02396-4}{Eur.\ Phys.\ J.\
  \textbf{C45} (2006) 97}, \href{http://arxiv.org/abs/hep-ph/0506026}{{\tt
  arXiv:hep-ph/0506026}}\relax
\mciteBstWouldAddEndPuncttrue
\mciteSetBstMidEndSepPunct{\mcitedefaultmidpunct}
{\mcitedefaultendpunct}{\mcitedefaultseppunct}\relax
\EndOfBibitem
\bibitem{Bargiotti:2007zz}
M.~Bargiotti and V.~Vagnoni, \ifthenelse{\boolean{articletitles}}{\emph{{Heavy
  quarkonia sector in PYTHIA 6.324: Tuning, validation and perspectives at
  LHC(b)}}, }{}
  \href{http://cdsweb.cern.ch/search?p=CERN-LHCb-2007-042&f=reportnumber&action_search=Search&c=LHCb+Notes}
  {CERN-LHCb-2007-042}\relax
\mciteBstWouldAddEndPuncttrue
\mciteSetBstMidEndSepPunct{\mcitedefaultmidpunct}
{\mcitedefaultendpunct}{\mcitedefaultseppunct}\relax
\EndOfBibitem
\bibitem{Allison:2006ve}
Geant4 collaboration, J.~Allison {\em et~al.},
  \ifthenelse{\boolean{articletitles}}{\emph{{Geant4 developments and
  applications}}, }{}\href{http://dx.doi.org/10.1109/TNS.2006.869826}{IEEE
  Trans.\ Nucl.\ Sci.\  \textbf{53} (2006) 270}\relax
\mciteBstWouldAddEndPuncttrue
\mciteSetBstMidEndSepPunct{\mcitedefaultmidpunct}
{\mcitedefaultendpunct}{\mcitedefaultseppunct}\relax
\EndOfBibitem
\bibitem{Agostinelli:2002hh}
Geant4 collaboration, S.~Agostinelli {\em et~al.},
  \ifthenelse{\boolean{articletitles}}{\emph{{Geant4: A simulation toolkit}},
  }{}\href{http://dx.doi.org/10.1016/S0168-9002(03)01368-8}{Nucl.\ Instrum.\
  Meth.\  \textbf{A506} (2003) 250}\relax
\mciteBstWouldAddEndPuncttrue
\mciteSetBstMidEndSepPunct{\mcitedefaultmidpunct}
{\mcitedefaultendpunct}{\mcitedefaultseppunct}\relax
\EndOfBibitem
\bibitem{LHCb-PROC-2011-006}
M.~Clemencic {\em et~al.}, \ifthenelse{\boolean{articletitles}}{\emph{{The
  \lhcb simulation application, Gauss: Design, evolution and experience}},
  }{}\href{http://dx.doi.org/10.1088/1742-6596/331/3/032023}{{J.\ Phys.\ Conf.\
  Ser.\ } \textbf{331} (2011) 032023}\relax
\mciteBstWouldAddEndPuncttrue
\mciteSetBstMidEndSepPunct{\mcitedefaultmidpunct}
{\mcitedefaultendpunct}{\mcitedefaultseppunct}\relax
\EndOfBibitem
\bibitem{Barschel:1693671}
C.~Barschel, {\em {Precision luminosity measurement at LHCb with beam-gas
  imaging}}, PhD thesis, RWTH Aachen, 2014,
  {\href{https://cds.cern.ch/record/1693671}{CERN-THESIS-2013-301}}\relax
\mciteBstWouldAddEndPuncttrue
\mciteSetBstMidEndSepPunct{\mcitedefaultmidpunct}
{\mcitedefaultendpunct}{\mcitedefaultseppunct}\relax
\EndOfBibitem
\bibitem{LHCb-PAPER-2014-047}
LHCb collaboration, R.~Aaij {\em et~al.},
  \ifthenelse{\boolean{articletitles}}{\emph{{Precision luminosity measurements
  at LHCb}}, }{}\href{http://dx.doi.org/10.1088/1748-0221/9/12/P12005}{JINST
  \textbf{9} (2014) P12005}, \href{http://arxiv.org/abs/1410.0149}{{\tt
  arXiv:1410.0149}}\relax
\mciteBstWouldAddEndPuncttrue
\mciteSetBstMidEndSepPunct{\mcitedefaultmidpunct}
{\mcitedefaultendpunct}{\mcitedefaultseppunct}\relax
\EndOfBibitem
\bibitem{Skwarnicki:1986xj}
T.~Skwarnicki, {\em {A study of the radiative cascade transitions between the
  Upsilon-prime and Upsilon resonances}}, PhD thesis, Institute of Nuclear
  Physics, Krakow, 1986,
  {\href{http://inspirehep.net/record/230779/}{DESY-F31-86-02}}\relax
\mciteBstWouldAddEndPuncttrue
\mciteSetBstMidEndSepPunct{\mcitedefaultmidpunct}
{\mcitedefaultendpunct}{\mcitedefaultseppunct}\relax
\EndOfBibitem
\bibitem{Santos:2013gra}
D.~Mart\'{i}nez~Santos and F.~Dupertuis,
  \ifthenelse{\boolean{articletitles}}{\emph{{Mass distributions marginalized
  over per-event errors}},
  }{}\href{http://dx.doi.org/10.1016/j.nima.2014.06.081}{Nucl.\ Instrum.\
  Meth.\  \textbf{A764} (2014) 150}, \href{http://arxiv.org/abs/1312.5000}{{\tt
  arXiv:1312.5000}}\relax
\mciteBstWouldAddEndPuncttrue
\mciteSetBstMidEndSepPunct{\mcitedefaultmidpunct}
{\mcitedefaultendpunct}{\mcitedefaultseppunct}\relax
\EndOfBibitem
\bibitem{LHCb-DP-2013-002}
LHCb collaboration, R.~Aaij {\em et~al.},
  \ifthenelse{\boolean{articletitles}}{\emph{{Measurement of the track
  reconstruction efficiency at LHCb}},
  }{}\href{http://dx.doi.org/10.1088/1748-0221/10/02/P02007}{JINST \textbf{10}
  (2015) P02007}, \href{http://arxiv.org/abs/1408.1251}{{\tt
  arXiv:1408.1251}}\relax
\mciteBstWouldAddEndPuncttrue
\mciteSetBstMidEndSepPunct{\mcitedefaultmidpunct}
{\mcitedefaultendpunct}{\mcitedefaultseppunct}\relax
\EndOfBibitem
\bibitem{Jacob:1959at}
M.~Jacob and G.~C. Wick, \ifthenelse{\boolean{articletitles}}{\emph{{On the
  general theory of collisions for particles with spin}},
  }{}\href{http://dx.doi.org/10.1016/0003-4916(59)90051-X}{Annals Phys.\
  \textbf{7} (1959) 404}, [Annals Phys. \textbf{281} (2000) 774]\relax
\mciteBstWouldAddEndPuncttrue
\mciteSetBstMidEndSepPunct{\mcitedefaultmidpunct}
{\mcitedefaultendpunct}{\mcitedefaultseppunct}\relax
\EndOfBibitem
\bibitem{Pivk:2004ty}
M.~Pivk and F.~R. Le~Diberder,
  \ifthenelse{\boolean{articletitles}}{\emph{{sPlot: A statistical tool to
  unfold data distributions}},
  }{}\href{http://dx.doi.org/10.1016/j.nima.2005.08.106}{Nucl.\ Instrum.\
  Meth.\  \textbf{A555} (2005) 356},
  \href{http://arxiv.org/abs/physics/0402083}{{\tt
  arXiv:physics/0402083}}\relax
\mciteBstWouldAddEndPuncttrue
\mciteSetBstMidEndSepPunct{\mcitedefaultmidpunct}
{\mcitedefaultendpunct}{\mcitedefaultseppunct}\relax
\EndOfBibitem
\bibitem{ATLAS-CONF-2015-030}
ATLAS collaboration, G.~Aad {\em et~al.},
  \ifthenelse{\boolean{articletitles}}{\emph{{Measurement of the differential
  non-prompt $J/\psi$ production fraction in $ \sqrt{s}= 13$ TeV $pp$
  collisions at the ATLAS experiment}}, }{}
  \href{http://cdsweb.cern.ch/search?p=ATLAS-CONF-2015-030&f=reportnumber&action_search=Search&c=LHCb+Notes}
  {ATLAS-CONF-2015-030}\relax
\mciteBstWouldAddEndPuncttrue
\mciteSetBstMidEndSepPunct{\mcitedefaultmidpunct}
{\mcitedefaultendpunct}{\mcitedefaultseppunct}\relax
\EndOfBibitem
\bibitem{PDG2012}
Particle Data Group, J.~Beringer {\em et~al.},
  \ifthenelse{\boolean{articletitles}}{\emph{{\href{http://pdg.lbl.gov/}{Review
  of particle physics}}},
  }{}\href{http://dx.doi.org/10.1103/PhysRevD.86.010001}{Phys.\ Rev.\
  \textbf{D86} (2012) 010001}\relax
\mciteBstWouldAddEndPuncttrue
\mciteSetBstMidEndSepPunct{\mcitedefaultmidpunct}
{\mcitedefaultendpunct}{\mcitedefaultseppunct}\relax
\EndOfBibitem
\bibitem{Cacciari:2015fta}
M.~Cacciari, M.~L. Mangano, and P.~Nason,
  \ifthenelse{\boolean{articletitles}}{\emph{{Gluon PDF constraints from the
  ratio of forward heavy quark production at the LHC at $\sqrt{s}=7$ and $13$
  TeV}}, }{}\href{http://arxiv.org/abs/1507.06197}{{\tt
  arXiv:1507.06197}}\relax
\mciteBstWouldAddEndPuncttrue
\mciteSetBstMidEndSepPunct{\mcitedefaultmidpunct}
{\mcitedefaultendpunct}{\mcitedefaultseppunct}\relax
\EndOfBibitem
\bibitem{Shao:2014yta}
H.-S. Shao {\em et~al.}, \ifthenelse{\boolean{articletitles}}{\emph{{Yields and
  polarizations of prompt $J/\psi$ and $\psi(2S)$ production in hadronic
  collisions}}, }{}\href{http://dx.doi.org/10.1007/JHEP05(2015)103}{JHEP
  \textbf{05} (2015) 103}, \href{http://arxiv.org/abs/1411.3300}{{\tt
  arXiv:1411.3300}}\relax
\mciteBstWouldAddEndPuncttrue
\mciteSetBstMidEndSepPunct{\mcitedefaultmidpunct}
{\mcitedefaultendpunct}{\mcitedefaultseppunct}\relax
\EndOfBibitem
\end{mcitethebibliography}
\clearpage
\section*{Appendix}
\subsection*{Change of efficiency with respect to polarization}
The detection efficiency is affected by the polarisation, especially by the polarisation parameter
$\lambda_\theta$. Zero polarisation is assumed in these simulations since there
is no prior knowledge of the polarisation of \jpsi mesons in $pp$ collisions at 13$\TEV$, and only small polarisations have been
found in all \lhc quarkonia polarisation analyses~\cite{Abulencia:2007us, Abelev:2011md, Chatrchyan:2013cla, LHCb-PAPER-2013-008}. 
In Table~\ref{tab:PolarizationEffect20}, the increase of the total efficiency is given in bins of $(\pt,y)$ of the \jpsi meson for a polarisation of
$\lambda_\theta=-20\%$, compared to zero polarisation. 
This information facilitates the extrapolation of the cross-sections measured assuming zero polarisation to other 
polarisation values. The relative change in efficiency is linear, to 5\% accuracy, between polarisation values of 
zero and 20\%.

\begin{table}[!bp]
\caption{The relative increase of the total efficiency (in \%), for a $-20\%$ polarisation rather than zero, in different bins of \pt and $y$.}
\centering
\scalebox{0.9}{
\begin{tabular}{c|ccccc}
\hline
$\pt\,[\!\gevc]$& $2.0<y<2.5$& $2.5<y<3.0$& $3.0<y<3.5$& $3.5<y<4.0$& $4.0<y<4.5$\\
\hline
$ 0- 1$&$6.24\pm 0.35  $& $4.89\pm 0.10  $& $3.45\pm 0.11  $& $3.31\pm 0.09  $& $4.66\pm 0.17$\\
 $ 1- 2$&$5.58\pm 0.18  $& $4.30\pm 0.07  $& $2.94\pm 0.06  $& $2.55\pm 0.03  $& $2.82\pm 0.12$\\
 $ 2- 3$&$4.88\pm 0.14  $& $3.47\pm 0.06  $& $1.97\pm 0.04  $& $1.52\pm 0.06  $& $1.65\pm 0.13$\\
 $ 3- 4$&$4.77\pm 0.14  $& $3.39\pm 0.06  $& $1.94\pm 0.04  $& $1.17\pm 0.07  $& $1.13\pm 0.15$\\
 $ 4- 5$&$4.68\pm 0.14  $& $3.34\pm 0.08  $& $1.97\pm 0.04  $& $1.20\pm 0.07  $& $0.73\pm 0.14$\\
 $ 5- 6$&$4.43\pm 0.12  $& $3.28\pm 0.10  $& $2.03\pm 0.06  $& $1.42\pm 0.06  $& $0.75\pm 0.14$\\
 $ 6- 7$&$4.21\pm 0.09  $& $3.03\pm 0.12  $& $2.05\pm 0.08  $& $1.57\pm 0.04  $& $0.77\pm 0.14$\\
 $ 7- 8$&$3.88\pm 0.04  $& $2.81\pm 0.15  $& $1.98\pm 0.10  $& $1.69\pm 0.05  $& $0.74\pm 0.14$\\
 $ 8- 9$&$3.59\pm 0.15  $& $2.65\pm 0.20  $& $1.81\pm 0.11  $& $1.65\pm 0.11  $& $1.01\pm 0.13$\\
 $ 9-10$&$3.53\pm 0.18  $& $2.44\pm 0.24  $& $1.81\pm 0.15  $& $1.68\pm 0.16  $& $1.17\pm 0.14$\\
 $10-11$&$3.39\pm 0.27  $& $2.30\pm 0.26  $& $1.88\pm 0.22  $& $1.73\pm 0.26  $& $1.26\pm 0.14$\\
 $11-12$&$3.09\pm 0.32  $& $2.18\pm 0.38  $& $1.47\pm 0.18  $& $1.65\pm 0.27  $& $1.35\pm 0.43$\\
 $12-13$&$3.25\pm 0.45  $& $1.65\pm 0.32  $& $1.93\pm 0.36  $& $1.49\pm 0.26  $& $1.48\pm 0.21$\\
 $13-14$&$2.72\pm 0.58  $& $1.68\pm 0.32  $& $1.71\pm 0.38  $& $1.17\pm 0.27  $& $1.36\pm 0.51$\\
 \hline
\end{tabular}\label{tab:PolarizationEffect20}
}
\end{table}

\clearpage
\newpage
\cleardoublepage
\newpage

\centerline{\large\bf LHCb collaboration}
\begin{flushleft}
\small
R.~Aaij$^{38}$, 
B.~Adeva$^{37}$, 
M.~Adinolfi$^{46}$, 
A.~Affolder$^{52}$, 
Z.~Ajaltouni$^{5}$, 
S.~Akar$^{6}$, 
J.~Albrecht$^{9}$, 
F.~Alessio$^{38}$, 
M.~Alexander$^{51}$, 
S.~Ali$^{41}$
% ...

{\footnotesize \it
% ...
$ ^{5}$Clermont Universit\'{e}, Universit\'{e} Blaise Pascal, CNRS/IN2P3, LPC, Clermont-Ferrand, France\\
$ ^{6}$CPPM, Aix-Marseille Universit\'{e}, CNRS/IN2P3, Marseille, France\\
% ...
$ ^{9}$Fakult\"{a}t Physik, Technische Universit\"{a}t Dortmund, Dortmund, Germany\\
% ...
$ ^{37}$Universidad de Santiago de Compostela, Santiago de Compostela, Spain\\
$ ^{38}$European Organization for Nuclear Research (CERN), Geneva, Switzerland\\
% ...
$ ^{41}$Nikhef National Institute for Subatomic Physics, Amsterdam, The Netherlands\\
% ...
$ ^{46}$H.H. Wills Physics Laboratory, University of Bristol, Bristol, United Kingdom\\
% ...
$ ^{51}$School of Physics and Astronomy, University of Glasgow, Glasgow, United Kingdom\\
$ ^{52}$Oliver Lodge Laboratory, University of Liverpool, Liverpool, United Kingdom\\
% ...
}
\end{flushleft}

\end{document}